\documentclass[submission, PhysCodeb]{SciPost}

\newcommand\cmsout{\bgroup\markoverwith{\textcolor{blue}{\rule[0.5ex]{2pt}{0.4pt}}}\ULon}
\newcommand\mssout{\bgroup\markoverwith{\textcolor{olive}{\rule[0.5ex]{2pt}{0.4pt}}}\ULon}

\hypersetup{
    colorlinks,
    linkcolor={red!50!black},
    citecolor={blue!50!black},
    urlcolor={blue!80!black}
}

\usepackage[bitstream-charter]{mathdesign}
\DeclareSymbolFont{usualmathcal}{OMS}{cmsy}{m}{n}
\DeclareSymbolFontAlphabet{\mathcal}{usualmathcal}

\usepackage{listings}
\usepackage{soul}
\usepackage[finalizecache=false,frozencache=true,cachedir=minted-cache]{minted} 
\makeatletter
\let\c@lofdepth\relax
\let\c@lotdepth\relax
\makeatother
\usepackage{subfigure}
\usepackage{courier}
\usepackage{comment}
\usepackage[capitalise]{cleveref}
\crefname{section}{Sec.}{Sections}
\Crefname{section}{Section}{Sections}
\crefname{lstlisting}{Listing}{listings}
\Crefname{lstlisting}{Listing}{Listings}
\crefname{listing}{Listing}{listings}
\Crefname{listing}{Listing}{Listings}

\usepackage{letltxmacro}
\newcommand*{\SavedLstInline}{}
\LetLtxMacro\SavedLstInline\lstinline
\DeclareRobustCommand*{\lstinline}{%
  \ifmmode
    \let\SavedBGroup\bgroup
    \def\bgroup{%
      \let\bgroup\SavedBGroup
      \hbox\bgroup
    }%
  \fi
  \SavedLstInline
}

\definecolor{codegreen}{rgb}{0,0.6,0}
\definecolor{codegray}{rgb}{0.5,0.5,0.5}
\definecolor{codepurple}{rgb}{0.58,0,0.82}
\definecolor{backcolour}{rgb}{0.95,0.95,0.95}

\lstdefinestyle{mystyle}{
    backgroundcolor=\color{backcolour},   
    commentstyle=\color{codegreen},
    keywordstyle=\color{magenta},
    numberstyle=\tiny\color{codegray},
    stringstyle=\color{codepurple},
    basicstyle=\ttfamily\footnotesize,
    breakatwhitespace=false,         
    breaklines=true,                 
    captionpos=b,                    
    keepspaces=true,                 
    showspaces=false,                
    showstringspaces=false,
    showtabs=false,                  
    tabsize=2,
    frame=single,
    frameround = tttt,
    rulecolor=\color{gray}
}

\graphicspath{ {figures/} }

\begin{document}

\begin{center}{\Large \textbf{
The Cytnx Library for Tensor Networks\\
}}\end{center}


\begin{center}
Kai-Hsin Wu\textsuperscript{1,$^\ast$},
Chang-Teng Lin \textsuperscript{2,3},
Ke Hsu\textsuperscript{2.3},
Hao-Ti Hung\textsuperscript{2,3},
Manuel Schneider \textsuperscript{4},
Chia-Min Chung \textsuperscript{5,6,7},
Ying-Jer Kao \textsuperscript{2,3,8$^\dagger$},
Pochung Chen \textsuperscript{6,9,10$^\star$}
\end{center}

\begin{center}
{\bf 1} Department of Physics, Boston University, Boston 02215, United States of America
\\
{\bf 2} Department of Physics, National Taiwan University, Taipei 10607, Taiwan
\\
{\bf 3} Center for Theoretical Physics, National Taiwan University, Taipei 10607, Taiwan
\\
{\bf 4} Institute of Physics, National Yang Ming Chiao Tung University, Hsinchu 30010, Taiwan
\\
{\bf 5} Department of Physics, National Sun Yat-Sen University, Kaohsiung 80424, Taiwan
\\
{\bf 6} Physics Division, National Center for Theoretical Sciences, Taipei 10617, Taiwan
\\
{\bf 7} Center for Theoretical and Computational Physics, National Sun Yat-Sen University, Kaohsiung 80424, Taiwan
\\
{\bf 8} Center for Quantum Science and Technology,  National Taiwan University, Taipei 10607, Taiwan
\\
{\bf 9} Department of Physics, National Tsing Hua University, Hsinchu 30013, Taiwan
\\
{\bf 10} Frontier Center for Theory and Computation, National Tsing Hua University, Hsinchu 30013, Taiwan
\\
${}^\dagger$ {\small \sf yjkao@phys.ntu.edu.tw}
${}^\star$ {\small \sf pcchen@phys.nthu.edu.tw}
${}^\ast$ {\small \sf kaihsinwu@gmail.com}
\end{center}

\begin{center}
\today
\end{center}

\section*{Abstract}

We introduce a tensor network library designed for classical and quantum physics simulations called Cytnx (pronounced as \textit{sci-tens}). This library provides an almost identical interface and syntax for both C++ and Python, allowing users to effortlessly switch between the two languages. Aiming at a quick learning process for new users of tensor network algorithms, the interfaces resemble the popular Python scientific libraries like NumPy, Scipy, and PyTorch. Not only multiple global Abelian symmetries can be easily defined and implemented, Cytnx also provides a new tool called \lstinline{Network} that allows users to store large tensor networks and perform tensor network contractions in an optimal order automatically. With the integration of cuQuantum, tensor calculations can also be executed efficiently on GPUs. 
We present benchmark results for tensor operations on both devices, CPU and GPU. 
We also discuss features and higher-level interfaces to be added in the future.

\vspace{10pt}
\noindent\rule{\textwidth}{1pt}
\tableofcontents\thispagestyle{fancy}
\noindent\rule{\textwidth}{1pt}
\vspace{10pt}

\section{Introduction}\label{sec:intro}
Tensor networks have been well established as an important tool in various applications including, but not limited to, many-body systems in condensed matter physics\cite{ORUS2014117,SCHOLLWOCK201196,RevModPhys.93.045003,Verstraete2023,introduction_ManyParticlePhysics,introductionOverview,openQuantumSystems,McCulloch_2007, Silvi.2019} and for quantum chemistry\cite{doi:10.1146/annurev-physchem-032210-103338,chan2016matrix,quantumChemistry_review,quantumChemistry_review_White,quantumChemistry_DMRG_recentReview,quantumChemistry_DMRG_bookReview,quantumChemistry_postDMRG}. In high energy physics, tensor networks can provide an alternative where established Monte Carlo methods can not be used~\cite{Banuls_2020,HEP_TN_QCD_overview,HEP_TNS_review,HEP_TNS_QS_QC_overview_MariCarmen}. Recently, tensor network methods  have also been applied to machine learning \cite{NIPS2016_5314b967,PhysRevE.107.L012103,Chen_2021,Chen_2022}, computational fluid dynamics~\cite{kiffner2023tensor,2023.Peddinti} and have emerged as an efficient method to simulate quantum circuits~\cite{liao2023simulation,2023fast,tindall2023efficient}.
Therefore, having a tensor library is crucial to accelerate the development of tensor network algorithms. With such a tool,  users can focus on the applications and use state-of-the-art, pre-tested, and optimized tensor operations provided by the library, instead of investing time in implementing and testing basic tensor functionalities. Here, we present a new open-source tensor network library  -- Cytnx~\cite{cytnx,cytnxDoc} (pronounce as \textit{sci-tens})\footnote{The name Cytnx is derived from the following keywords, C++(C), Python(y), and  {T}ensor {N}etworks (tnx).}, which provides an easy-to-use framework for beginners with extensive and powerful functionality  for advanced tensor network applications~\cite{cytnx}.

\subsection{Overview of the Cytnx Design}
There  exist several tensor libraries, each addressing various aspects of the tensor computation~\cite{itensor,MOTOYAMA2022108437,10.21468/SciPostPhysLectNotes.5,syten,roberts2019tensornetwork,TensorKit,gray2018quimb,Kao:2015gb, Weichselbaum:2012gs, Weichselbaum.2020}. Cytnx is highly influenced by the design of Uni10 \cite{Kao:2015gb} with improved API design and performance optimization. In Cytnx, we envision a framework designed and optimized for higher dimensional tensor networks, such as projected entangled pair states (PEPS) or multi-scale entanglement renormalization ansatz (MERA). Compared to the matrix product state (MPS), these tensor networks have higher connectivity, which makes keeping track of tensor indices and connectivity cumbersome. Therefore, we design a \lstinline{Network} class to store the blueprint of a tensor network. This enables  under-the-hood optimizations for tensor contractions. Moreover, the \lstinline{Network} can be reused for further contractions by loading a new set of tensors. 

We make Cytnx easy to interface with Python packages such as NumPy and Scipy and machine learning frameworks such as PyTorch. This allows Cytnx users to take advantage of the available Python libraries supported by the community. Moreover, new users can easily port their existing Python code to take advantage of the high-level design of Cytnx.

For the performance minded users, Cytnx keeps most of the Application Programming Interfaces (APIs) exactly the same for both C++ and Python; therefore, users can quickly prototype their algorithms in Python, and switch to C++ with minimal re-writing of the code base for further optimization. In Cytnx, unlike in many libraries where the Python interface serves as a wrapper,  both C++ and Python interfaces have direct access to the underlying tensor containers, giving users greater flexibility.

Cytnx also supports GPU computing. When creating a tensor, Cytnx allows users to specify the device for storing the data. Similar to PyTorch, users can use the same API calls regardless of the devices where the tensors are kept. This allows users to benefit from the GPU acceleration  with minimal changes for heterogeneous computing environments. In addition to the traditional CUDA libraries, Cytnx also incorporates functions from cuQuantum, allowing more efficient usage of GPU when performing tensor contractions.

Another  easy-to-use feature of Cytnx is the ability to access indices of tensors by user-defined labels. Once meaningful index labels are set, the ordering and permutation of the indices in tensors can be handled by  the library automatically. This way, users can focus on their application instead of dealing with the technical details of the tensors and how the information is stored. 

Moreover, Abelian symmetries and quantum numbers can be easily included in tensors. With symmetric tensors, Cytnx ensures that all operations respect the symmetries of the system, and makes use of an efficient block structure of the tensors. Directional bonds are required for symmetric tensors, but can also be defined without being linked to symmetries. Cytnx ensures that only bonds with opposite directions can be contracted, which makes the code more resilient against errors. 

\subsection{How to use this document}
This document aims to give an overview of the functionality and usage of Cytnx. Users can find more information about the library in the user guide~\cite{cytnxDoc}. Also, an API documentation of the classes and functions in Cytnx is available online~\cite{cytnxAPI}. 

Here we present  example codes in Python, with some comments on C++ in \cref{sec:design}. Advanced users can refer to the online documentation~\cite{cytnxDoc} for C++ example codes. 
All code examples in this document assume a successful installation of Cytnx. In order to run the Python examples, the user needs to insert this line before the example code:
\begin{lstlisting}[language=python]
import cytnx
\end{lstlisting}
Building from the source code and Conda installation of Cytnx are both available. See the online documentation for detailed installation instructions~\cite{cytnxDoc}. The explanations here refer to version 1.0.0 of the Cytnx library.

This article is organized as follows. In \cref{sec:ex} we introduce the graphical tensor notation and provide a simple example showing how to use Cytnx to implement a tensor contraction. \Cref{sec:design} explains basic concepts and general naming conventions of the library. \cref{sec:bond,sec:uniten,sec:symm} show the creation and manipulation of tensors and their indices, where \cref{sec:symm} introduces symmetries and symmetric tensors. Tensor contractions are covered in detail in \cref{sec:contraction,sec:network}. In \cref{sec:decomp} we explain how to decompose tensors and use linear algebra functions. The performance of Cytnx is benchmarked against ITensor\cite{itensor} and PEPS-Torch~\cite{peps-torch} in \cref{sec:benchmark}. Finally, we give a summary and discuss the future road map of Cytnx in \cref{sec:summary}.

\section{Sneak preview: tensors and tensor contraction}
\label{sec:ex}

In this sneak preview, we start from the standard mathematical symbolic notation of a tensor and introduce the graphical notation. With this, we show how the tensor contraction can be represented transparently. Finally, we demonstrate how to convert the graphical notation into corresponding code in Cytnx and perform tensor contractions.

A graphical notation for tensors and tensor networks is very useful to visualize the tensor and the connectivity in tensor network algorithms. We introduce our graphical convention in the following.
A tensor can be thought of as a multi-dimensional array. For example, a rank-0 tensor is a scalar, a rank-1 tensor is a vector and a rank-2 tensor is a matrix. Mathematically, a rank-$N$ tensor can be written in symbolic notation as:
\begin{equation}
T_{i_1,i_2 \cdots, i_N},
\end{equation}
where $i_1,i_2,\cdots,i_N$ are the indices of the tensor. This notation is standard, but for a tensor network that involves a series of tensor contractions, the expression soon becomes complicated and difficult to read. This is where the graphical tensor notation is useful.

\begin{figure}[ht]
    \centering
    \includegraphics[scale=0.45]{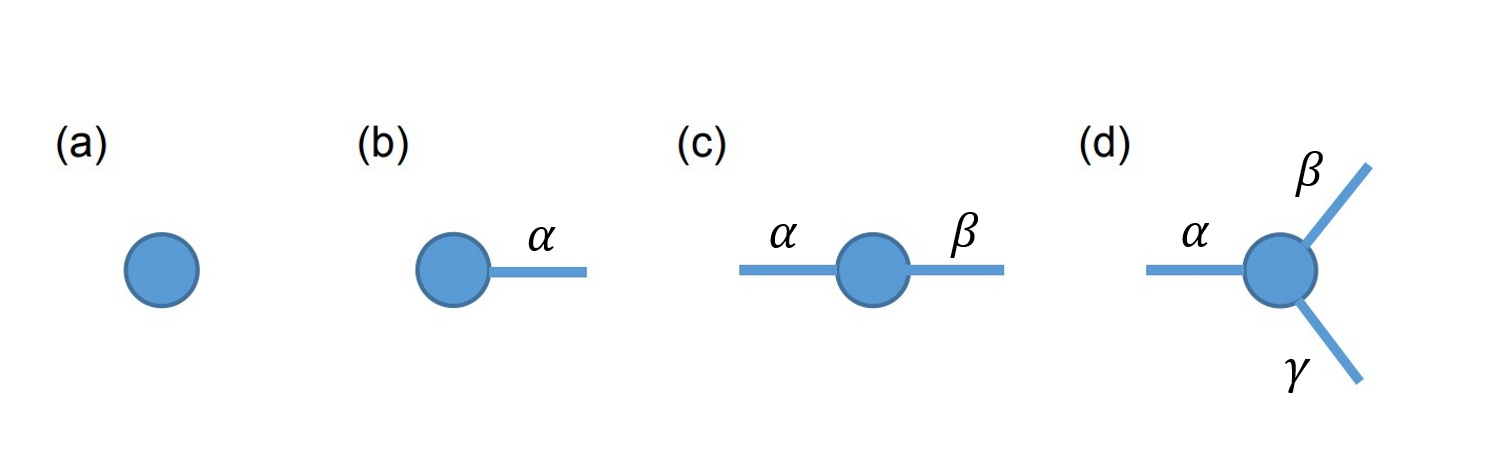}
    \caption{Graphical notation of tensors. (a) scalar; (b) vector; (c) matrix; (d) rank-3 tensor.}
    \label{fig:graphical_notation}
\end{figure}

In the graphical notation, each tensor is represented by a node (sometimes it is also called a vertex in graph theory) and several bonds (legs) attached to it. 
Each bond corresponds to an index of the tensor. The number of bonds equals the rank of a tensor. \cref{fig:graphical_notation} shows examples of several tensors.

\begin{figure}[ht]
    \centering
    \includegraphics[scale=0.4]{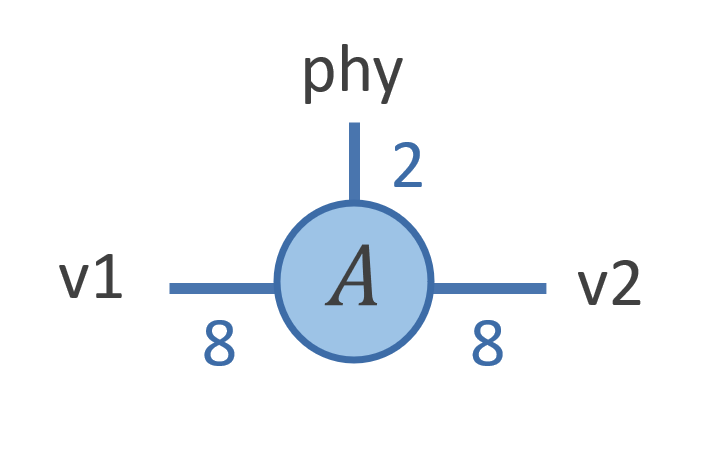}
    \caption{Visualization of the tensor constructed in \cref{code:rank3tensor}, with the dimension and label of each leg.}
    \label{fig:tensor}
\end{figure}

The elements of a tensor and its shape can be stored in a multidimensional array, as implemented in \lstinline{NumPy.array} or \lstinline{PyTorch.tensor}. However, for tensor network calculations it is also important to store metadata such as the name, labels, and quantum number information, among others. Similar to the UNI10\cite{Kao:2015gb} library, in Cytnx, the \lstinline{UniTensor} object is used to represent a tensor with all the necessary metadata, while the \lstinline{Tensor} object contains only the elements and the shape, similar to \lstinline{NumPy.array} and \lstinline{PyTorch.tensor}.
A user will typically create and manipulate \lstinline{UniTensor} objects with or without symmetries. This is discussed in more detail in \cref{sec:uniten,sec:uniten_symm} respectively.

Consider a rank-3 tensor \lstinline{A} as shown in \cref{fig:tensor} with indices labeled by ``v1'',``phys'',``v2'' and dimensions $8,2,8$ respectively. 
The following code example (Listing \ref{code:rank3tensor}) shows how to create a \lstinline{UniTensor} to represent such a tensor and initialize all the tensor elements to one. 

\begin{lstlisting}[language=python,
    caption={Create a rank-3 tensor. Two styles are shown here. The first syntax is compatibility with C++ and is recommended. In the second syntax we use keyword arguments in Python.},
    label={code:rank3tensor}]
A = cytnx.UniTensor.ones([8,2,8]).relabel(["v1","phy","v2"])\
                         .set_name("A")
# or, using argument names:
A = cytnx.UniTensor.ones([8,2,8], labels=["v1","phy","v2"], name="A")
\end{lstlisting}

\begin{figure}[ht]
    \centering
    \includegraphics[scale=1]{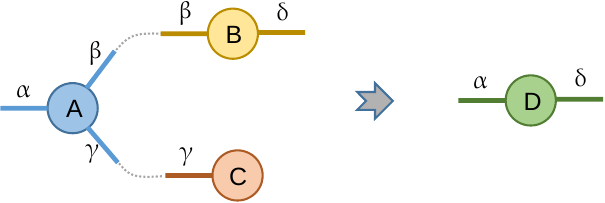}
    \caption{Graphical notation for the tensor contraction in \cref{eq:contract}}
    \label{fig:contract}
\end{figure}

One of the most important operations in tensor network algorithms is the tensor contraction. It is an operation that sums the product of tensor elements over some common indices. For example, consider the contraction of three tensors $A$, $B$, and $C$:
\begin{equation}
    D_{\alpha\delta} = \sum_{\beta\gamma} A_{\alpha\beta\gamma}B_{\beta\delta}C_{\gamma}
    \label{eq:contract}
\end{equation}
In this example, $A$ has three indices labeled by $\alpha$, $\beta$, and $\gamma$. $B$ has two indices labeled by $\beta$ and $\delta$, and $C$ has one index labeled by $\gamma$. \Cref{eq:contract} defines a tensor contraction that sums over the common indices $\beta$ and $\gamma$, and the result is a tensor $D$ with two indices labeled by $\alpha$ and $\delta$. The corresponding graphical notation is shown in \cref{fig:contract}. Indices that connect two nodes are assumed to be summed over. The dashed lines indicate the connection of two bonds with the same indices to be summed.

Cytnx provides several methods for tensor contractions.
One possibility is to use \lstinline{Contract} to perform a simple contraction of two or more tensors. It sums over common labels on these tensors via utilizing the metadata. \lstinline{Contract} is easy to use and the code is readable as demonstrated in the following example, where we show how to contract three tensors \lstinline{A}, \lstinline{B} and \lstinline{C}. We note in passing that users need to take care of the correct index labeling in advance. See \cref{sec:contraction} for details.

\begin{lstlisting}[language=python,
                   caption={Using \lstinline{Contract} to perform tensor contractions.},
                   label={code:contract}]
# Initialize UniTensors and their labels
A = cytnx.UniTensor.ones([2,2,2])\
                       .set_name("A").relabel(["alpha","beta","gamma"])
B = cytnx.UniTensor.ones([2,2])\
                       .set_name("B").relabel(["beta","delta"])
C = cytnx.UniTensor.ones([2])\
                       .set_name("C").relabel(["gamma"])
# Do the contractions
AB = cytnx.Contract([A, B])
D = cytnx.Contract([AB, C])
# Similarly, in one step:
D = cytnx.Contract([A,B,C])
\end{lstlisting}

In a realistic algorithm, the number of tensors in a tensor network contraction can be quite many. The overall contraction is typically implemented in a sequence of binary contractions of tensors. However, the order of these binary contractions is not unique, and the computational costs depend strongly on the contraction order. In the example above, when one contracts three tensors in one step, Cytnx will first search for the optimal contraction order. While the contraction order does not affect the overall efficiency of an algorithm for small tensors, it is essential to use the optimal contraction order for more complex tensor network contractions and large bond dimensions.

In Cytnx, we introduce a powerful \lstinline{Network} contraction method, which allows one to define a tensor contraction in an abstract way. The actual tensors to be contracted do not need to have a certain index order or specific labels. Consequently, the same \lstinline{Network} object can be reused for successive contractions. Moreover, the contraction order can be optimized automatically. When a \lstinline{Network} contraction is launched, it computes an optimized contraction order by a dynamic programming algorithm. The optimized order can be saved and re-used for the subsequent contraction of the same \lstinline{Network} with different tensors.

Using \lstinline{Network} to perform a tensor contraction includes three steps: 1) define a network, 2) load the tensors by \lstinline{PutUniTensor}, and 3) \lstinline{launch} the contraction. In what follows we briefly introduce how to use \lstinline{Network}. Further details and a more complicated example of a contraction from an actual tensor network algorithm are discussed in \cref{sec:network}.

\begin{figure}[ht]
    \centering
    \includegraphics[scale=0.5]{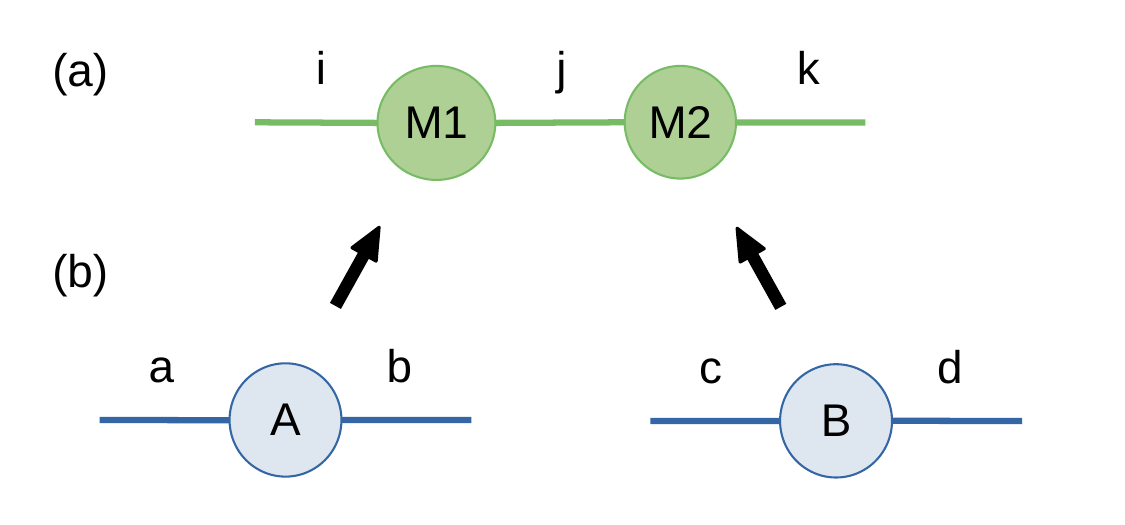}
    \caption{(a) Graphical notation for a matrix-matrix multiplication. (b) Specifying each tensor by an actual tensor object.}
    \label{fig:matdot}
\end{figure}

In this example, the goal is to use a \lstinline{Network} contraction to multiply the tensors $M_1$ and $M_2$ and sum over the common index $j$, as shown in \cref{fig:matdot}. One first creates a file \lstinline{matmul.net} with content as shown in \cref{code:netfile}.
Note that a \lstinline{Network} is just a blueprint of the tensor contractions.
In the code example (\cref{code:MM}) below, we first create a \lstinline{Network} object using the file \lstinline{matmul.net}. 
Then, we initialize two rank-2 \lstinline{UniTensor} \lstinline{A} and \lstinline{B}.
We specify the \lstinline{UniTensor} objects that are actually to be contracted, by loading them into the \lstinline{Network} and matching the indices by their labels.
In this step, the indices have to be assigned in the right order.
In particular, we specify \lstinline{A} (\lstinline{B}) as \lstinline{M1} (\lstinline{M2}) in the network. The indices \lstinline{a} and \lstinline{b} (\lstinline{c} and \lstinline{d}) are associated to the indices \lstinline{i} and \lstinline{j} (\lstinline{j} and \lstinline{k}). After loading the tensors, we launch the contraction and receive the result as a tensor \lstinline{AB} whose indices are labeled by \lstinline{i} and \lstinline{k}.

Note that the indices \lstinline{a}, \lstinline{b}, \lstinline{c}, \lstinline{d} of the actual \lstinline{UniTensor} objects are independent of the indices \lstinline{i}, \lstinline{j}, \lstinline{k} defined in the network. With this design, one can implement an algorithm with meaningful index names, independent of the actual tensors to be contracted later on. On the other hand, a \lstinline{Network} can be reused and loaded with different tensors and label conventions. As shown in code example \cref{code:M2M1}, 
the same \lstinline{Network} object can be used to perform the multiplication \lstinline{BA} instead of \lstinline{AB} without relabeling.

\begin{lstlisting}[language=python, caption={Network file "matmul.net" defining a matrix-matrix multiplication.}, label={code:netfile}]
# Network definition saved in file "matmul.net"
M1:  i, j
M2:  j, k
TOUT: i, k
\end{lstlisting}

\begin{lstlisting}[language=python, 
                   caption={Network contraction for matrix-matrix multiplication \lstinline{AB}.},
                   label={code:MM}]
# Create a Network object from (*\color{codegreen}\cref{code:netfile}*)
net = cytnx.Network("matmul.net")
# Initialize the tensors and their labels
A = cytnx.UniTensor.ones([2,2]).relabel(["a","b"]).set_name("A")
B = cytnx.UniTensor.ones([2,2]).relabel(["c","d"]).set_name("B")
# Put the tensors into the network
net.PutUniTensor("M1", A, ["a", "b"])
net.PutUniTensor("M2", B, ["c", "d"])
# Launch the contraction
AB = net.Launch()
\end{lstlisting}

\begin{lstlisting}[language=python, 
                   caption={Network contraction for matrix-matrix multiplication \lstinline{BA}.}, 
                   label={code:M2M1}]
# Supposes net is a Network that is already created as in (*\color{codegreen}\cref{code:MM}*)
net.PutUniTensor("M1", B, ["c", "d"])
net.PutUniTensor("M2", A, ["a", "b"])
BA = net.Launch()
\end{lstlisting}

\section{Conventions and object behavior}
\label{sec:design}

In this section we discuss the naming conventions of Cytnx and the object behavior.

\subsection{\label{sec:naming}Naming convention}

Generally, the function names in Cytnx follow these rules:
\begin{itemize}
    \item If a function is acting on objects, its name starts with a capital letter. 
        For example, \lstinline{Contract(A,B)} contracts two tensors \lstinline{A} and \lstinline{B}.
    \item If a function is a member function or a generating function, it starts with a lowercase letter. For example, \lstinline{UniTensor.permute([1,0])} permutes the indices of a \lstinline{UniTensor} object, where \lstinline{permute} is a member function. \lstinline{uT=cytnx.UniTensor.zeros([2,2])} generates a \lstinline{UniTensor uT}, so \lstinline{zeros} is a generating function for the \lstinline{UniTensor} object.
    \item Types defined in Cytnx always start with capital letters. For example \lstinline{bd=cytnx.Bond(10)} creates an object of \lstinline{Bond} type.
    \item If a function name does not end with an underscore, it will not change the input object. For example, \lstinline{B=A.permute([1,0])} creates a new object \lstinline{B} that is \lstinline{A} after permutation, while \lstinline{A} remains unpermuted.
    \item A function that ends with an underscore indicates one of the following scenarios:
    \begin{enumerate}
        \item The input object will be changed. For example, \lstinline{cytnx.linalg.Pow_(A,2)} replaces \lstinline{A} by $\lstinline{A}^2$. Compare with \lstinline{Asq = cytnx.linalg.Pow(A,2)}, which creates a new tensor $\lstinline{Asq} = \lstinline{A}^2$ and leaves \lstinline{A} unchanged.
        \item For a member function of a class, it is an in-place operation. For example, \lstinline{A.permute_([1,0])} changes \lstinline{A} by permuting its indices. Member functions which end with an underscore return a reference to the object.
    \end{enumerate}
   
\end{itemize}

\subsection{Object behavior}

In order to maintain a similar syntax and API in C++ and Python, and to reduce redundant memory allocations, all objects in Cytnx are references. The only exceptions are \lstinline{Accessor} and \lstinline{LinOp}. In Python, all objects are references by default already. On the C++ side, however, the reference character is implemented by Cytnx explicitly. For example, consider a \lstinline{UniTensor} object in Python and C++ respectively.
\begin{lstlisting}[language=Python, 
                   caption={Python example.}]
# Python:
A = cytnx.UniTensor.zeros([2,3])
B = A
print(B is A) # Output: True
\end{lstlisting}
\begin{lstlisting}[language=C++, 
                            caption={C++ example.}]
// C++
auto A = cytnx::UniTensor::zeros({2,3});
auto B = A;
cout << cytnx::is(B,A) << endl; // Output: true
\end{lstlisting}
It is apparent that a line to line translation between C++ and Python can be done with minimal effort. To achieve this, we implement the assignment "\lstinline{=}" on the C++ side such that $B$ is a reference to $A$. This also avoids copying of data. Additionally, Cytnx implements the \lstinline{is} clause in the C++ API. This way, one can use \lstinline{is} to check if two objects reference to the same memory. These additional implementations ensure that C++ and Python can have the same behavior. We note that this Pythonic design for the C++ API may not comply with common C++ behavior, and caution is necessary when using the C++ API.

In case where a copy of an object is needed, one can use \lstinline{clone} as shown below.
The functions \lstinline{is} and \lstinline{clone} support all objects of Cytnx.

\begin{lstlisting}[language=Python, caption={Clone an object in Python.}]
# Python
A = cytnx.UniTensor.zeros([2,3])
B = A
C = A.clone()
print(B is A) # Output: True 
print(C is A) # Output: False
\end{lstlisting}

\begin{lstlisting}[language=C++, caption={Clone an object in C++.}]
// C++
auto A = cytnx::UniTensor::zeros({2,3});
auto B = A;
auto C = A.clone();
cout << cytnx::is(B,A) << endl; // Output: true 
cout << cytnx::is(C,A) << endl; // Output: false
\end{lstlisting}

Member functions which only change the metadata of a tensor (e.g. \lstinline{permute}, \lstinline{reshape}, \lstinline{relabel}) create a new tensor with independent metadata. However, the new and old tensors share the same memory for the tensor elements. Thus, changing the elements of the new tensor will change the elements of the old one as well and vice versa. Changes to the metadata do not affect the other tensor though. If the tensor elements of the old and new tensors shall be independent of each other, use \lstinline{clone}. Member functions which end with an underscore (e.g. \lstinline{permute_}, \lstinline{reshape_}, \lstinline{relabel_}) change the tensor directly and return a reference to it, so no independent metadata is created. For example, we permute a tensor and change an element to see the behavior of the functions:
\begin{lstlisting}[language=Python, caption={References and copies of tensor elements and metadata when using member functions.}]
# Python
A = cytnx.UniTensor.zeros([2,3])
B = A.permute([1,0]) # A and B share tensor elements, different metadata
B[0,1] = 1.
C = A.clone().permute([1,0]) # C is independent of A
C[0,1] = 2.
print(A) # Original shape, element A[1,0] == 1.
print(B) # Permuted shape, element B[0,1] == 1.
print(C) # Permuted shape, element C[0,1] == 2.
D = A.permute_([1,0])   # D is a reference to A
print(A) # Permuted shape, element A[0,1] == 1.
print(D is A) # Output: True 
\end{lstlisting}

\section{Non-symmetric Bonds}
\label{sec:bond}

In Cytnx, a \lstinline{Bond} represents an index (or axis, as in the NumPy convention) of a tensor. A \lstinline{Bond} carries information about the dimension of a given index, which is the number of possible values the index can take. Moreover, bonds can have a direction, distinguishing incoming and outgoing indices of a tensor. When symmetries shall be preserved, the \lstinline{Bond} objects carry additional information like quantum numbers associated with the symmetries. Bonds with symmetries are explained in \cref{sec:bond_symm}. Here, we focus on bonds without symmetry properties. \lstinline{Bond} objects can be used to initialize a \lstinline{UniTensor}, see \cref{sec:uniten}.

For a bond without quantum numbers, one can define a \lstinline{Bond} object simply by specifying its dimension. Additionally, bonds can be directed and either be incoming or outgoing. 
If only the bond dimension is specified, the type is always \lstinline{REGULAR} and thus un-directed. A directed bond can be created with \lstinline{cytnx.BD_IN} or \lstinline{cytnx.BD_OUT}.
In this case, an outgoing bond can only be contracted with an incoming bond with the same size. When using symmetries, the bonds must be directed.

\begin{lstlisting}[language=Python, caption={Create an undirected bond or an outgoing bond.}]
# Create an undirected bond with dimension 10
bond = cytnx.Bond(10)
print(bond)     # Output: Dim = 10 |type: REGULAR
# Create a directed (incoming) bond with dimension 20
bondIn = cytnx.Bond(20, cytnx.BD_IN)
print(bondIn)   # Output: Dim = 20 |type: |IN (KET)> 
\end{lstlisting}

\paragraph{Combining bonds.\label{sec:combining_nonsymm}}
One can combine two bonds to create a new bond.
The dimension of the new bond is the product of the dimensions of the input bonds,
as illustrated below (\cref{code:cd}):
\begin{lstlisting}[language=python, caption={Combine two bonds to create a new bond.}, label={code:cd}]
bond1 = cytnx.Bond(10)
bond2 = cytnx.Bond(2)
# Combine bond1 and bond2
bond12 = bond1.combineBond(bond2)
print(bond1)    # Output: Dim = 10 |type: REGULAR
print(bond2)    # Output: Dim = 2 |type: REGULAR
print(bond12)   # Output: Dim = 20 |type: REGULAR
\end{lstlisting}
Moreover, an existing bond can be modified by merging another bond:
\begin{lstlisting}[language=python, caption={Merge one bond into another bond.}, label={code:merge_bd}]
bond1 = cytnx.Bond(10)
bond2 = cytnx.Bond(2)
# Merge bond2 into bond1
bond1.combineBond_(bond2)
print(bond1)    # Dim = 20 |type: REGULAR
\end{lstlisting}
It is also possible to combine several bonds to create a new bond,
or to modify an existing bond by merging several bonds. 
In this case, a list of \lstinline{Bond} objects is used as the argument,
as illustrated below (\cref{code:bonds}):
\begin{lstlisting}[language=python, 
    caption={Merge multiple bonds.},
    label={code:bonds}]
bond1 = cytnx.Bond(2)
bond2 = cytnx.Bond(2)
bond3 = cytnx.Bond(2)
# Merge bond2 and bond 3 with bond1
bond123 = bond1.combineBond([bond2,bond3])
print(bond123)  # Output: Dim = 8 |type: REGULAR
bond1.combineBond_([bond2,bond3])
print(bond1)    # Output: Dim = 8 |type: REGULAR
\end{lstlisting}

\begin{figure}[ht]
  \centering
  \includegraphics[scale=0.5]{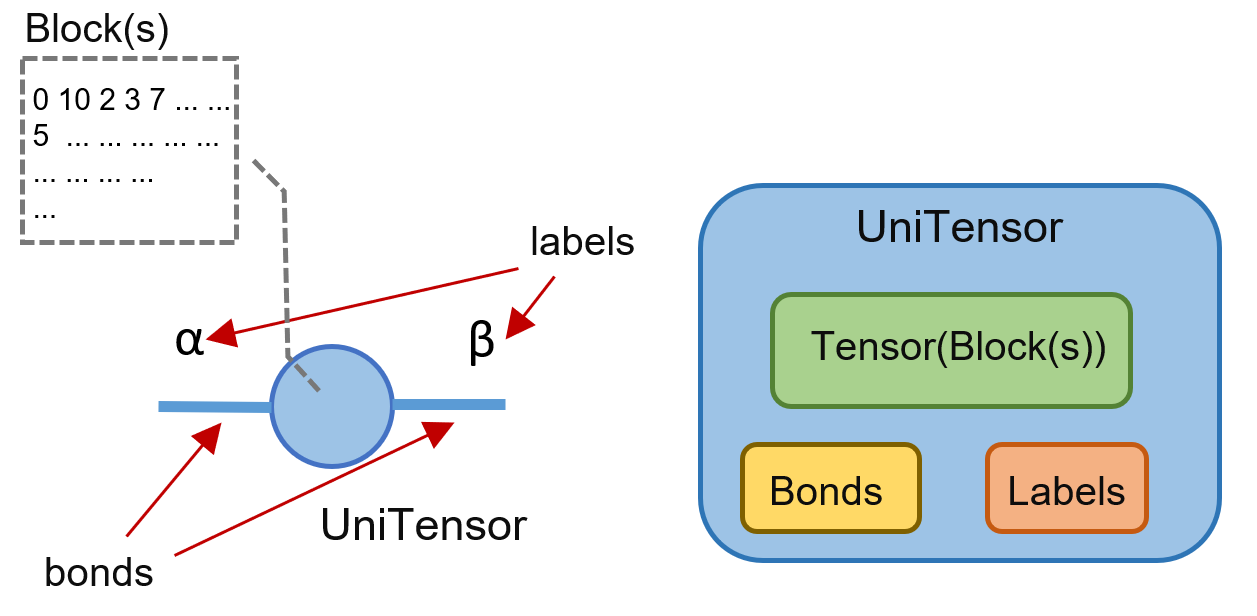}
  \caption{Structure of a \lstinline{UniTensor}.}
  \label{fig:structureUniTensor}
\end{figure}

\section{Non-Symmteric UniTensor}
\label{sec:uniten}

\lstinline{UniTensor} is the central object in Cytnx. 
It is designed to represent all necessary information of a tensor so that tensor manipulations and calculations can be easily done. 
As indicated in \cref{fig:structureUniTensor}, a \lstinline{UniTensor} consists of three parts: Block(s), Bond(s), and Label(s). 
Additionally, each tensor can be given a name.
Blocks contain the value of the tensor elements. Internally they are represented by \lstinline{Tensor} objects (see \cref{sec:tensor}), which are very similar to NumPy.array or PyTorch.tensor. However, in typical situations the user does not need to interact directly with \lstinline{Tensor} objects. 
Bonds are represented by \lstinline{Bond} objects that define dimensions of the indices and can be directed as incoming or outgoing (see \cref{sec:bond}). 

\subsection{Creating a non-symmetric UniTensor}

\begin{figure}
    \centering
    \includegraphics[scale=0.3]{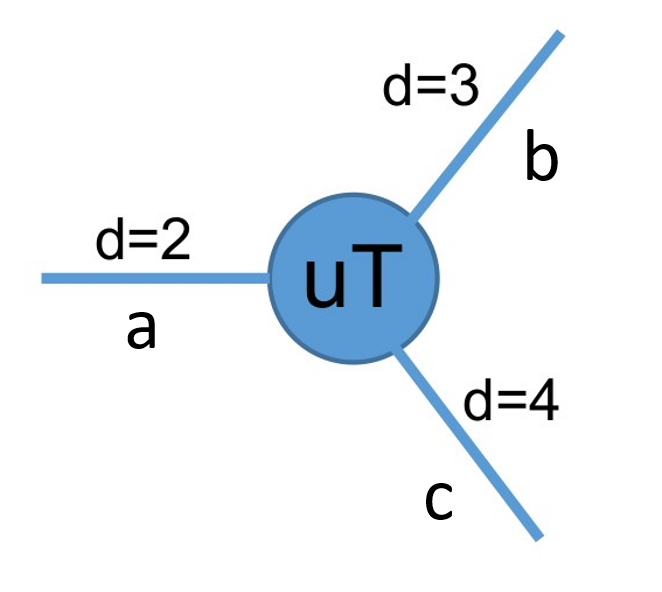}
    \caption{Graphical representation of a tensor. Each line corresponds to one index. The indices have the labels "a", "b" and "c" with dimensions d=2, d=3 and d=4 respectively. "uT": tensor name; \cref{code:UniTensorCreation} shows how to create such a tensor with Cytnx.}
    \label{fig:untag}
\end{figure}

One can create a \lstinline{UniTensor} through generators such as \lstinline{zero}, \lstinline{ones} or \lstinline{eye}. The first argument provides shape information, which is used to construct the \lstinline{Bond} objects and to determine the rank -- the number of tensor indices. Labels can be specified when creating a \lstinline{UniTensor}, otherwise they are set to be \lstinline{"0"}, \lstinline{"1"}, \lstinline{"2"}, \dots by default. 
One can also use similar generators to create a \lstinline{Tensor} (\lstinline{T} in \cref{code:UniTensorCreation}), and then convert it to a \lstinline{UniTensor} (\lstinline{uT2} in \cref{code:UniTensorCreation}).

Consider a rank-3 tensor as shown in \cref{fig:untag}. The corresponding \lstinline{UniTensor} can be created by the code example below (\cref{code:UniTensorCreation}).
\begin{lstlisting}[language=python, 
                   caption={Create a \lstinline{UniTensor} by using generators. The elements of \lstinline{uT} and \lstinline{uT2} are initialized in different ways.},
                   label={code:UniTensorCreation}]
# Create a rank-3 UniTensor with shape [2,3,4]
uT = cytnx.UniTensor.ones([2,3,4])\
                            .relabel(["a","b","c"])\
                            .set_name("tensor uT")
# Create a rank-3 Tensor with shape [2,3,4]
T = cytnx.arange(2*3*4).reshape(2,3,4)
# Create a UniTensor from a Tensor
uT2 = cytnx.UniTensor(T).relabel(["a","b","c"]).set_name("tensor uT")   
\end{lstlisting}

Alternatively, we can also create a \lstinline{UniTensor} by using \lstinline{Bond} objects as shown in \cref{code:UniTensorByBond}.
When a \lstinline{UniTensor} is created from \lstinline{Bond} objects, it is initialized with zeros by default.
A \lstinline{Bond} within a \lstinline{UniTensor} can be accessed by \lstinline{UniTensor.bond}.
\begin{lstlisting}[language=python,
                caption={Create a \lstinline{UniTensor} from \lstinline{Bond} objects.}, 
                label={code:UniTensorByBond}]
# Define bonds by their dimensions
bond1 = cytnx.Bond(2)
bond2 = cytnx.Bond(3)
bond3 = cytnx.Bond(4)
# Create a UniTensor by bonds
uT3 = cytnx.UniTensor([bond1,bond2,bond3])\
                      .relabel(["a","b","c"]).set_name("uT3")
# or, alternative initialization:
uT4 = cytnx.UniTensor([bond1,bond2,bond3], labels=["a","b","c"],\
                      name="uT4")
# Get a bond from a UniTensor
bonda = uT3.bond("a")
\end{lstlisting}
In this example we show two ways to initialize the attributes of the \lstinline{UniTensor}. We can either use the methods \lstinline{relabel} to define the labels and \lstinline{set_name} to set the name of a tensor as for \lstinline{uT2} in \cref{code:UniTensorByBond}. Or we set these properties with the arguments \lstinline{labels} and \lstinline{name} in the initialization method as for \lstinline{uT3} in \cref{code:UniTensorByBond}. We recommend the former way where methods are used subsequently. This way, the code can be ported between Python and C++ easily. In some cases (for example in \cref{code:UniTensorCreation}), the tensor has to be reshaped as well before the labels can be set, which is straightforward if the properties of the tensor are set by methods instead of initializer arguments. We note that the name can be any string and the default value is an empty string. The name is stored in the metadata associated with the \lstinline{UniTensor} object and is independent of the variable name of the object.

\paragraph{Complex elements and data types.}

We can create a \lstinline{UniTensor} with a different data type, for example, with complex elements. This is achieved by setting the \lstinline{dtype} argument when creating a \lstinline{UniTensor}:

\begin{lstlisting}[language=python, 
                   caption={Create a \lstinline{UniTensor} with complex elements.}, 
                   label={code:UniTensorComplex}]
# 1. Create a complex valued UniTensor by initializer
uTc = cytnx.UniTensor.zeros([2,2], dtype=cytnx.Type.ComplexDouble)\
                            .relabel(["a","b"]).set_name("uT complex")
# 2. Create a complex valued UniTensor by Bond
bond1 = cytnx.Bond(2)
bond2 = cytnx.Bond(2)
uTc2 = cytnx.UniTensor([bond1,bond2], dtype=cytnx.Type.ComplexDouble)\
                       .relabel(["bd1","bd2"]).set_name("uT complex 2")
\end{lstlisting}

\paragraph{Random elements.}
A \lstinline{UniTensor} with random elements can be created easily. 
The code example below (\cref{code:UniTensorRandom}) demonstrates how to initialize a \lstinline{UniTensor} 
with elements uniformly in the range [0, 1] or  with elements from a Gaussian distribution with mean value 0 and standard deviation 1 respectively.
\begin{lstlisting}[language=python,
                   caption={Create \lstinline{UniTensor} with random elements.},
                   label={code:UniTensorRandom}]
# Create a random tensor with elements uniformly in the range [0, 1]
rT = cytnx.UniTensor.uniform([2,3,2], low=0., high=1.)
# Create a random tensor with elements from a Gaussian distribution
# with mean value 0 and standard deviation 1
nT = cytnx.UniTensor.normal([2,3,2], mean=0., std=1.)
\end{lstlisting}
Moreover, one can randomize an existing \lstinline{UniTensor}. 
This is particularly useful if a \lstinline{UniTensor} is created from \lstinline{Bond} objects, as it is typically the case for symmetric tensors.
To obtain a \lstinline{UniTensor} with random elements, 
one first creates a \lstinline{UniTensor}, and then applies \lstinline{random.uniform_} or \lstinline{random.normal_} to it, as shown in the example below.

\begin{lstlisting}[language=python,
                   caption={Randomize the elements of a \lstinline{UniTensor}.},
                   label={code:UniTensorRandom_}]
# Supposes uT is a UniTensor that is already created as in (*\color{codegreen}\cref{code:UniTensorCreation}*)
# Randomize the elements of uT uniformly in the range [-1, 1]
cytnx.random.uniform_(uT, low=-1., high=1.)
# Randomize the elements of uT according to a normal distribution
# with mean value 2 and standard deviation 3; set a random number seed
cytnx.random.normal_(uT, mean=2., std=3., seed=10)

\end{lstlisting}
We note that \lstinline{random.uniform_} and \lstinline{random.normal_} work with a \lstinline{UniTensor} with or without symmetry. Real and complex date types of the \lstinline{UniTensor} are supported. For complex data types, the real and imaginary part are randomized independently of each other. A random number seed can be set in all cases above by the argument \lstinline{seed} to ensure reproducibility.

\subsection{Arithmetic operations and tensor operations}

Cytnx supports basic addition (\lstinline{+}), subtraction (\lstinline{-}), multiplication (\lstinline{*}), and division (\lstinline{/}) operations for \lstinline{UniTensor}s. Note that all of these operators are \emph{element-wise}, which means that the operation is applied between each pair of elements from the two \lstinline{UniTensor}s. Obviously, the two operated \lstinline{UniTensor}s must have the same shape. Other useful operations are \lstinline{UniTensor.Norm},  which returns the two-norm of the tensor and \lstinline{UniTensor.Conj}, which takes the complex conjugate of all elements (see \cref{sec:UniTensor_change_bond_direction} for an example). An extensive list of tensor operations and linear algebra functions is provided in the user guide~\cite{cytnxDoc_LinearAlgebra}. Further linear algebra functions are discussed in \cref{sec:decomp}. In the following example, we first create a tensor with all elements set to 1. 
Then, we demonstrate tensor arithmetics which create tensors with all elements equal to 5 and 0.2 respectively.
To add noise to an existing \lstinline{UniTensor}, one can clone and randomize it, and then add the randomized tensor back to the original \lstinline{UniTensor} as shown in the last line in \cref{code:Arithmetic}.

\begin{lstlisting}[language=python, 
                   caption={Arithmetic operations and tensor operations.}, 
                   label={code:Arithmetic}]
uTa = cytnx.UniTensor.ones([2,3]) # all the elements of uTa equal 1.0
uTb = uTa*3 + 2 # all the elements of uTb equal 5.0
uTc = uTa/uTb # all the elements of uTc equal 0.2
# randomize the cloned UniTensor then add it back
uT_noise = uT + cytnx.random.normal_(uT.clone(), mean=2., std=3.)
\end{lstlisting}

\subsection{Viewing/Changing  labels}
\label{sec:relabel}

Indices can be addressed by their labels. To view the current labels and their internal order, one can use \lstinline{UniTensor.labels}. Moreover, it might be necessary to change the label(s) of some bond(s), which can be done with \lstinline{UniTensor.relabel_}. 
In this code example, we show how to relabel one bond or multiple bonds and verify that the labels are changed correctly:

\begin{lstlisting}[language=python, 
                   caption={Relabel the bond(s) in a \lstinline{UniTensor}.}, 
                   label={code:UniTensoRelabel}]
# Create a rank-4 UniTensor with shape [2,3,4,5]
uT = cytnx.UniTensor.ones([2,3,4,5])\
                          .relabel(["a","b","c","d"]).set_name("uT")
print(uT.labels()) # Output: ['a', 'b', 'c', 'd']                     
# Change the label "a" to "i"
uT.relabel_(old_label="a", new_label="i")
print(uT.labels()) # Output: ['i', 'b', 'c', 'd']                     
# Change multiple labels
uT.relabel_(["b","c"], ["j","k"])
print(uT.labels()) # Output: ['i', 'j', 'k', 'd']
\end{lstlisting}

We may want to create a \lstinline{UniTensor} from an existing one without modifying the original tensor, but we want to use a different set of labels. A typical situation where this is useful is the contraction of tensors by summing over common indices (see \cref{sec:contraction}). Creating a copy of the tensor data is not desired, since it would double the memory usage. In such cases one can use \lstinline{relabel} without underscore. This returns a new \lstinline{UniTensor} object with different metadata (in this case only the labels are changed), but the actual memory block(s) are still referring to the old ones. 

\begin{lstlisting}[language=python]
uT_new = uT.relabel(old_label="d", new_label="l")
\end{lstlisting}
Users need to pay attention when using \lstinline{relabel} for creating a \lstinline{UniTensor}. Since the old and new tensors share the same memory for the data, any change of the elements in the new \lstinline{UniTensor} affect the original \lstinline{UniTensor} as well and vice versa. One can check if two \lstinline{UniTensor}s share the same data with

\begin{lstlisting}[language=python]
print(uT_new.same_data(uT)) # Output: True
\end{lstlisting}
which will print \lstinline{True} (\lstinline{False}) if they share (do not share) the same memory for the data.

\subsection{Permute}
When a \lstinline{UniTensor} is created, two important attributes are defined for each index: its position and label. We consider the code example  \cref{code:UniTensorCreation}. The three indices \lstinline{bond1}, \lstinline{bond2} and \lstinline{bond3} have positions 0, 1, and 2, and labels \lstinline{"a"}, \lstinline{"b"} and \lstinline{"c"} respectively. The library is designed such that users do not need to worry about the index positions in most applications. Instead, indices can be addressed by their labels. For example, tensor contractions can be defined purely by the index labels. However, the actual data in a \lstinline{UniTensor} is stored as a multi-dimensional array (a \lstinline{Tensor} object, see \cref{sec:tensor}) which has a specific order for the index positions. Conceptually, if one changes the index order of a \lstinline{UniTensor}, it still represents the same tensor, but the way that the data is stored in memory is changed.

To specify the internal order of indices in a \lstinline{UniTensor}, one can use \lstinline{UniTensor.permute}. The new order can be defined by the index labels (strings) or the index positions (integers), as shown in the code example below (\cref{code:UniTensorPermute}). There, we use \lstinline{labels} to see the new index order after calling \lstinline{UniTensor.permute}. Note that the same labels can still be used to find the correct index even after a permutation. Therefore, all methods that use index labels do not need to be changed after a permutation.

\begin{lstlisting}[language=python, 
                   caption={Permute the indices of a \lstinline{UniTensor}.},
                   label={code:UniTensorPermute}]
# Create a UniTensor
T = cytnx.arange(2*3*4).reshape(2,3,4)
uT = cytnx.UniTensor(T).relabel(["a","b","c"]).set_name("uT") 
print(uT.labels())  # Output: ['a', 'b', 'c']
print(uT.shape())   # Output: [2, 3, 4]
# Permute by index labels
uT2 = uT.permute(["b","c","a"])
print(uT2.labels()) # Output: ['b', 'c', 'a']
print(uT2.shape())  # Output: [3, 4, 2]
# Permute by index positions
uT3 = uT.permute([2,0,1])
print(uT3.labels()) # Output: ['c', 'a', 'b']
print(uT3.shape())  # Output: [4, 2, 3]
\end{lstlisting}

\subsection{Tensor information}
\label{sec:print}
Cytnx provides different ways to print information about a \lstinline{UniTensor}.
For the meta information but not the value of the elements,
\lstinline{print_diagram} can be used to visualize a \lstinline{UniTensor} as a diagram.
Consider the following code example and corresponding output (\cref{code:print_diagram}):

\begin{lstlisting}[language=python,
                caption={Diagram representation of a \lstinline{UniTensor}.},
                label={code:print_diagram}]
# Supposes uT is a UniTensor that is already created as in (*\color{codegreen}\cref{code:UniTensorCreation}*)
# Print the diagram of a UniTensor
uT.print_diagram()
# Change the rowrank
uT.set_rowrank_(2)
uT.print_diagram()
''' -------- Output ---------
tensor Name : uT
tensor Rank : 3
block_form  : False
is_diag     : False
on device   : cytnx device: CPU
          ---------     
         /         \    
   a ____| 2     3 |____ b
         |         |    
         |       4 |____ c
         \         /    
          ---------     
-----------------------
tensor Name : uT
tensor Rank : 3
block_form  : False
is_diag     : False
on device   : cytnx device: CPU
          ---------     
         /         \    
   a ____| 2     4 |____ c
         |         |    
   b ____| 3       |        
         \         /    
          ---------     

'''
\end{lstlisting}
The command \lstinline{UniTensor.print_diagram} shows information about the name, rank, block structure, diagonal form, and device.\footnote{name: can be set with \lstinline{set_name}; rank: number of indices; block structure: used for a symmetric \lstinline{UniTensor}, see \cref{sec:symm}; diagonal form: used to efficiently store diagonal matrices like the identity, singular values or eigenvalues; device: the data can be stored on the CPU or GPU memory, see \cref{sec:gpu} for details.} Moreover, it prints a diagram of the tensor. The diagram provides visual information about labels, shapes, and \lstinline{rowrank}. The diagram shows that the indices are labeled by \lstinline{["a", "b", "c"]} with this internal order. They are ordered from upper left to lower left, and then from upper right to lower right. Typically, users do not need to know the internal order of the indices since bonds can be accessed by their labels instead. However, the index order can be used explicitly if needed. The integers inside the tensor show the dimension of the bond, such that one can read the shape \lstinline{[2, 3, 4]} of the tensor from the diagram. In the first output in \cref{code:print_diagram}, the first output bond~\lstinline{"a"} is attached to the left and bonds~\lstinline{"b", "c"} are attached to the right of the box which represents the tensor. In the second output, the bonds~\lstinline{"a", "b"} are attached to the left and the bond~\lstinline{"c"} is attached to the right. This splitting is controlled by a property called \lstinline{rowrank}, which can be changed via \lstinline{UniTensor.set_rowrank_}. It determines how to interpret a tensor as a matrix: The combination of the first \lstinline{rowrank} bonds are interpreted as the row index, while the combination of the remaining bonds is interpreted as the column index. In the example above, setting rowrank to 2 means that if we want to interpret this rank-3 tensor as a matrix, we get $uT_{abc} \rightarrow [M]_{(ab),(c)}$, where the matrix $[M]$ has row index $(ab)$ and column index $(c)$.

When a \lstinline{UniTensor} is created, it acquires certain default \lstinline{rowrank}. This plays a role {\em only if} a \lstinline{UniTensor} is used in a linear algebra function, since it needs to interpret a tensor as a matrix. Consequently, before calling a tensor decomposition function, one may need to permute the labels and set the proper \lstinline{rowrank} to ensure the correction mapping to a matrix. More details will be discussed in \cref{sec:decomp}.

Moreover, one can use \lstinline{print} to see the tensor elements as an $N$-dimensional array:
\begin{lstlisting}[language=python,
                   caption={Print a \lstinline{UniTensor}.},
                   label={Print the elements of a UniTensor.}]
# Supposes uT is a UniTensor that is already created as in (*\color{codegreen}\cref{code:UniTensorCreation}*)
print(uT)

''' -------- Output ---------
Tensor name: tensor uT
is_diag    : False
contiguous : True

Total elem: 24
type  : Double (Float64)
cytnx device: CPU
Shape : (2,3,4)
[[[0.00000e+00 1.00000e+00 2.00000e+00 3.00000e+00 ]
  [4.00000e+00 5.00000e+00 6.00000e+00 7.00000e+00 ]
  [8.00000e+00 9.00000e+00 1.00000e+01 1.10000e+01 ]]
 [[1.20000e+01 1.30000e+01 1.40000e+01 1.50000e+01 ]
  [1.60000e+01 1.70000e+01 1.80000e+01 1.90000e+01 ]
  [2.00000e+01 2.10000e+01 2.20000e+01 2.30000e+01 ]]]
'''
\end{lstlisting}\

\subsection{Getting/Setting elements}\label{sec:get_set}
\paragraph{\lstinline{UniTensor.at()}.}

To get or set a single element in a \lstinline{UniTensor}, one can call \lstinline{UniTensor.at}. It returns a \emph{proxy} which contains a \emph{reference} to the element, and therefore can be used to get or set the element.
The first argument of \lstinline{UniTensor.at} contains the labels of the bonds in the \lstinline{UniTensor}, and the second argument contains the corresponding index values for the element one wants to access. We note that the internal index order of the tensor is irrelevant if one accesses elements by their labels. Consider a rank-3 tensor $T_{a,b,c}$. The example in \cref{code:AccessElement} shows how to set (1) $T_{a=0,b=1,c=2}=-1$, (2) $T_{b=1,a=0,c=2}=-2$, (3) $T_{c=2,a=0,b=1}=-3$ respectively. We print $T_{a=0,b=1,c=2}$ to check the results. It is also possible to specify the index values only when calling \lstinline{UniTensor.at}, with no labels given. In this case the internal order of indices is used.

\begin{lstlisting}[language=python, 
                   caption={Get or set a single element.}, 
                   label={code:AccessElement}]
# Create a UniTensor
uT=cytnx.UniTensor.zeros([3,3,3]).relabel(["a","b","c"]).set_name("uT")
# Print an element of the UniTensor
print(uT.at(["a","b","c"],[0,1,2]).value) # Output: 0.0
# Set the element to a new value and see the change
uT.at(["a","b","c"], [0,1,2]).value = -1
# Access an element by the internal index order
print(uT.at([0,1,2]).value) # Output: -1.0
# Access an element by labels
print(uT.at(["a","b","c"],[0,1,2]).value) # Output: -1.0
# Any label order is possible
uT.at(["b","a","c"], [1,0,2]).value = -2
print(uT.at(["a","b","c"],[0,1,2]).value) # Output: -2.0
uT.at(["c","b","a"], [2,1,0]).value = -3
print(uT.at(["a","b","c"],[0,1,2]).value) # Output: -3.0
\end{lstlisting}
For symmetric tensors, some of the elements may not be stored due to symmetry constraints which forces them to be be zero. In those cases, the reference can be used to check the existence of the element as explained in \cref{sec:get_set_sym}.
 
\paragraph{Slicing. }
To get or set a range of elements in a \lstinline{UniTensor} one can use slicing. In this case, the internal order of indices is used.

\begin{lstlisting}[language=python, 
                   caption={Get or set a slice.}, 
                   label={code:AccessElements}]
# Create a UniTensor
uT = cytnx.UniTensor.zeros([2,3,4]).set_name("uT")\
                    .relabel(["a", "b", "c"])
# set and get an element
uT[1,1,2] = 3.0
print(uT[1,1,2])    # Output: 3.0
# get a slice of a UniTensor
uTpart = uT[0,:,1:3]
# set a slice of a UniTensor
uT[0,:,1:3]=cytnx.UniTensor.ones([3,2])
# print the original slice
print(uTpart)
# print the UniTensor
print(uT)

''' -------- Output ---------
-------- start of print ---------
Tensor name: 
is_diag    : False
contiguous : True

Total elem: 1
type  : Double (Float64)
cytnx device: CPU
Shape : (1)
[3.00000e+00 ]

-------- start of print ---------
Tensor name: 
is_diag    : False
contiguous : True

Total elem: 6
type  : Double (Float64)
cytnx device: CPU
Shape : (3,2)
[[0.00000e+00 0.00000e+00 ]
 [0.00000e+00 0.00000e+00 ]
 [0.00000e+00 0.00000e+00 ]]

-------- start of print ---------
Tensor name: uT
is_diag    : False
contiguous : True

Total elem: 24
type  : Double (Float64)
cytnx device: CPU
Shape : (2,3,4)
[[[0.00000e+00 1.00000e+00 1.00000e+00 0.00000e+00 ]
  [0.00000e+00 1.00000e+00 1.00000e+00 0.00000e+00 ]
  [0.00000e+00 1.00000e+00 1.00000e+00 0.00000e+00 ]]
 [[0.00000e+00 0.00000e+00 0.00000e+00 0.00000e+00 ]
  [0.00000e+00 0.00000e+00 3.00000e+00 0.00000e+00 ]
  [0.00000e+00 0.00000e+00 0.00000e+00 0.00000e+00 ]]]
'''
\end{lstlisting}

\subsection{Accessing  data in a block}\label{sec:block}
The elements of a \lstinline{UniTensor} are stored in a \lstinline{Tensor} object. One can access and manipulate the \lstinline{Tensor} as described in \cref{sec:tensor} and read and change the data of the \lstinline{UniTensor} this way.

\paragraph{Getting a block.}
The \lstinline{Tensor} can be accessed by \lstinline{UniTensor.get_block} and \lstinline{UniTensor.get_block_}, where the first method returns a copy of the \lstinline{Tensor} object in the \lstinline{UniTensor}, while the second method returns a reference. As an example, we manipulate the \lstinline{Tensor} in order to change an element of the \lstinline{UniTensor}:

\begin{lstlisting}[language=python, caption={Get a block (\lstinline{Tensor}) from a non-symmetric \lstinline{UniTensor}.}, label={code:get_block_}]
uT = cytnx.UniTensor.ones([3,3]).set_name("uT")
B = uT.get_block_()
B[0,0] = 0.
print(uT)

''' -------- Output ---------
Tensor name: uT
is_diag    : False
contiguous : True

Total elem: 9
type  : Double (Float64)
cytnx device: CPU
Shape : (3,3)
[[0.00000e+00 1.00000e+00 1.00000e+00 ]
 [1.00000e+00 1.00000e+00 1.00000e+00 ]
 [1.00000e+00 1.00000e+00 1.00000e+00 ]]
'''
\end{lstlisting}
A \lstinline{UniTensor} without symmetries contains only one block, while several blocks exist when symmetries are involved, see \cref{sec:block_symm}.

\paragraph{Putting a block.}

The methods \lstinline{UniTensor.put_block} and \lstinline{UniTensor.put_block_} replace the data of a \lstinline{UniTensor} by a \lstinline{Tensor}. For example, after getting a block from a \lstinline{UniTensor}, we can manipulate the corresponding \lstinline{Tensor} object and write it back to the \lstinline{UniTensor}:

\begin{lstlisting}[language=python, caption={Replace a block of a \lstinline{UniTensor} by a \lstinline{Tensor}.}]
uT = cytnx.UniTensor.ones([3,3]).set_name("uT")
B = uT.get_block()
# Manipulate the Tensor
B[0,0] = 0.
# Put the Tensor back to the UniTensor
uT.put_block(B)
print(uT)
\end{lstlisting}
We get the same output as in \cref{code:get_block_}, where we manipulated the reference to the block directly. Note that the shape of the \lstinline{Tensor} has to match that of the \lstinline{UniTensor} whose data shall be replaced.

\subsection{Converting to/from a NumPy array}\label{sec:to_from_numpy}
It is very easy to create a \lstinline{UniTensor} from a NumPy array and vice versa. This is particularly useful for including existing code or when one wants to quickly test an algorithm using Python and then convert to Cytnx for acceleration, quantum numbers, or larger-scale applications. Such flexibility also allows users to make use of the tools from other packages like NumPy, SciPy and PyTorch. Since all popular tensor libraries allow for conversion of their containers to a NumPy array, the latter serves as a bridge to Cytnx. In the future, we plan to support PyTorch directly as the underlying container of a tensor in Cytnx. This way, users will be able to use automatic differentiation directly in Cytnx.

In \cref{code:to_from_numpy} we show an example where we create a NumPy array and then convert it to a \lstinline{UniTensor}. The tensor corresponds to the matrix product operator that defines the Hamiltonian of the transverse-field Ising model~\cite{MPO}.

\begin{lstlisting}[language=python, 
                   caption={Create the tensor of a matrix-product-operator for the transverse-field Ising model as a NumPy array and then convert it to a UniTensor.},
                   label={code:to_from_numpy}]
import numpy as np
# Define operators
Sx = np.array([[0.,1.],[1.,0.]])
Sz = np.array([[1.,0.],[0.,-1.]])
I  = np.array([[1.,0.],[0.,1.]])
# Define MPO tensor
hx, hz = 0.5, 0.5       # longitudinal and transverse fields
M = np.zeros((3,3,2,2)) # The first two indices are internal MPO
                        # indices, and the last two are physical indices
M[0,0] = I
M[2,2] = I
M[1,0] = Sx
M[2,1] = -Sx
M[2,0] = -hz*Sz - hx*Sx
M = cytnx.from_numpy(M) # NumPy array to a Tensor (see (*\color{codegreen}\cref{sec:numpy}*))
M = cytnx.UniTensor(M)  # Tensor to a UniTensor
M.set_name("M").relabel_(["mpoL", "mpoR", "physKet", "physBra"])
M.print_diagram()       # Check diagram
# Create a NumPy array from a non-symmetric UniTensor
T = M.get_block().numpy()
\end{lstlisting}

\section{UniTensor with Symmetries}
\label{sec:symm}
In physics, systems are often symmetric under certain transformations. Exploiting such symmetries can be advantageous in many cases. Cytnx allows for the incorporation of the symmetries on the level of the tensors directly. We explain the basic concepts that are used in Cytnx. Further details on symmetries in tensor network algorithms are found in~\cite{Silvi.2019, Weichselbaum:2012gs, Weichselbaum.2020, symm_U1,symm_decomposition}.

According to Noether's theorem, symmetries imply conserved quantities, for example, the total charge in the system. In that case, the Hamiltonian will have a block-diagonal structure where each symmetry sector has a definite charge (quantum number), which is caused by the fact that different charge sectors do not mix with each other. When the symmetric structure is imposed on the level of the tensors which describe the system, they become block-diagonal as well. This substantially reduces the memory costs and the number of variational parameters in numerical algorithms. Thus, larger system sizes or larger bond dimensions are accessible with the same computational costs. Moreover, numerical errors contributed by the elements that do not conserve symmetry (and hence are zero) can be avoided. Finally, simulations can be restricted to certain symmetry sectors if needed.

\begin{figure}[ht]
    \centering
    \includegraphics[scale=0.7]{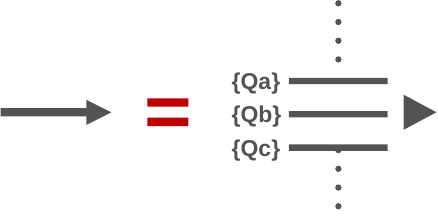}
    \caption{Sketch of a bond with quantum numbers.}
    \label{fig:qn}
\end{figure}

A quantum-number conserving tensor can be understood in the following way. By definition, each element in a tensor can be labeled by its index value for each index. For example \lstinline{T[0,0,0]} is an element with index values 0 for all three indices. Consider that in addition to the integer, each index value also carries a quantum number. This is shown in \cref{fig:qn}, where bonds are indicated by directed arrows, and each index value, represented by a horizontal line, has an associated quantum number.

\begin{figure}[ht]
    \centering
    \includegraphics[scale=0.6]{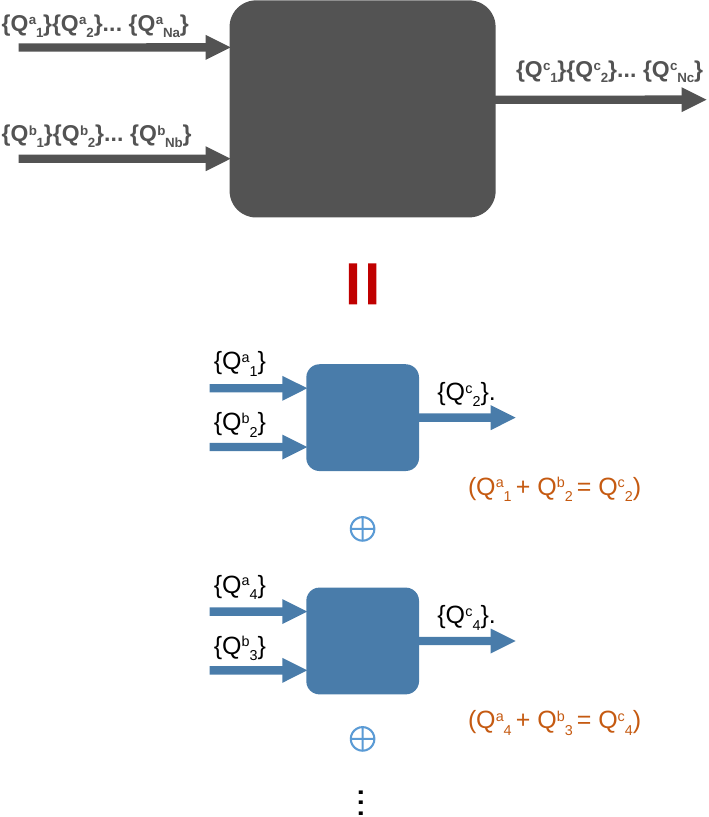}
    \caption{Block structure for a tensor with quantum numbers.}
    \label{fig:qnblock}
\end{figure}

One can then define a \emph{quantum-number flux} for each element by adding all the incoming quantum numbers and subtracting all the outgoing quantum numbers.
Note that the rules of "addition" and "subtraction" depend on the symmetry group associated with the quantum number. For charge conservation, which is associated to a $U(1)$ symmetry, addition and subtraction follow the same rules as for integers.
The conservation of a quantum number is ensured by restricting the tensors to only have non-zero elements with \emph{zero flux}. For these elements, the total quantum number flowing into the tensor has to be equal to the total quantum number flowing out.
{\lstset{breakatwhitespace}All other elements that do not conserve the charge must be zero. For a \lstinline{UniTensor}, these zero-elements are not stored in memory. Therefore, a symmetric tensor can be thought of as a direct sum of blocks of different quantum-number sectors, as represented in \cref{fig:qnblock}.}

In Cytnx, symmetric \lstinline{UniTensor}s contain \lstinline{Bond} objects which carry quantum numbers. The bonds themselves contain a \lstinline{Symmetry} object describing all the symmetries involved. Currently, abelian symmetries are implemented in Cytnx. Non-abelian symmetries like $SU(2)$ are planed to be supported in the future~\cite{symm_SU2}. Cytnx has predefined symmetry objects for $U(1)$ and $Z_n$ symmetries. Once a symmetric \lstinline{UniTensor} is defined, it can be used just like before, with typically no changes in the API of functions. In particular, the linear algebra operations respect the block structure automatically.

Sometimes a quantum number flux which differs from zero can be desired. For example, an external field can change the flux locally. In these cases, an additional bond can be attached to the \lstinline{UniTensor}, which has a dimension of one and carries the quantum number of the flux change. The code in \cref{code:create_Hamiltonian_symmetric} shows an example where this is used for particle number changing operators.

\subsection{Symmetry and quantum numbers}\label{sec:symmetries}

The symmetry properties are contained in the \lstinline{Symmetry} class, and  \lstinline{Symmetry} objects are used in the construction of a \lstinline{Bond} with quantum numbers. When a \lstinline{Symmetry} object is printed, the name and type of the symmetry are displayed, as well as the combine rule and the reverse rule. The combine rule defines how two quantum numbers are merged to a new quantum number. For the additive quantum numbers of $U(1)$, the combine rule is simply the sum of the individual quantum numbers. The reverse rule defines how a quantum number is transformed when the bond is redirected. Quantum numbers on outgoing bonds have the reverse rule applied when they are combined with other quantum numbers. With this procedure, the symmetry preserving entries of a tensor are those where the combination of all quantum numbers is zero. For example, the inverse rule of $U(1)$ takes the negative value of the original quantum number.

For example, we can create a \lstinline{Symmetry} describing the group $U(1)$ or $Z_3$:

\begin{lstlisting}[language=Python,
                   caption={Create a $U(1)$ \lstinline{Symmetry} object.},
                   label={code:SymU1}]
U1 = cytnx.Symmetry.U1()
print(U1)

''' -------- Output ---------
[Symmetry]
type : Abelian, U1
combine rule : Q1 + Q2
reverse rule : Q*(-1) 
'''
\end{lstlisting}

\begin{lstlisting}[language=Python,
                   caption={Create a $Z_3$ \lstinline{Symmetry} object.},
                   label={code:SymZ3}]
Z3 = cytnx.Symmetry.Zn(3)
print(Z3)

''' -------- Output ---------
[Symmetry]
type : Abelian, Z(3)
combine rule : (Q1 + Q2)%3
reverse rule : Q*(-1) 
'''
\end{lstlisting}

\subsection{Bond with quantum numbers}\label{sec:bond_symm}

In Cytnx, the quantum number information is stored in the \lstinline{Bond} objects. A \lstinline{Bond} with quantum numbers must have the following properties defined: 
\begin{enumerate}
    \item A direction. The bond is either incoming (\lstinline{BD_IN}) or outgoing (\lstinline{BD_OUT}).
    \item A list of quantum numbers and their dimensions. Quantum numbers are integers.
    \item The corresponding symmetry group(s), see \cref{sec:symmetries}.
\end{enumerate}
A direction is essential for any \lstinline{Bond} in a symmetric \lstinline{UniTensor}. Symmetry preserving tensors have a block structure, which is determined by the requirement that for non-zero elements the total quantum numbers of the incoming bonds equals the total quantum number of the outgoing bonds. Thus, all entries that do not fulfill this condition have to vanish and the corresponding blocks do not need to be stored explicitly. By defining bonds with quantum numbers, a \lstinline{UniTensor} created with these bonds will automatically have the desired block structure. 

\subsubsection{Creating bonds with quantum numbers}

We consider as an example a tensor with $U(1)$ symmetry. We would like to construct a bond for which the subspace with quantum number $q=2$ has a dimension (or degeneracy) of 3, while the $q=4$ subspace has dimension 5. We can create the bond using a list to describe the quantum numbers \lstinline{[[2],[4]]} and another list containing the dimensions \lstinline{[3,5]}. Note that the quantum number is enclosed in a list to accommodate multiple symmetries. Instead of the two lists, we can also use a single list of tuples where each tuple combines the information of quantum number and dimension \lstinline{[([2],3),([4],5)]}. Cytnx also includes a helper function \lstinline{Qs} to allow for a simpler and well readable syntax. The code \lstinline{[cytnx.Qs(2)>>3, cytnx.Qs(4)>>5]} generates the aforementioned list of tuples. 

\begin{lstlisting}[language=python, 
                   caption={Create a bond with a $U(1)$ quantum number.}]
#Creating a Bond from a list of quantum numbers and a list of dimensions
bond = cytnx.Bond(cytnx.BD_IN, [[2],[4]], [3,5], [cytnx.Symmetry.U1()])
#Creating the same Bond from a tuple of quantum numbers and dimensions
bond = cytnx.Bond(cytnx.BD_IN, [([2],3),([4],5)], [cytnx.Symmetry.U1()])
#Creating the same Bond with the helper function Qs
bond = cytnx.Bond(cytnx.BD_IN,\
                  [cytnx.Qs(2)>>3,cytnx.Qs(4)>>5],[cytnx.Symmetry.U1()])
\end{lstlisting}

The symmetries and the corresponding quantum numbers are encoded by lists. This way, multiple symmetries can be present at the same time. Consider a tensor with $U(1) \otimes Z_2$ symmetry. We specify the symmetry by \lstinline{[Symmetry.U1(), Symmetry.Zn(2)])} and the quantum numbers by \lstinline{[q_U1, q_Z2]}. We can use \lstinline{Qs(q_U1, q_Z2)>>dim} to construct the tuple containing the quantum numbers and the dimension of the subspace. 
One can print a \lstinline{bond} to check its quantum numbers.

\begin{lstlisting}[language=python, 
                   caption={Create a bond with two quantum numbers.}, 
                   label={code:BondU1Z2}]
# This creates an incoming Bond
# with U(1)xZ2 quantum numbers qn_U1 and qn_Z2.
# The sector with qn_U1=2 and qn_Z2=0 has dimension 3.
# The sector with qn_U1=4 and qn_Z2=1 has dimension 5.
bond = cytnx.Bond(cytnx.BD_IN,\
                  [cytnx.Qs(2,0)>>3, cytnx.Qs(4,1)>>5],\
                  [cytnx.Symmetry.U1(), cytnx.Symmetry.Zn(2)])
print(bond)

''' -------- Output ---------
Dim = 8 |type: |IN (KET)>     
 U1::   +2  +4
 Z2::   +0  +1
Deg>>    3   5
'''
\end{lstlisting}

\subsubsection{Combining bonds with quantum numbers}

Bonds with the same type and symmetries can be combined with the same API as for combining bonds without symmetries, see \cref{sec:combining_nonsymm}. The new quantum numbers are the combination of the two input quantum numbers. The actual combine rule depends on the corresponding symmetry group. Considering the $U(1)$ symmetry as an example, two quantum numbers are combined by adding them.
In general, different pairs of quantum numbers can be combined to the same quantum number. For example, both $0+3$ and $1+2$ give the quantum number $3$ in the $U(1)$ case. Therefore, the output bond in general can have several quantum-number sectors with the same quantum number. By default, the sectors with the same quantum numbers will be grouped into a single sector with accordingly increased dimension.\footnote{If necessary, users can choose not to group them by setting the argument \lstinline{is_grp=False}.} Here is an example of how to combine two bonds: 

\begin{lstlisting}[language=python, caption={Combine bonds with quantum numbers.}]
bond1 = cytnx.Bond(cytnx.BD_IN,\
                [cytnx.Qs(0)>>1,cytnx.Qs(1)>>1], [cytnx.Symmetry.U1()])
bond2 = cytnx.Bond(cytnx.BD_IN,\
                [cytnx.Qs(2)>>1,cytnx.Qs(3)>>1], [cytnx.Symmetry.U1()])
bond12 = bond1.combineBond(bond2)
print('bond1:\n',bond1)
print('bond2:\n',bond2)
print('bond12:\n',bond12)

''' -------- Output ---------
bond1:
 Dim = 2 |type: |IN (KET)>     
 U1::   +0  +1
Deg>>    1   1

bond2:
 Dim = 2 |type: |IN (KET)>     
 U1::   +2  +3
Deg>>    1   1

bond12:
 Dim = 4 |type: |IN (KET)>     
 U1::   +2  +3  +4
Deg>>    1   2   1
'''
\end{lstlisting}
In this example, we print the \lstinline{Bond} objects to check their quantum numbers and the corresponding dimensions. We can see that the output quantum numbers are the sums of the input ones, and the sectors with the same quantum numbers are grouped.

Note that only \lstinline{Bond}s with the same direction can be combined, and the resulting \lstinline{Bond} will have the same direction. In a physical interpretation, an incoming or outgoing \lstinline{Bond} can be used to define the Hilbert space for a ket or a bra state, respectively. In this sense, combining two \lstinline{Bond}s represents a product of Hilbert spaces for the two states, which is well defined only if both states are ket or bra states.

\subsubsection{Redirecting bonds}
One can change the bond direction by using \lstinline{Bond.redirect} or \lstinline{Bond.redirect_}:

\begin{lstlisting}[language=python, caption={Redirect a \lstinline{Bond}.}]
# Supposes bond is a Bond that is already created as in (*\color{codegreen}\cref{code:BondU1Z2}*)
print(bond)     # Output: Dim = 8 |type: |IN (KET)>
# Create a new bond that has the direction changed.
bond2 = bond.redirect()
# Change the direction of a bond
bond.redirect_()
print(bond)     # Output: Dim = 8 |type: < OUT (BRA)|
print(bond2)    # Output: Dim = 8 |type: < OUT (BRA)|
\end{lstlisting}

\subsection{Symmetric UniTensor}\label{sec:uniten_symm}

\subsubsection{Creating a symmetric UniTensor}
A symmetric \lstinline{UniTensor} is created with \lstinline{Bond} objects which carry quantum numbers. Conceptually, one needs to identify the symmetries (for example, $U(1)$ symmetry) in the system, and then specify the dimensions of all quantum number sectors for the bonds. 
As an example, we consider a rank-3 tensor with $U(1)$ symmetry. The first two bonds are incoming bonds with two quantum number sectors $1$ and $-1$, both with dimension $1$. The third bond is an outgoing bond with three quantum number sectors $2$, $0$, and $-2$, with dimensions $1$, $2$, and $1$ respectively. We then initialize a \lstinline{UniTensor} using these three bonds, 
where all elements are set to be zero. As shown in the example, one can multiply and divide a symmetric \lstinline{UniTensor} by a scalar. Adding or subtraction a scalar, however, might break the block structure and is therefore not allowed.
If one tries to add/substrate a scalar, an error is thrown.

\begin{lstlisting}[language=python, 
                   caption={Create a symmetric \lstinline{UniTensor} using bonds with quantum numbers.}, 
                   label={code:symmTensor}]
# Define the bonds with quantum numbers
bond1 = cytnx.Bond(cytnx.BD_IN,\
                [cytnx.Qs(1)>>1, cytnx.Qs(-1)>>1],[cytnx.Symmetry.U1()])
bond2 = cytnx.Bond(cytnx.BD_IN,\
                [cytnx.Qs(1)>>1, cytnx.Qs(-1)>>1],[cytnx.Symmetry.U1()])
bond3 = cytnx.Bond(cytnx.BD_OUT,\
                [cytnx.Qs(2)>>1, cytnx.Qs(0)>>2, cytnx.Qs(-2)>>1],\
                [cytnx.Symmetry.U1()])
# Create a symmetric UniTensor with these bonds
uTsym = cytnx.UniTensor([bond1, bond2, bond3]).relabel(["a","b","c"])\
                        .set_name("uTsym")
uTsym.print_diagram()
uTsym + 1.0
''' -------- Output ---------
tensor Name : uTsym
tensor Rank : 3
contiguous  : True
valid blocks : 4
is diag   : False
on device   : cytnx device: CPU
      row           col 
         -----------    
         |         |    
   a  -->| 2     4 |-->  c
         |         |    
   b  -->| 2       |        
         |         |    
         -----------    

RuntimeError: 
# Cytnx error occur at virtual void cytnx::BlockUniTensor::Add_(const cytnx::Scalar&)
# error: [ERROR] cannot perform elementwise arithmetic '+' btwn Scalar and BlockUniTensor.
 This operation will destroy block structure. [Suggest] using get/set_block(s) to do operation on the block(s). 
'''
                        
\end{lstlisting}

\begin{figure}[ht]
\centering
\includegraphics[scale=0.7]{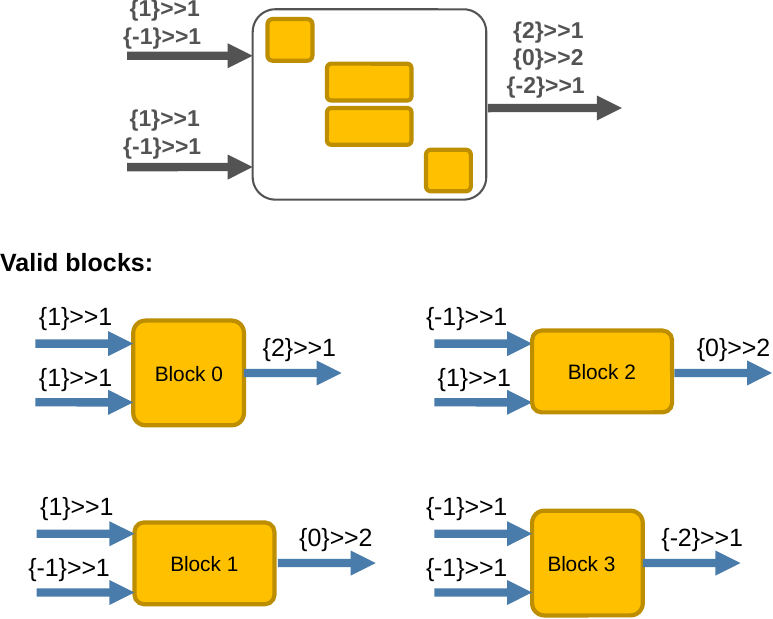}
\caption{Schematic representation of the block structure of a symmetric tensor. See \cref{code:symmTensor} for the creation of this tensor in Cytnx.}
\label{fig:blocks}
\end{figure}

For this \lstinline{UniTensor}, there are 4 valid blocks that carry zero-flux and thus can have non-zero elements that do not break the symmetry. \Cref{fig:blocks} shows the block structure and the corresponding quantum numbers for each block. We can use \lstinline{UniTensor.print_blocks} to see all the blocks and their corresponding quantum numbers:
\begin{lstlisting}[language=python, 
                   caption={Print the blocks of a symmetric \lstinline{UniTensor}.}, 
                   label={code:symmTensorBlocks}]
# Supposes uTsym is a UniTensor that is already created as in (*\color{codegreen}\cref{code:symmTensor}*)
uTsym.print_blocks()

''' -------- Output ---------
-------- start of print ---------
Tensor name: uTsym
braket_form : True
is_diag    : False
[OVERALL] contiguous : True
========================
BLOCK [#0]
 |- []   : Qn index 
 |- Sym(): Qnum of correspond symmetry
                 -----------
                 |         |
   [0] U1(1)  -->| 1     1 |-->  [0] U1(2)
                 |         |
   [0] U1(1)  -->| 1       |
                 |         |
                 -----------

Total elem: 1
type  : Double (Float64)
cytnx device: CPU
Shape : (1,1,1)
[[[0.00000e+00 ]]]

========================
BLOCK [#1]
 |- []   : Qn index 
 |- Sym(): Qnum of correspond symmetry
                  -----------
                  |         |
   [0] U1(1)   -->| 1     2 |-->  [1] U1(0)
                  |         |
   [1] U1(-1)  -->| 1       |
                  |         |
                  -----------

Total elem: 2
type  : Double (Float64)
cytnx device: CPU
Shape : (1,1,2)
[[[0.00000e+00 0.00000e+00 ]]]

========================
BLOCK [#2]
 |- []   : Qn index 
 |- Sym(): Qnum of correspond symmetry
                  -----------
                  |         |
   [1] U1(-1)  -->| 1     2 |-->  [1] U1(0)
                  |         |
   [0] U1(1)   -->| 1       |
                  |         |
                  -----------

Total elem: 2
type  : Double (Float64)
cytnx device: CPU
Shape : (1,1,2)
[[[0.00000e+00 0.00000e+00 ]]]

========================
BLOCK [#3]
 |- []   : Qn index 
 |- Sym(): Qnum of correspond symmetry
                  -----------
                  |         |
   [1] U1(-1)  -->| 1     1 |-->  [2] U1(-2)
                  |         |
   [1] U1(-1)  -->| 1       |
                  |         |
                  -----------

Total elem: 1
type  : Double (Float64)
cytnx device: CPU
Shape : (1,1,1)
[[[0.00000e+00 ]]]
'''
\end{lstlisting}

The \emph{block index} is printed as \lstinline{BLOCK [#]}, and the number in the square bracket \lstinline{[]} beside each bond is called a \emph{Qn index} (quantum number index). Qn indices are used to access the sectors in a bond. Consider the third bond (the out-going bond) as an example. It has three possible values for the quantum number: $2$, $0$, and $-2$. It is created with this particular ordering of the quantum numbers. Hence, Qn=0 corresponds to the quantum number $2$, Qn=1 to the quantum number $0$, and Qn=2 to the quantum number $-2$. Both the block index and the Qn indices can be used to access a block in a \lstinline{UniTensor} as it will be described below in \cref{sec:block_symm}.

\subsubsection{Accessing elements of a symmetric tensor}\label{sec:get_set_sym}
The \lstinline{UniTensor.at} method works in the same way as without symmetries (see \cref{sec:get_set}).
For a symmetric \lstinline{UniTensor}, some elements may not exist due to the quantum number constraint. If one tries to access an element that does not correspond to a valid block, an error is thrown:

\begin{lstlisting}[language=python, caption={Example of accessing a non-existing element in a \lstinline{UniTensor}. An error is thrown in this case.}]
# Supposes uTsym is a UniTensor that is already created as in (*\color{codegreen}\cref{code:symmTensor}*)
# Print an invalid element. This will raise an error.
print(uTsym.at([0,0,1]).value)

''' -------- Output ---------
ValueError: [ERROR] trying access an element that is not exists!, using T.if_exists = sth or checking with T.exists() to verify before access element!
'''
\end{lstlisting}
However, one can use the proxy to check whether an element corresponds to a valid block with symmetries:

\begin{lstlisting}[language=python,caption={Check if an element is valid in a symmetric \lstinline{UniTensor}.}]
# Supposes uTsym is a UniTensor that is already created as in (*\color{codegreen}\cref{code:symmTensor}*)
print(uTsym.at([0,0,0]).exists()) # Output: True
print(uTsym.at([0,0,1]).exists()) # Output: False
\end{lstlisting}
This way, one can make sure that an element is part of a valid block before the data is accessed.

\subsubsection{Accessing blocks of a symmetric tensor}\label{sec:block_symm}
\paragraph{Getting a block.}

A symmetric \lstinline{UniTensor} contains multiple blocks for different quantum number sectors in general. We can use the methods \lstinline{get_block} and \lstinline{get_block_} to access the blocks. Compared to the non-symmetric case in \cref{sec:block}, here we have to specify the block for a symmetric \lstinline{UniTensor}. Note that \lstinline{get_block} returns a copy of the \lstinline{Tensor} object, while \lstinline{get_block_} returns a reference.
We use the $U(1)$ symmetric \lstinline{UniTensor} from the example in \cref{code:symmTensor} to demonstrate how to access blocks.
There are two ways to get a certain block from a symmetric \lstinline{UniTensor}: by its Qn indices or by its block index. When using Qn indices, the labels of the indices can be given as the first argument of \lstinline{get_block}, followed by a list of the Qn indices associated with the block.

\begin{lstlisting}[language=python, 
                   caption={Access a block of a symmetric \lstinline{UniTensor}.}]
# Supposes uTsym is a UniTensor that is already created as in (*\color{codegreen}\cref{code:symmTensor}*)
# Get a block by its Qn indices
# The first argument specifies the labels of the indices,
# and the second argument the corresponding Qn indices
B1 = uTsym.get_block_(["a","b","c"],[0,1,1])
# Get a block by its block index
B2 = uTsym.get_block_(1)
\end{lstlisting}
It is also possible to access all valid blocks in a symmetric \lstinline{UniTensor} with \lstinline{get_blocks} or \lstinline{get_blocks_}. These methods return a list (or a vector in C++) of blocks, where each block is a \lstinline{Tensor} object. The blocks are ordered with ascending block index, corresponding to the order in which they are stored in the \lstinline{UniTensor}. This order can be checked by \lstinline{UniTensor.print_blocks()}.

\begin{lstlisting}[language=python, 
                   caption={Get all the blocks in a \lstinline{UniTensor}.}]
# Supposes uTsym is a UniTensor that is already created as in (*\color{codegreen}\cref{code:symmTensor}*)
# Get all the blocks in a UniTensor
Blks = uTsym.get_blocks_()
# Print the number of blocks
print(len(Blks))    # Output: 4
# Print each block
print( *Blks )

''' -------- Output ---------
4

Total elem: 1
type  : Double (Float64)
cytnx device: CPU
Shape : (1,1,1)
[[[0.00000e+00 ]]]

Total elem: 2
type  : Double (Float64)
cytnx device: CPU
Shape : (1,1,2)
[[[0.00000e+00 0.00000e+00 ]]]

Total elem: 2
type  : Double (Float64)
cytnx device: CPU
Shape : (1,1,2)
[[[0.00000e+00 0.00000e+00 ]]]

Total elem: 1
type  : Double (Float64)
cytnx device: CPU
Shape : (1,1,1)
[[[0.00000e+00 ]]]
'''
\end{lstlisting}

\paragraph{Putting a block.}
Similarly to the non-symmetric case, we can put a \lstinline{Tensor} as a block to a \lstinline{UniTensor}. In the symmetric case, one can put a \lstinline{Tensor} to a particular block by specifying the Qn indices or the block index:

\begin{lstlisting}[language=python, caption={Put a block into a \lstinline{UniTensor}.}]
# Supposes uTsym is a UniTensor that is already created as in (*\color{codegreen}\cref{code:symmTensor}*)
# Get a block as a Tensor object by Qn indices
B1 = uTsym.get_block([0,1,1])
# Manipulate the Tensor
B1[0,0,0] = 1.
# Put the Tensor to a block by Qn indices
uTsym.put_block(B1,[0,1,1])
# Put the Tensor to a block by block index
uTsym.put_block(B1,1)
print(uTsym.get_block(1))

''' -------- Output ---------
Total elem: 2
type  : Double (Float64)
cytnx device: CPU
Shape : (1,1,2)
[[[1.00000e+00 0.00000e+00 ]]]
'''
\end{lstlisting}

\subsection{Change the bond direction}\label{sec:UniTensor_change_bond_direction}

The direction of all bonds can be changed with \lstinline{UniTensor.Transpose} and \lstinline{UniTensor.Dagger}. The latter returns the Hermitian conjugate of a \lstinline{UniTensor}, with all bond directions inverted and the complex conjugate of all elements taken. For tensors with non-directional bonds (\lstinline{REGULAR} type), the bonds remain non-directional after the operation. If only the complex conjugate of a tensor is desired, \lstinline{UniTensor.Conj} can be used. If the \lstinline{dtype} of the tensor corresponds to real valued elements, the complex conjugation has no effect. In line with the general naming convention, \lstinline{UniTensor.Transpose_}, \lstinline{UniTensor.Conj_} and \lstinline{UniTensor.Dagger_} change the tensor in-place. The following code shows an example for the \lstinline{UniTensor.Transpose}, \lstinline{UniTensor.Conj} and \lstinline{UniTensor.Dagger} operations:

\begin{lstlisting}[language=python, 
                   caption={Transpose, conjugate and Hermitian conjugate of a tensor.} %\lstinline{UniTensor.Dagger} returns the complex conjugate of the original tensor, and all the bonds are redirected.}
                   ]
# Create bonds
bond1 = cytnx.Bond(1,cytnx.BD_IN)
bond2 = bond1.redirect()
# Create a symmetric UniTensor with these bonds
uTc = cytnx.UniTensor([bond1,bond2], dtype=cytnx.Type.ComplexDouble)\
                      .relabel(["a", "b"]).set_name("uTc")
# Set an element
uTc.at([0,0]).value = 1.+2.j
# Change the direction of the bonds
uTtransp = uTc.Transpose().set_name("uTc transpose")
# Conjugate the elements
uTconj = uTc.Conj().set_name("uTc conjugate")
# Take the Hermitian conjugate
uTdag = uTc.Dagger().set_name("uTc Hermitian conjugate")
# Equivalently:
uTdag2 = uTc.Conj().Transpose().set_name("uTc conjugate transpose")
print(uTc)      # Part of output: [[1.00000e+00+2.00000e+00j ]]
print(uTconj)   # Part of output: [[1.00000e+00-2.00000e+00j ]]
print(uTtransp) # Part of output: [[1.00000e+00+2.00000e+00j ]]
print(uTdag)    # Part of output: [[1.00000e+00-2.00000e+00j ]]
\end{lstlisting}

\subsection{Conversion between UniTensor with and without symmetries}

One can use \lstinline{UniTensor.convert_from} to perform conversion between \lstinline{UniTensor}s with and without symmetries. If one tries to convert a non-symmetric \lstinline{UniTensor} with non-zero elements in the symmetry violating blocks to a symmetric one, an error is thrown. It is possible to overwrite the check by using keyword argument \lstinline{force=True}. In this case, the elements in the symmetry violating blocks are ignored.

As an example, we first create a non-symmetric \lstinline{UniTensor}, which represents the two-site Hamiltonian $H = \sigma^+_1 \sigma^-_2 + \sigma^-_1 \sigma^+_2 + \sigma^z_1 \sigma^z_2$. We initialize the $\sigma$-matrices, take Kronecker products of pairs of them, and then sum the terms to get the Hamiltonian:
\begin{lstlisting}[language=python, 
                   caption={Create a two-site Hamiltonian  $H = \sigma^+_1 \sigma^-_2 + \sigma^-_1 \sigma^+_2 + \sigma^z_1 \sigma^z_2$ as a non-symmetric \lstinline{UniTensor}.},
                   label={code:create_Hamiltonian_dense}]
# Create sigma matrices as non-symmetric UniTensors
Sp = cytnx.UniTensor.zeros([2,2]) # sigma^+
Sp[0, 1] = +1.
Sm = cytnx.UniTensor.zeros([2,2]) # sigma^-
Sm[1, 0] = +1.
Sz = cytnx.UniTensor.zeros([2,2]) # sigma^z
Sz[0, 0] = +1.
Sz[1, 1] = -1.
# Create two copies of all sigma matrices for the two sites
Sp1 = Sp.clone().set_name("Sp1").relabel(["i1","j1"])
Sp2 = Sp.clone().set_name("Sp2").relabel(["i2","j2"])
Sm1 = Sm.clone().set_name("Sm1").relabel(["i1","j1"])
Sm2 = Sm.clone().set_name("Sm2").relabel(["i2","j2"])
Sz1 = Sz.clone().set_name("Sz1").relabel(["i1","j1"])
Sz2 = Sz.clone().set_name("Sz2").relabel(["i2","j2"])
# Kronecker product of two sigma matrices at different sites
Sp1Sm2 = cytnx.Contract(Sp1,Sm2).set_name("Sp1Sm2")
Sp1Sm2.permute_(["i1","i2","j1","j2"])
Sm1Sp2 = cytnx.Contract(Sm1,Sp2).set_name("Sm1Sp2")
Sm1Sp2.permute_(["i1","i2","j1","j2"])
Sz1Sz2 = cytnx.Contract(Sz1,Sz2).set_name("Sz1Sz2")
Sz1Sz2.permute_(["i1","i2","j1","j2"])
# Create Hamiltonian as non-symmetric UniTensor
H = Sp1Sm2 + Sm1Sp2 + Sz1Sz2
H.set_name("H non-Symmetric").relabel_(["i1","i2","j1","j2"])
# Print in matrix form to check
print(H.reshape(4,4))
\end{lstlisting}

This non-symmetric \lstinline{UniTensor} can be converted to a symmetric one, which has to be initialized with the correct quantum numbers. In our case, we create a $U(1)$ symmetric \lstinline{UniTensor} and copy data from the non-symmetric UniTensor:
\begin{lstlisting}[language=python, 
                   caption={Convert a non-symmetric \lstinline{UniTensor} to a symmetric \lstinline{UniTensor}.},
                   label={code:convert_dense_to_sym}]
# Supposes H is a UniTensor that is already created as in (*\color{codegreen}\cref{code:create_Hamiltonian_dense}*)
# Create bonds with U(1) symmetry
bi = cytnx.Bond(cytnx.BD_IN, [cytnx.Qs(+1)>>1, cytnx.Qs(-1)>>1],\
                [cytnx.Symmetry.U1()])
bo = cytnx.Bond(cytnx.BD_OUT, [cytnx.Qs(+1)>>1, cytnx.Qs(-1)>>1],\
                [cytnx.Symmetry.U1()])
# Initialize U(1) symmetric UniTensor
H_sym = cytnx.UniTensor([bi,bi,bo,bo]).set_name("H symmetric")\
                        .relabel(["i1","i2","j1","j2"])
# Copy data from the non-symmetric UniTensor H
H_sym.convert_from(H)
print(H_sym) # Check the blocks
\end{lstlisting}

Similarly, we can directly create a symmetric \lstinline{UniTensor}, which represents the previous Hamiltonian. This time, the $\sigma$-matrices need to be symmetric tensors which carry quantum numbers. The $\sigma^+$ and $\sigma^-$ operators change the quantum number locally. In this case, we can see them as creation and annihilation operators, which change the particle number. Thus, the corresponding tensors would naively not conserve the quantum number. However, we can attach an additional bond with only one index to these $\sigma$-matrices, whose quantum number corresponds to the local particle number change (\lstinline{bqo} or \lstinline{bqi} in \cref{code:create_Hamiltonian_symmetric}). We observe that the terms of the Hamiltonian only contain pairs which involve one $\sigma^+$ and one $\sigma^-$ operator. The Hamiltonian thus conserves the particle number, which makes it a good quantum number. We can use this to make sense of the previously introduced quantum number changing bonds: each $\sigma^+$ tensor needs to have such an additional bond, which is contracted with the additional bond of a $\sigma^-$ tensor according to the terms of the Hamiltonian. The following code example demonstrates this:

\begin{lstlisting}[language=python, 
                   caption={Create a two-site Hamiltonian $H = \sigma^+_1 \sigma^-_2 + \sigma^-_1 \sigma^+_2 + \sigma^z_1 \sigma^z_2$ as symmetric \lstinline{UniTensor}.},
                   label={code:create_Hamiltonian_symmetric}]
# Supposes H_sym is a symmetric UniTensor that is already created
# as in (*\color{codegreen}\cref{code:convert_dense_to_sym}*)
# Create bonds with U(1) symmetry
bi = cytnx.Bond(cytnx.BD_IN, [cytnx.Qs(+1)>>1, cytnx.Qs(-1)>>1],\
                [cytnx.Symmetry.U1()])
bo = cytnx.Bond(cytnx.BD_OUT, [cytnx.Qs(+1)>>1, cytnx.Qs(-1)>>1],\
                [cytnx.Symmetry.U1()])
# Bond between sigma^+ and sigma^- needs to carry a quantum number
bqo = cytnx.Bond(cytnx.BD_OUT, [cytnx.Qs(+2)>>1],\
                [cytnx.Symmetry.U1()])
bqi = cytnx.Bond(cytnx.BD_IN, [cytnx.Qs(+2)>>1],\
                [cytnx.Symmetry.U1()])
# Create sigma matrices as symmetric UniTensors
Sp = cytnx.UniTensor([bi,bo,bqo])   # sigma^+
Sp.at([0, 1, 0]).value = +1.
Sm = cytnx.UniTensor([bqi,bi,bo])   # sigma^-
Sm.at([0, 1, 0]).value = +1.
Sz = cytnx.UniTensor([bi,bo])       # sigma^z
Sz.at([0, 0]).value = +1.
Sz.at([1, 1]).value = -1.
# Create two copies of all sigma matrices for the two sites
Sp1 = Sp.clone().set_name("Sp1").relabel(["i1","j1","q"])
Sp2 = Sp.clone().set_name("Sp2").relabel(["i2","j2","q"])
Sm1 = Sm.clone().set_name("Sm1").relabel(["q","i1","j1"])
Sm2 = Sm.clone().set_name("Sm2").relabel(["q","i2","j2"])
Sz1 = Sz.clone().set_name("Sz1").relabel(["i1","j1"])
Sz2 = Sz.clone().set_name("Sz2").relabel(["i2","j2"])
# Product of two sigma matrices at different sites;
# labels "q" are summed over
Sp1Sm2 = cytnx.Contract(Sp1,Sm2).set_name("Sp1Sm2")
Sp1Sm2.permute_(["i1","i2","j1","j2"])
Sm1Sp2 = cytnx.Contract(Sm1,Sp2).set_name("Sm1Sp2")
Sm1Sp2.permute_(["i1","i2","j1","j2"])
Sz1Sz2 = cytnx.Contract(Sz1,Sz2).set_name("Sz1Sz2")
Sz1Sz2.permute_(["i1","i2","j1","j2"])
# Create Hamiltonian as symmetric UniTensor
H2_sym = Sp1Sm2 + Sm1Sp2 + Sz1Sz2
H2_sym.set_name("H2 symmetric").relabel_(["i1","i2","j1","j2"])
# Check the blocks
print(H2_sym - H_sym) # Output: all elements are zero
\end{lstlisting}

Finally, we can convert the symmetric \lstinline{UniTensor} into a non-symmetric one and compare to our previous initialization of the Hamiltonian:
\begin{lstlisting}[language=python, 
                   caption={Convert a symmetric \lstinline{UniTensor} to a non-symmetric \lstinline{UniTensor}.},
                   label={code:convert_sym_to_dense}]
# Supposes H2_sym is a symmetric UniTensor that is already created
# as in (*\color{codegreen}\cref{code:create_Hamiltonian_symmetric}*)
H2=cytnx.UniTensor.zeros([2,2,2,2]).relabel((["i1","i2","j1","j2"]))\
                         .set_name("H2")
H2.convert_from(H2_sym)
# Print the Hamiltonian in matrix form
print(H2.reshape(4,4)) # Output: same data as in  (*\color{codegreen}\cref{code:create_Hamiltonian_dense}*)
\end{lstlisting}

The conversion between symmetric and non-symmetric can be very useful, for example for conveniently checking the elements of a symmetric \lstinline{UniTensor} in matrix form. We note, however, that performing operations on non-symmetric tensors and then converting them to symmetric ones can lead to problems. This can, for example, happen, when linear algebra routines are performed on non-symmetric tensors. Numerical errors or simply the fact that symmetry violating elements are not explicitly forbidden can lead to non-zero elements in the wrong blocks. Even when these are ignored using the argument \lstinline{force=True}, errors can accumulate in tensor network algorithms. We therefore suggest to avoid \lstinline{convert_from} when possible and only use it with care. Symmetric tensors should be used wherever symmetries shall be preserved. Cytnx provides linear algebra routines that preserve the symmetries of the tensors. If operations are needed that are not natively supported by Cytnx, these should be operating on the level of the blocks to make sure symmetry violating elements can not arise.

Besides avoiding numerical errors, symmetric tensors also help in the code development by ensuring that only operations are allowed which respect the symmetries. For example, the $\sigma^+$ and $\sigma^-$ operators in \cref{code:create_Hamiltonian_symmetric} can only come in pairs, such that their additional indices are contracted. Otherwise, open indices arise that still need to be contracted with suitable operators. This behavior is much less error prone than using non-symmetric tensors which allow for any operation, including symmetry violating ones.

\section{Tensor contraction}
\label{sec:contraction}

All tensor network algorithms include tensor contractions, where we multiply tensors and sum over shared indices. Cytnx provides three different ways to perform these contractions. A simple way to contract tensors is provided by \lstinline{Contract}. It contracts two or more \lstinline{UniTensor} objects by summing over all indices that have the same labels. We also support the API of \lstinline{ncon}~\cite{pfeifer2015ncon} which allows users to contract tensors by defining the connectivity and contraction order of the bonds. We refer to the user guide~\cite{cytnxDoc_LinearAlgebra} for details. For using \lstinline{ncon}, the user needs to specify the bonds by their indices instead of their labels, so the index order matters. This can be error prone when users have to keep track of the change of the index order in complicated tensor network algorithms. In Cytnx we introduce an advanced method for tensor network contractions via a \lstinline{Network} object. It is particularly useful for contractions involving many tensors and reusable contraction diagrams. The \lstinline{Network} based contraction will be explained in more details in \cref{sec:network}.

The efficiency of tensor network contractions depends crucially on the contraction order. In all the contraction methods mentioned above, users can explicitly specify the contraction order. However, it often desirable that the optimal contraction order can be automatically generated, as pioneered in Ref.~\cite{pfeifer2015ncon, Pfeifer:2014bw}. The optimal contraction order can be automatically generated in Cytnx for \lstinline{Contract}, \lstinline{ncon}, and \lstinline{Network} by setting a flag. For this, we implemented the dynamical programming algorithm with the cost capping strategy proposed in Ref.~\cite{Pfeifer:2014bw}. In this way users can focus on designing the tensor network without worrying about finding the optimal contraction order.


In the following we explain how to use \lstinline{Contract}, to contract two or more tensors. In this function, common indices of the tensors are summed over. A new \lstinline{UniTensor} is created to represent the result, while the input tensors are not modified. 
The labels of the output tensor are inherited from the input tensors. In order to avoid potential mistakes, Cytnx checks that the dimensions of the two contracted indices are the same before running a contraction. For symmetric tensors, Cytnx also checks the bond direction, the quantum numbers, and the dimensions of each symmetry block. In the code example in \cref{code:contractTwoTensors} we show how to contract two tensors.

\begin{lstlisting}[language=python,
                   caption={Contract two tensors with \lstinline{Contract}.},
                   label={code:contractTwoTensors}]
# Define two UniTensor A and B
A = cytnx.UniTensor.ones([2,3,5]).relabel(["i","j","l"]).set_name("A")
B = cytnx.UniTensor.ones([3,1,5,4]).relabel(["j","k","l","m"])\
                         .set_name("B")
# Contract A and B by their common indices and store the result in AB
# Indices "j" and "l" are contracted
# Output tensor has indices "i", "k", and "m"
AB = cytnx.Contract(A, B)
AB.set_name("AB")
# One can check the result with print_diagram
A.print_diagram()
B.print_diagram()
AB.print_diagram()
\end{lstlisting}

\begin{figure}[ht]
  \centering
  \includegraphics[scale=0.5]{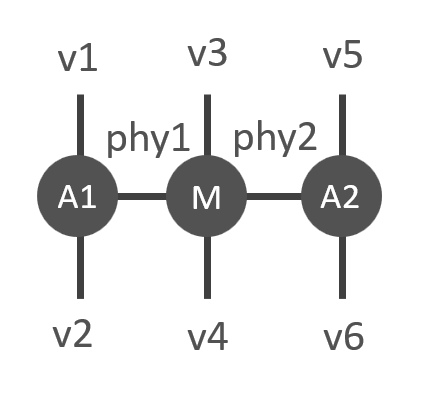}
  \caption{Diagram for contraction of three tensors.}
  \label{fig:TensorContracts}
\end{figure}

To contract multiple tensors, one uses a list containing all \lstinline{UniTensors}  as the first argument.
There are two further, optional arguments: \lstinline{order} (string), which specifies a desired contraction order, 
and \lstinline{optimal} (\lstinline{True} or \lstinline{False}; default \lstinline{True}), for specifying if one wants to use an automatically optimized contraction order. 
Since, the contraction order is given in terms of the names of the \lstinline{UniTensor}s, the names have to be set for all tensors involved in the contraction; otherwise a run-time error is raised. If \lstinline{optimal=True}, a contraction order is computed in every function call, which creates some unnecessary overhead if a similar contraction is executed several times. In these cases, we recommend using a \lstinline{Network} object for the contraction (see \cref{sec:network}), where the optimal order can be calculated in the first run and be reused for consecutive contractions. When a specific order is given, the \lstinline{optimal} argument should be set to \lstinline{False}.
As an example, consider a contraction consisting of tensors \lstinline{A1}, \lstinline{A2} and \lstinline{M} as sketched in \cref{fig:TensorContracts}.
The corresponding code example is shown in \cref{code:UniTensorContracts}.

In certain scenarios, one may need to change the labels of the input tensors for the contraction, while the labels of the original tensors shall stay unchanged. In this case, one can use \lstinline{UniTensor.relabel} to obtain a \lstinline{UniTensor} with new labels but without actually copying the data (see \cref{sec:relabel}). 
Alternatively, a \lstinline{Network} can be used where no initial relabeling is needed, see \cref{sec:network}.

\begin{lstlisting}[language=python,
                   caption={Contract three tensors according to \cref{fig:TensorContracts} with \lstinline{Contracts}.},
                   label={code:UniTensorContracts}]
# Create tensors A1, A2, M
A1 = cytnx.UniTensor.ones([2,8,8]).relabel(["phy1","v1","v2"])\
                          .set_name("A1")
A2 = cytnx.UniTensor.ones([2,8,8]).relabel(["phy2","v5","v6"])\
                          .set_name("A2")
M = cytnx.UniTensor.ones([2,2,4,4]).relabel(["phy1","phy2","v3","v4"])\
                          .set_name("M")
# Contract tensors using the auto-optimized contraction order
res = cytnx.Contract([A1,M,A2]).set_name("res")
# Contract tensors using a user-defined contraction order
res2 = cytnx.Contract([A1,M,A2], order="(M,(A1,A2))", optimal=False)
res2.set_name("res2")
# Check result
res.print_diagram()
\end{lstlisting}

\section{Network}
\label{sec:network}

Cytnx makes it easy to design, implement, and optimize the contraction of sophisticated tensor networks via the \lstinline{Network} object, which contains the blueprint of the corresponding tensor network. A \lstinline{Network} can be created from a string or a plain text file, which contains the description of the network in a human readable format. The \lstinline{Network} can be displayed as well to check the implementation. One can contract multiple tensors at once with a \lstinline{Network}. Furthermore, the contraction order can either be specified by the user, or an optimal order can be generated automatically. 
A \lstinline{Network} is particularly useful when the same kind of tensor network contraction has to be performed repeatedly, but with different tensor objects each time. We can define the \lstinline{Network} once and reuse it for several contractions. The contraction order can be optimized once after the first initialization with tensors. In the proceeding steps, this optimized order can be reused. This re-usability is one of the main advantages of \lstinline{Network} over \lstinline{ncon}.
Another main difference is that \lstinline{Network} is label based, while \lstinline{ncon} is index based.
User can use physically meaningful strings to label the tensors and the tensor network. This makes the code more human readable and easier to debug. Furthermore, users do not need to worry about the internal ordering of the indices, which avoids errors due to incorrect permutations of the indices.

Using \lstinline{Network} to perform a tensor contraction includes three steps. First, we need to define the target tensor network and create a \lstinline{Network} object. Second, we need to load (\lstinline{put}) the actual \lstinline{UniTensor}s that are to be contracted. Finally, we launch the contraction. In the following we explain these steps in more detail. 

\subsection{Creating a Network}\label{sec:creatnet}
\begin{figure}[ht]
  \centering
  \includegraphics[scale=0.6]{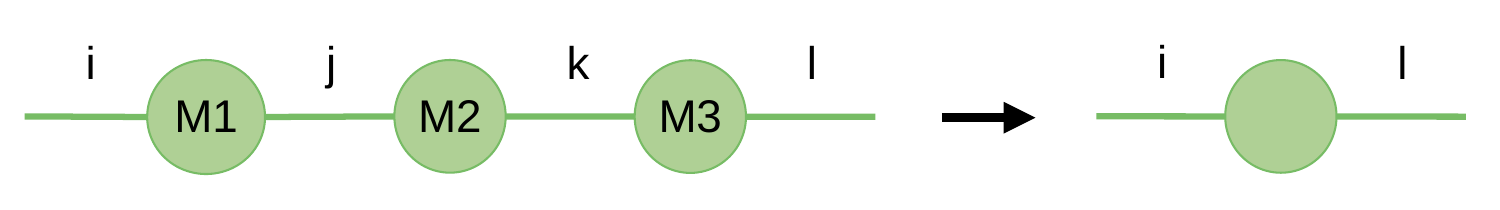}
  \caption{Network diagram for the the three matrices multiplication.}
  \label{fig:MMM}
\end{figure}

The creation of a \lstinline{Network} object is demonstrated with an example here: we use a \lstinline{Network} to perform two matrix-matrix multiplications for three rank-2 tensors $M_1$, $M_2$, and $M_3$ at once. 
The outcome is also a rank-2 tensor. The network diagram and its output tensor are shown in \cref{fig:MMM}. The code example in \cref{code:NetworkMMM} shows how to create a corresponding \lstinline{Network} object from a string.
One can print the network to check the structure.

\begin{lstlisting}[language=python,
                   caption={Create a Network for the multiplication of three matrices according to \cref{fig:MMM}.},
                   label={code:NetworkMMM}]
# Define a Network object for the contraction task in (*\color{codegreen}\cref{fig:MMM}*)
net = cytnx.Network()
net.FromString(["M1: i, j",\
                "M2: j, k",\
                "M3: k, l",\
                "TOUT: i, l",\
                "ORDER: ((M1,M2),M3)"])
print(net)

''' -------- Output ---------
==== Network ====
[x] M1 : i j 
[x] M2 : j k 
[x] M3 : k l 
TOUT : i ; l 
ORDER : ((M1,M2),M3)
=================
'''
\end{lstlisting}

The contraction is defined by the rule that all the common indices are summed over, and the remaining indices will be the indices of the output tensor. The keyword \lstinline{TOUT} is used to specify the \emph{order} of the remaining indices that are not summed over.
In this example, indices labeled by \lstinline{i} and \lstinline{j} are not contracted and they become the first and second index of the output tensor.
Consequently, if the output tensor represents a scalar (rank-0 tensor), no labels should be given after \lstinline{TOUT:}. 
The keyword \lstinline{ORDER} is used to specify a user-defined contraction order. If nothing is specified after \lstinline{ORDER:}, the default order is contracting the tensors in the order of their appearance in the creation of the \lstinline{Network}.
One can modify the \lstinline{ORDER} after the creation of the \lstinline{Network} or let Cytnx find an optimal contraction order after all tensors are loaded, see \cref{sec:network_optical}.

One can also create a \lstinline{Network} object from a file. For this, the same content as in the aforementioned network creation string (\cref{code:network_file}) is written to a plain text file. Here, we name the file \lstinline{example.net}. Such a network file can then be loaded with the \lstinline{Network} initializer or by using \lstinline{Network.Fromfile}. One can also write a \lstinline{Network} object to a network file using \lstinline{Network.Savefile}. The code example below (\cref{code:NetworkFile}) illustrates loading and saving a network file.

\begin{lstlisting}[language=,
                   caption={Network file "example.net", corresponding to the network in \cref{code:NetworkMMM}.},
                   label={code:network_file}]
M1: i, j
M2: j, k
M3: k, l
TOUT: i, l
ORDER: ((M1,M2),M3)
\end{lstlisting}


\begin{lstlisting}[language=Python,
                   caption={Create a Network from a file.},
                   label={code:NetworkFile}]
# Create a Network object from the network file in (*\color{codegreen}\cref{code:network_file}*)
net = cytnx.Network("example.net")
# Write the current network to a network file
net.Savefile("savefile")
# Read a network file
net.Fromfile("savefile.net")
\end{lstlisting}

\begin{figure}[ht]
  \centering
  \includegraphics[scale=0.65]{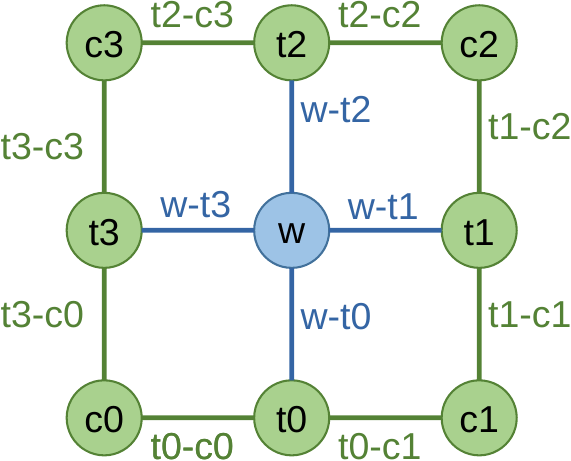}
  \caption{Network diagram for the corner transfer matrix algorithm.}
  \label{fig:CTM}
\end{figure}

To further demonstrate the power of \lstinline{Network}, we consider a contraction that appears in the corner transfer matrix algorithm~\cite{CTRG_original,CTRG_improved}. The tensor network diagram is sketched in \cref{fig:CTM}.
In the code example below (\cref{code:CTM}), we create a \lstinline{Network} from a string for this case.
The example demonstrates that it is also straightforward to translate a more complicated network diagram into a string, which defines the corresponding \lstinline{Network}. The \lstinline{Network} object is thus particularly useful for algorithms in higher dimensions, where many tensors need to be contracted. 
We note that in the case shown here, \lstinline{TOUT:} is empty because the outcome tensor is a scalar.

\begin{lstlisting}[language=Python, 
                   caption={Create a \lstinline{Network} for the corner transfer matrix method as shown in \cref{fig:CTM}.},
                   label={code:CTM}]
net = cytnx.Network()
net.FromString(["c0: t0-c0, t3-c0",\
                "c1: t1-c1, t0-c1",\
                "c2: t2-c2, t1-c2",\
                "c3: t3-c3, t2-c3",\
                "t0: t0-c1, w-t0, t0-c0",\
                "t1: t1-c2, w-t1, t1-c1",\
                "t2: t2-c3, w-t2, t2-c2",\
                "t3: t3-c0, w-t3, t3-c3",\
                "w: w-t0, w-t1, w-t2, w-t3",\
                "TOUT:",\
                "ORDER: ((((((((c0,t0),c1),t3),w),t1),c3),t2),c2)"])
\end{lstlisting}

We emphasize that the labels used in the \lstinline{Network} are dummy indices only for the network, and are completely independent of the labels in the \lstinline{UniTensor} objects that are going to be contracted. This design gives us the flexibility to define a \lstinline{Network} object without knowing the labeling conventions of the \lstinline{UniTensor}s to be contracted. In particular, the \lstinline{Network} can be loaded with different \lstinline{UniTensor}s for each contraction and be reused.

\subsection{Putting UniTensors and performing a contraction}
A \lstinline{Network} is a skeleton that does not contain the actual tensor objects to be contracted. To perform the contraction, one needs to specify a \lstinline{UniTensor} object for each tensor defined in the \lstinline{Network}, and make a connection between the dummy indices in the \lstinline{Network} and the actual indices of the \lstinline{UniTensor}s. This is done by using \lstinline{Network.PutUniTensor}. To give an example, we consider the \lstinline{Network} defined in \cref{code:NetworkMMM} and the following \lstinline{UniTensor}s which shall be contracted:

\begin{figure}[ht]
  \centering
  \includegraphics[scale=0.65]{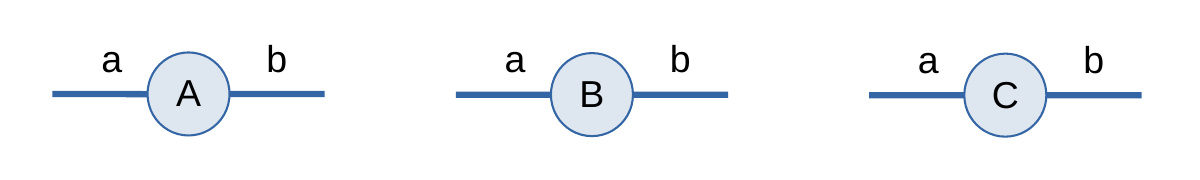}
  \caption{UniTensors to be put into the Network.}
  \label{fig:UniTensorNetwork}
\end{figure}

Note that the indices of \lstinline{A}, \lstinline{B}, and \lstinline{C} have common index labels $a$ and $b$. This is often the case in certain tensor network applications. 
We load the \lstinline{Network} with the specific \lstinline{UniTensor}s by using \lstinline{Network.PutUniTensor} for each tensor individually.
The first argument is the name of the tensor to be specified in the network, and the second argument is the \lstinline{UniTensor} object as a realization of that tensor. The third argument is a list of the labels in the \lstinline{UniTensor} in the order of the indices of the \lstinline{Network}. 
For example, \lstinline{A} is a realization of \lstinline{M1}, and the indices \lstinline{a} and \lstinline{b} in \lstinline{A} correspond to the indices \lstinline{i} and \lstinline{j} in \lstinline{M1}.
Similarly, \lstinline{B(C)} is a realization of \lstinline{M2(M3)}, and the indices \lstinline{a} and \lstinline{b} in \lstinline{B(C)} correspond to the indices \lstinline{j(k)} and \lstinline{k(l)} in \lstinline{M2(M3)}.
When all tensors in the network are specified, one can perform the contraction by simply calling \lstinline{Network.Launch} to obtain the result. Note that the output tensor inherits the index order and labels from the specifications following \lstinline{TOUT:}.

\begin{lstlisting}[language=python,
                   caption={Load tensors into the network and launch contraction.},
                   label={code:PutTensor}]
# Supposes net is a Network that is already created as in (*\color{codegreen}\cref{code:NetworkMMM}*)
# Create UniTensors that are going to be contracted
A = cytnx.UniTensor.ones([2,2]).relabel(["a","b"]).set_name("A")
B = cytnx.UniTensor.ones([2,2]).relabel(["a","b"]).set_name("B")
C = cytnx.UniTensor.ones([2,2]).relabel(["a","b"]).set_name("C")
# Put UniTensors into a Network object
net.PutUniTensor("M1", A, ["a","b"])
net.PutUniTensor("M2", B, ["a","b"])
net.PutUniTensor("M3", C, ["a","b"])
# Launch the contraction
Tout = net.Launch()
\end{lstlisting}

\subsection{Setting the contraction order}
\label{sec:network_optical}

The efficiency of a tensor network contraction depends on the order of the contractions. 
This order can be specified when the \lstinline{Network} is defined (see \cref{sec:creatnet}). However, one can also reset the contraction order using the method \lstinline{Network.setOrder}. A certain contraction order can be forced in the same way as explained in \cref{sec:creatnet}. Alternatively, one can let Cytnx compute the optimal order by setting \lstinline{Network.setOrder(optimal=True)}. Note that in this case, the optimal order will be computed (and stored) \emph{every time} when a contraction is launched. This is necessary because the optimal order depends on the bond dimensions of the input tensors, which can be different for each contraction. In practice, one often increases the dimensions of the indices during the simulations until they are kept constant when they reach a certain value. Then, the optimal contraction order typically does not change anymore. In this cases, one can set \lstinline{Network.setOrder(optimal=True)} up to some large enough bond dimensions, and then set \lstinline{Network.setOrder(optimal=False)} such that the \lstinline{Network} will no longer compute the optimal order before the contractions, but just use the stored contraction order. Similarly, the dimensions do not change when a network contraction is used in an iterative solver many times. Then, the contraction order only needs to be optimized in the first iteration. One can check the current contraction order using \lstinline{Network.getOrder}:

\begin{lstlisting}[language=python, caption={Set and get the contraction order.}]
# Supposes net is a Network that is already created as in (*\color{codegreen}\cref{code:NetworkMMM}*)
# and all tensors are loaded (put) as in (*\color{codegreen}\cref{code:PutTensor}*)
# Calculate the optimal contraction order for the loaded tensors
net.setOrder(optimal=True)
# Set a specific contraction order
net.setOrder(optimal=False, contract_order='(M2,(M1,M3))')
# Print the current contraction order
print(net.getOrder()) # Output: (M2,(M1,M3))
\end{lstlisting}

\subsection{Advanced example: Expectation value of a Projected Entangled Pair State}
\begin{figure}[ht]
    \centering
    \includegraphics[scale=0.5]{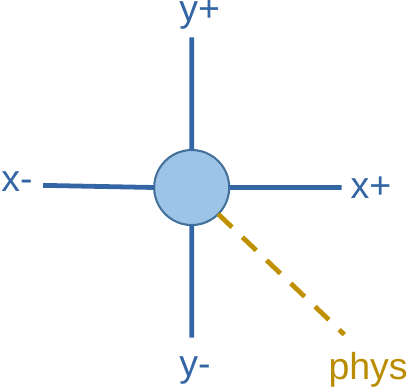}
    \caption{PEPS tensor and labeling convention.}
    \label{fig:peps_tensor}
\end{figure}

\begin{figure}[ht]
    \centering
    \includegraphics[scale=0.5]{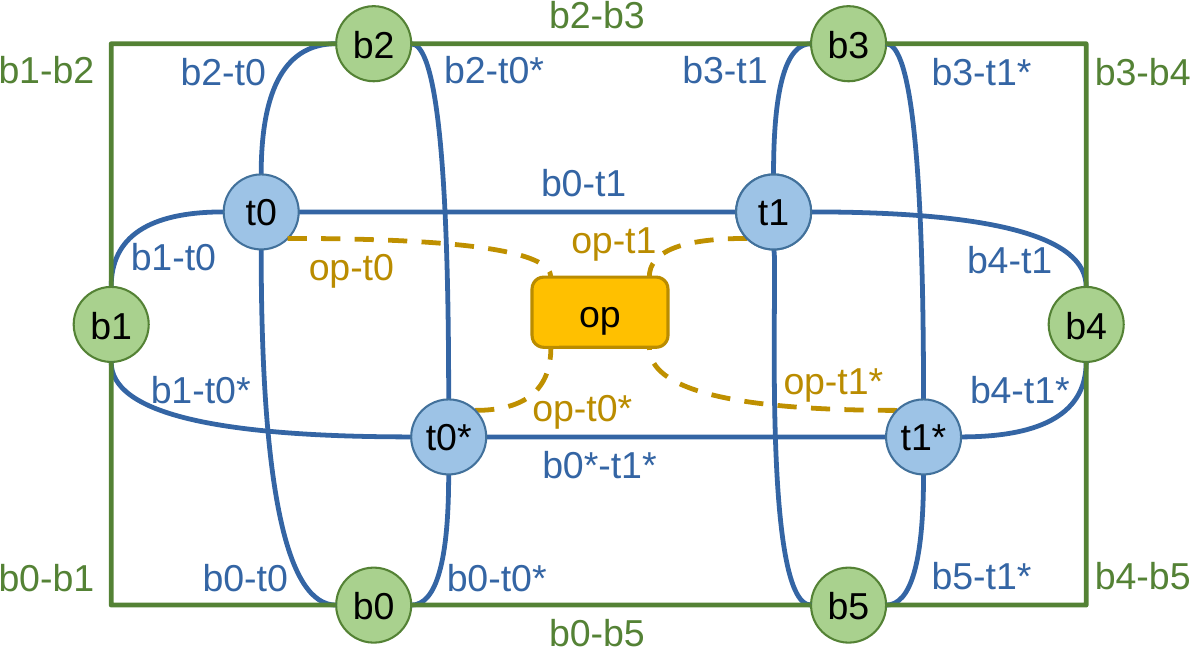}
    \caption{Expectation value for a nearest neighbor operator acting on a PEPS. The boundary MPS approximation is used. Symbology -- yellow: operator; blue: PEPS tensors; starred blue: corresponding complex conjugate PEPS tensors; green: boundary matrix product state, an approximation of the remaining sites. Adapted from Ref.~\cite{Schneider2022}.}
    \label{fig:peps}
\end{figure}

In this advanced example, we show how a more complicated tensor network contraction can be implemented by using \lstinline{Network}. 
Specifically, we consider the contraction for the expectation value of a two-dimensional tensor network state.
Physical states in two spatial dimensions, in particular ground states, can be approximated by a Projected Entangled Pair State (PEPS)~\cite{peps}. This tensor network consists of one tensor at each lattice site. One index of each tensor corresponds to the physical degrees of freedom at this site (dotted lines in \cref{fig:peps_tensor,fig:peps}). Further internal bonds connect each tensor to its nearest neighbors. \cref{fig:peps_tensor} shows a PEPS tensor in two spatial dimensions.

A typical PEPS algorithms prepares a desired state, for example the ground state of a Hamiltonian. Then, expectation values are measured as the quantities of interest for physics~\cite{ORUS2014117}. The boundary matrix product state (boundary MPS) method is used to efficiently calculate expectation values. See Ref.~\cite{PhysRevB.104.155118} as an example of a finite lattice PEPS calculation with more details on the algorithm.

The expectation value of a nearest neighbor operator $op$ corresponds to the tensor network in \cref{fig:peps}. 
Here, the blue tensors $t0$ and $t1$ are the PEPS tensors that the operator acts on, while the starred versions are there complex conjugate counterparts.
The green tensors $b0$ to $b5$ represent a boundary matrix product state, which is used to approximate the PEPS tensors at the remaining sites.

The diagram in \cref{fig:peps} can directly be translated into a \lstinline{Network} file. One writes the names of the tensors and their indices one by one to the file:

\begin{lstlisting}[language=python, label={code:peps_network}, caption={Network file \lstinline{peps_exp_val.net} describing the contraction in \cref{fig:peps}.}]
#network file peps_exp_val.net
b0:  b0-b5,  b0-b1,   b0-t0,  b0-t0*
b1:  b0-b1,  b1-b2,   b1-t0,  b1-t0*
b2:  b1-b2,  b2-b3,   b2-t0,  b2-t0*
b3:  b2-b3,  b3-b4,   b3-t1,  b3-t1*
b4:  b3-b4,  b4-b5,   b4-t1,  b4-t1*
b5:  b4-b5,  b0-b5,   b5-t1,  b5-t1*
t0:  op-t0,  t0-t1,   b0-t0,  b1-t0,  b2-t0
t0*: op-t0*, t0*-t1*, b0-t0*, b1-t0*, b2-t0*
t1:  op-t1,  t0-t1,   b3-t1,  b4-t1,  b5-t1
t1*: op-t1*, t0*-t1*, b3-t1*, b4-t1*, b5-t1*
op:  op-t0,  op-t0*,  op-t1,  op-t1*
TOUT:
ORDER: 
\end{lstlisting}
The actual contraction is done by loading the \lstinline{Network} file and using \lstinline{PutUniTensor} with the correct \lstinline{UniTensors}. 
This way, the expectation value can be calculated for tensors at different sites, and also other naming conventions of the indices could be used without changing the format of the \lstinline{Network} file. Our index labeling convention for the PEPS tensors can be seen in \cref{fig:peps_tensor}. The names of the boundary MPS tensors start with "bMPS" in our convention, and the index labels indicate to which kind of tensor the index connects to.

\begin{figure}[ht]
    \centering
    \includegraphics[scale=0.5]{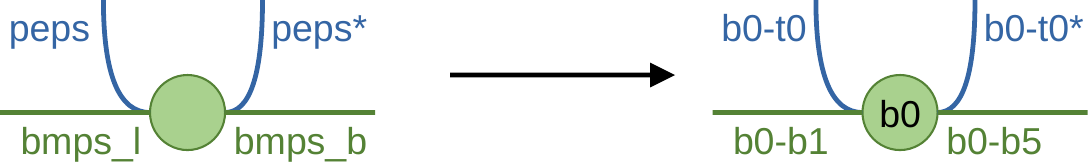}
    \caption{Example for the mapping of index labels. Left: \lstinline{UniTensor} \lstinline{bMPS_b0} with index labels. Right: Abstract tensor definition \lstinline{b0} in the \lstinline{Network}, see \cref{fig:peps}. The labels are matched by giving the bond labels as a third argument in \lstinline{Network.PutUniTensor}as in \cref{code:peps_contraction}.}
    \label{fig:peps_bmps}
\end{figure}

When loading the actual tensors, the indices of a \lstinline{UniTensor} have to be assigned to the corresponding ones in the Network. This can simply be done by the index labels in the correct order as the third argument of \lstinline{net.PutUniTensor}. \Cref{fig:peps_bmps} shows an example for this mapping. The code to put the \lstinline{UniTensors} and launch the contraction is shown below in \cref{code:peps_contraction}.

\begin{lstlisting}[language=python, label={code:peps_contraction}, caption={Put UniTensors and launch contraction. The index labels of the PEPS tensors are shown in \cref{fig:peps_tensor}. The boundary MPS tensors are named according to their position t(op), b(ottom), l(eft), r(ight). Their index labels indicate the kind of tensor they connect to. See \cref{fig:peps_bmps} for an example of the index convention and mapping.}]
# Supposes "peps_exp_val.net" is a Network that is already created as
# in (*\color{codegreen}\cref{code:peps_network}*) and the UniTensors to be put exist
ten0 = cytnx.UniTensor.ones([2,2,2,2,2]).relabel([
    "phys", "x+", "y-", "x-", "y+"
])
conjten0 = cytnx.UniTensor.ones([2,2,2,2,2]).relabel([
    "phys", "x+", "y-", "x-", "y+"
])
ten1 = cytnx.UniTensor.ones([2,2,2,2,2]).relabel([
    "phys", "x-", "y+", "x+", "y-"
])
conjten1 = cytnx.UniTensor.ones([2,2,2,2,2]).relabel([
    "phys", "x-", "y+", "x+", "y-"
])
bMPS_b0 = cytnx.UniTensor.ones([2,2,2,2]).relabel([
    "bmps_b", "bmps_l", "peps", "peps*"
])
bMPS_b1 = cytnx.UniTensor.ones([2,2,2,2]).relabel([
    "bmps_r", "bmps_b", "peps", "peps*"
])
bMPS_t0 = cytnx.UniTensor.ones([2,2,2,2]).relabel([
    "bmps_l", "bmps_t", "peps", "peps*"
])
bMPS_t1 = cytnx.UniTensor.ones([2,2,2,2]).relabel([
    "bmps_t", "bmps_r", "peps", "peps*"
])
bMPS_l = cytnx.UniTensor.ones([2,2,2,2]).relabel([
    "bmps_b", "bmps_t", "peps", "peps*"
])
bMPS_r = cytnx.UniTensor.ones([2,2,2,2]).relabel([
    "bmps_t", "bmps_b", "peps", "peps*"
])
op = cytnx.UniTensor.ones([2,2,2,2]).relabel([
    "peps_left", "peps_left*", "peps_right","peps_right*"
])

net = cytnx.Network("peps_exp_val.net")
net.PutUniTensor("t0",  ten0,     ["phys", "x+", "y-", "x-", "y+"])
net.PutUniTensor("t0*", conjten0, ["phys", "x+", "y-", "x-", "y+"])
net.PutUniTensor("t1",  ten1,     ["phys", "x-", "y+", "x+", "y-"])
net.PutUniTensor("t1*", conjten1, ["phys", "x-", "y+", "x+", "y-"])
net.PutUniTensor("b0",  bMPS_b0,  ["bmps_b", "bmps_l", "peps", "peps*"])
net.PutUniTensor("b5",  bMPS_b1,  ["bmps_r", "bmps_b", "peps", "peps*"])
net.PutUniTensor("b2",  bMPS_t0,  ["bmps_l", "bmps_t", "peps", "peps*"])
net.PutUniTensor("b3",  bMPS_t1,  ["bmps_t", "bmps_r", "peps", "peps*"])
net.PutUniTensor("b1",  bMPS_l,   ["bmps_b", "bmps_t", "peps", "peps*"])
net.PutUniTensor("b4",  bMPS_r,   ["bmps_t", "bmps_b", "peps", "peps*"])
net.PutUniTensor("op",  op,       ["peps_left", "peps_left*",
                                   "peps_right","peps_right*"])
exp_val = net.Launch().get_elem([0])
\end{lstlisting}

The diagram in \cref{fig:peps} is intended for calculating an expectation value of an operator that acts on a nearest-neighbor pair in x-direction. However, the same \lstinline{Network} file can be used also for operators acting in y-direction. 
Only the arguments of \lstinline{PutUniTensors} need to be adapted. This way, tensor diagrams can be implemented as network files and be used flexibly, without the need to rewrite them for different tensors to be loaded.
Also, the user does not need to care about the index order inside the \lstinline{UniTensor} if labeled indices are used.

\section{Tensor decomposition and linear algebra}\label{sec:decomp}

Tensor decompositions by linear algebra algorithms are essential ingredients in tensor network applications. 
Cytnx supports an extensive number of tensor operations and linear algebra functions and we refer to the user guide~\cite{cytnxDoc_LinearAlgebra} for the complete list and further documentation. Typically, linear algebra functions require matrices as input. Consequently, before calling a linear algebra function, Cytnx {\em internally} maps a \lstinline{UniTensor} to a matrix. In \cref{sec:matrix_mapping} we explain how to ensure that the correct mapping is carried out. Then, we show how to perform singular value decomposition (SVD) (\cref{sec:svd}), eigenvalue decomposition (\cref{sec:eigen}), and QR decomposition (\cref{sec:qr}) respectively.

\subsection{Mapping from tensor to matrix}
\label{sec:matrix_mapping}

Cytnx provides several linear algebra functions and tensor decomposition methods. The library calls established and well optimized linear algebra functions, which require matrices as input. Therefore, the tensors need to be represented by matrices in any of the methods. This is done by classifying each index in a tensor to be either part of the resulting row index or column index, and then combining all the row (column) indices into one row (column) index for the matrix representation.
The dimension of the final row (column) index is the product of the dimensions of all the original row (column) indices.

The \lstinline{UniTensor} objects have an attribute called \lstinline{rowrank} to control the mapping from a tensor to a matrix. In matrix representation, the first \lstinline{rowrank} indices are treated as row indices, and the remaining indices are treated as column indices.
There are different ways of mapping a tensor to a matrix depending on which indices are selected as row or column indices.
Before calling a tensor decomposition method, one needs to prepare the tensor such that it has the desired mapping. This is done by 1) permuting the indices such that all the row indices are in front of all the column indices, and 2) setting \lstinline{rowrank} to the number of row indices by using \lstinline{UniTensor.set_rowrank} or \lstinline{UniTensor.set_rowrank_}.

The following example shows how to prepare a specific matrix representation of a rank-3 \lstinline{UniTensor}, where one of the indices is interpreted as the row index and the other two indices form the column index:
\begin{lstlisting}[language=python,
    caption={Matrix form of a \lstinline{UniTensor}},
    label={code:matrixForm}]
# Create a rank-3 tensor with shape [2,2,6]
M = cytnx.UniTensor(cytnx.arange(2*2*6).reshape(2,2,6))\
                           .relabel(["a","b","c"]).set_name("M")
# 1. Permute the indices to ("c","a","b") ordering,
M.permute_(["c","a","b"])
# 2. Set rowrank=1, such that "c" is the row index,
# and "a", "b" are the column indices.
M.set_rowrank_(1)
print(M.rowrank()) # Output: 1
M.print_diagram()

''' -------- Output ---------
tensor Name : M
tensor Rank : 3
block_form  : False
is_diag     : False
on device   : cytnx device: CPU
          ---------     
         /         \    
   c ____| 6     2 |____ a
         |         |    
         |       2 |____ b
         \         /    
          ---------
'''
\end{lstlisting}
We see that \lstinline{M.rowrank} return the \lstinline{rowrank} information,
which is also encoded in the diagram created by \lstinline{UniTensor.print_diagram} -- the indices at the left (right) side are the row (column) indices.


\subsection{Singular value decomposition}\label{sec:svd}

\paragraph{Without truncation.}
The singular value decomposition (SVD) decomposes a matrix $M$ into the product of three matrices $U$, $S$, and $V^\dagger$ (i.e., $M=USV^\dagger$), where $S$ is a diagonal matrix and its non-negative elements are called singular values. $U$ and $V$ satisfy $U^\dagger U=I$ and $V^\dagger V=I$ with $I$ the identity matrix. In Cytnx, one can perform an SVD by using the function \lstinline{linalg.Svd}. There is an optional argument \lstinline{is_UvT} (default \lstinline{True}): if \lstinline{is_UvT=True}, the function returns three tensors $S$, $U$, and $V^\dagger$; if \lstinline{is_UvT=False}, the function returns only the singular-value tensor $S$. Note that this function does \emph{not} perform any truncation. We also note that $U$ inherits the row labels of $M$, while $V^\dagger$ inherits the column labels of $M$. The other labels are automatically generated in such a way that one can use \lstinline{Contract} to perform the matrix multiplication of $U$, $S$, and $V^\dagger$. However, the indices after the contraction with \lstinline{Contract} might have a different order and need to be permuted before comparing to the original tensor $M$. The following example shows how to perform a SVD of the previously created rank-3 tensor:

\begin{lstlisting}[language=python, caption={Singular value decomposition without truncation.}]
# Supposes M is a UniTensor that is already created as in (*\color{codegreen}\cref{code:matrixForm}*)
# Perform SVD on M
S, U, Vdag = cytnx.linalg.Svd(M)
# Compute the product of U*S*Vdag
Msvd = cytnx.Contract([U,S,Vdag])
Msvd.permute_(M.labels())
# Check that M = U*S*Vdag
Diff = M - Msvd
print(Diff.Norm()) # Output: 2.79115e-14
\end{lstlisting}

\paragraph{With truncation.}

In practice, one often uses an SVD as an approximating method by truncating the linear space based on the singular values. An SVD with truncation can be executed by calling \lstinline{linalg.Svd_truncate}. This function performs a full SVD of the initial tensors, and then truncates it to keep only the largest singular values. \lstinline{linalg.Svd_truncate} has several additional arguments:
\begin{itemize}
    \item \lstinline{keepdim} specifies the maximal dimension to be kept after the truncation.
    \item \lstinline{err} defines the threshold for singular values to be kept; the linear space with singular value smaller than \lstinline{err} will be truncated. Note that the singular values are not normalized before the truncation.
    \item \lstinline{return_err}; for \lstinline{return_err=1}, the function returns an additional one-element \lstinline{UniTensor} that stores the largest of the truncated singular values; for \lstinline{return_err=2}, a \lstinline{UniTensor} containing all the truncated singular values is returned instead; for \lstinline{return_err=0} (default), no truncation error is returned
    
\end{itemize}
Here is an example of an SVD with truncation:
\begin{lstlisting}[language=python, 
                   caption={Singular value decomposition with truncation.}]
# Supposes M is a UniTensor that is already created as in (*\color{codegreen}\cref{code:matrixForm}*)
# Perform SVD without truncation and return only the singular values
S_full = cytnx.linalg.Svd (M, is_UvT=False)
# Perform SVD with truncation
S_trunc, U, Vdag, s_err = cytnx.linalg.Svd_truncate\
                      (M, keepdim=3, err=1e-10, return_err=1)
# Compare the full sigular values with the kept and the truncated ones
print(S_full)   # [6.56235e+01 4.18988e+00 2.92323e-15 8.22506e-16 ]
print(S_trunc)  # [6.56235e+01 4.18988e+00 ]
print(s_err)    # [2.92323e-15 ]
\end{lstlisting}
\paragraph{SVD for symmetric UniTensor}

\lstinline{linalg.Svd} and \lstinline{linalg.Svd_truncate} also work with symmetric \lstinline{UniTensor}. In this case, returned $U$, $S$, and $V^\dagger$ are also symmetric \lstinline{UniTensor}s. In the fucntion \lstinline{linalg.Svd_truncate} a full SVD is still performed, then the singular values are sorted across all blocks and only the largest ones are kept. Consequently, some of the blocks might completely be discarded and do not contribute to the final tensors. In \cref{code:Svd_symm}, we show how to loop over all blocks in $S$ and print the singular values in each block. One finds that $S$ can be decomposed into three blocks, where the first/second/third block contains 2/4/2 singular values respectively.

\begin{lstlisting}[language=python, 
                   caption={Singular value decomposition without truncation.},
                   label={code:Svd_symm}]
# Define the bonds with quantum numbers
bond1 = cytnx.Bond(cytnx.BD_IN,\
                [cytnx.Qs(1)>>2, cytnx.Qs(-1)>>2],[cytnx.Symmetry.U1()])
bond2 = cytnx.Bond(cytnx.BD_IN,\
                [cytnx.Qs(1)>>2, cytnx.Qs(-1)>>2],[cytnx.Symmetry.U1()])
bond3 = cytnx.Bond(cytnx.BD_OUT,\
                [cytnx.Qs(2)>>2, cytnx.Qs(0)>>4, cytnx.Qs(-2)>>2],\
                [cytnx.Symmetry.U1()])
# Create a symmetric UniTensor with these bonds
uTsym = cytnx.UniTensor([bond1, bond2, bond3]).relabel(["a","b","c"])\
                        .set_name("uTsym")
uTsym.print_diagram()
# randomize the symmetric UniTensor
cytnx.random.uniform_(uTsym, low=-1., high=1., seed=4)

# Perform SVD on the symmetric UniTensor
S, U, Vdag = cytnx.linalg.Svd(uTsym)
for b in range(S.Nblocks()):
    print(S.get_block(b))

# Check truncation error by Norm(uTsym - U*S*Vdag)
truncerr = (uTsym - cytnx.Contract([U,S,Vdag])).Norm() # 2.77690e-15

''' -------- Output ---------
tensor Name : uTsym
tensor Rank : 3
contiguous  : True
valid blocks : 4
is diag   : False
on device   : cytnx device: CPU
      row           col 
         -----------    
         |         |    
   a  -->| 4     8 |-->  c
         |         |    
   b  -->| 4       |        
         |         |    
         -----------    

Total elem: 2
type  : Double (Float64)
cytnx device: CPU
Shape : (2)
[1.46508e+00 5.33174e-01 ]

Total elem: 4
type  : Double (Float64)
cytnx device: CPU
Shape : (4)
[2.30323e+00 1.78985e+00 1.59579e+00 1.13015e+00 ]

Total elem: 2
type  : Double (Float64)
cytnx device: CPU
Shape : (2)
[1.43749e+00 7.55563e-01 ]
'''

\end{lstlisting}

In \cref{code:Svd_truncate_symm}, we use \lstinline{linalg.Svd_truncate} and keep 4 singular values.
One finds that the first two blocks are truncated to 1 and 3 singular values respectively in this case. The third block is complete discarded. This can lead to spurious lost of whole symmetry sectors. In \cref{code:Svd_truncate_symm_min_blockdim} we show how this can be avoided by setting an optional argument \lstinline{min_blockdim}. For each block we can define a minimum number of singular values to be kept, even if they would be dropped by the other truncation criteria.

\begin{lstlisting}[language=python, 
                   caption={Singular value decomposition with truncation.},
                   label={code:Svd_truncate_symm}]
# Supposes uTsym is a symmetric UniTensor that is already created
# as in (*\color{codegreen}\cref{code:Svd_symm}*)
# Perform SVD with truncation on the symmetric UniTensor
S_trunc, U, Vdag, s_err = cytnx.linalg.Svd_truncate\
                      (uTsym, keepdim=4, err=1e-10, return_err=1)
for b in range(S_trunc.Nblocks()):
    print(S_trunc.get_block(b))

''' -------- Output ---------
Total elem: 1
type  : Double (Float64)
cytnx device: CPU
Shape : (1)
[1.46508e+00 ]

Total elem: 3
type  : Double (Float64)
cytnx device: CPU
Shape : (3)
[2.30323e+00 1.78985e+00 1.59579e+00 ]
```
\end{lstlisting}

\begin{lstlisting}[language=python, 
                   caption={Singular value decomposition with truncation and minimum block dimension.},
                   label={code:Svd_truncate_symm_min_blockdim}]
# Supposes uTsym is a symmetric UniTensor that is already created
# as in (*\color{codegreen}\cref{code:Svd_symm}*)
# Perform SVD with truncation on the symmetric UniTensor
# Force each block to contain at least one value by setting min_blockdim
S_trunc, U, Vdag, s_err = cytnx.linalg.Svd_truncate(uTsym, keepdim=4,\
                        min_blockdim=[1, 1, 1], err=1e-10, return_err=1)
for b in range(S_trunc.Nblocks()):
    print(S_trunc.get_block(b))

''' -------- Output ---------
Total elem: 1
type  : Double (Float64)
cytnx device: CPU
Shape : (1)
[1.46508e+00 ]

Total elem: 2
type  : Double (Float64)
cytnx device: CPU
Shape : (2)
[2.30323e+00 1.78985e+00 ]

Total elem: 1
type  : Double (Float64)
cytnx device: CPU
Shape : (1)
[1.43749e+00 ]
```
\end{lstlisting}
                   
We note in passing that the current implementation of \lstinline{linalg.Svd_truncate} first performs a full SVD, then does the truncation. There exists other algorithms which can perform a truncated SVD without a full SVD. One example is the Golub-Kahan-Lanczos bidiagonalization algorithm which builds a Krylov-like factorization of a general matrix~\cite{Golub1965}. Another example is the Randomized SVD(RSVD)~\cite{RSVD}.
These algorithms can be used to find the approximated dominant singular values and vectors. In general, they are less accurate but faster compared to truncation based on the full SVD, but the accuracy can be systematically improved at the cost of the exectution time. We plan to implement these algorithms in the future.

\subsection{Eigenvalue decomposition}\label{sec:eigen}

The eigenvalue decomposition, also known as matrix diagonalization, is defined for square matrices only. It decomposes a square matrix $M$ to $VDV^{-1}$, where $D$ is a diagonal matrix and its elements are called the eigenvalues. The columns of $V$ correspond to the right eivenvectors of $M$. If $M$ is a Hermitian matrix, all the eigenvalues must be real, and $V$ becomes a unitary matrix satisfying $V^\dagger V=VV^\dagger=I$. Consequently $M=VDV^\dagger$ for a Hermitian matrix.

One can perform the eigenvalue decomposition by using \lstinline{linalg.Eig} for a generic square matrix or \lstinline{linalg.Eigh} for a Hermitian matrix. Since the input matrix must be a square matrix, the row and the column dimensions in the matrix representation must be equal; otherwise, an error is raised. One can use the optional argument \lstinline{is_V} (default \lstinline{True}) to determine if the eigenvector matrix $V$ shall be returned. A complete description of the optional arguments can be found in the user guide~\cite{cytnxDoc}. Here is an example of an eigenvalue decomposition of a Hermitian matrix:

\begin{lstlisting}[language=python, 
                   caption={Eigenvalue decomposition.}]
# Create a randomly initialized Hermitian matrix
uT = cytnx.UniTensor.uniform([4,4], low=-1., high=1., seed=10)
uT = uT + uT.Conj().Transpose()
uT = uT.relabel(["a","b"]).set_name("uT").set_rowrank(1)
# Eigenvalue decomposition
eigvals, V = cytnx.linalg.Eigh(uT)
Vdag = V.Conj().Transpose()
# Set labels
V.relabel_(["a","_aux_L"]).set_name("V")
eigvals.relabel_(["_aux_L","_aux_R"]).set_name("eigvals")
Vdag.relabel_(["_aux_R","b"]).set_name("Vdag")
# Compare uT with V*eigvals*Vdag
Diff = uT - cytnx.Contract([V,eigvals,Vdag])
print(Diff.Norm()) # Output: 1.71297e-15
# Create identity matrix
eye = cytnx.UniTensor.zeros([4,4])
for i in range(4):
    eye[i,i] = 1
eye.relabel_(["_aux_L","_aux_R"]).set_name("eye")
# Compare eye with V*eye*Vdag
Diff = eye - cytnx.Contract([V,eye,Vdag])
print(Diff.Norm()) # Output: 7.67709e-16
\end{lstlisting}

If one is only interested in the lowest few eigenvalues, it can be very inefficient to first execute the contraction of a tensor network to obtain a matrix and then pass this matrix to the eigenvalue decomposition function. Methods like the Lancsoz solver for finding the smallest eigenvalues only require a linear operator, which defines how a vector is transformed. This can be a matrix multiplication, but also a more complicated tensor network contraction or any general linear transformation of the input vector. Cytnx provides a way to define such an operator as a \lstinline{LinOp} object and pass it to the \lstinline{Lanczos} linear algebra function to find the lowest few eigenvalues and corresponding eigenvectors of a Hermitian linear operator. More details and example code can be found in the user guide~\cite{cytnxDoc_IterativeSolver}.

\subsection{QR decomposition}\label{sec:qr}

The QR decomposition decomposes a matrix $M$ to the form $M = QR$, where $Q$ is a column-orthogonal matrix ($Q^T Q = I$), and $R$ is an upper-right triangular matrix. One can perform a QR decomposition by using \lstinline{cytnx.linalg.Qr}:

\begin{lstlisting}[language=python, 
                   caption={QR decomposition.}]
# Initial matrix
uT = cytnx.UniTensor.arange(5*4).reshape(5,4).relabel(["a","d"])\
                          .set_name("uT").set_rowrank(1)
# QR decomposition
Q, R = cytnx.linalg.Qr(uT)
# Compare uT with Q*R
Diff = uT - cytnx.Contract(Q,R)
print(Diff.Norm()) # Output: 1.14885e-14
# Check properties of Q,R
print(R) # upper triangular
QT = Q.Transpose().relabel_(["a","b"])
Q.relabel_(["b","c"])
eye = cytnx.UniTensor.zeros([4,4])
for i in range(4):
    eye[i,i] = 1
Diff = cytnx.Contract(QT, Q) - eye
print(Diff.Norm()) # Output: 6.13418e-16
\end{lstlisting}

\section{Device and GPU support}
\label{sec:gpu}

In Cytnx, the tensor elements can either be stored in the memory accessible to CPUs or GPUs. The use of GPUs can speed up tensor contractions and therefore shorten the run-time of tensor network algorithms. Using GPUs is very simple in Cytnx: After creating a tensor directly on a GPU or moving a tensor from CPU to GPU, the functions and methods provided by Cytnx can be used without any changes of the API. Therefore, tensor network algorithms can easily be ported to GPUs by only changing the initialization step of all tensors.
For example:

\begin{lstlisting}[language=python, label={code:gpu_example}, caption={Create and contract two tensors on a GPU.}]
# Create two UniTensors on a GPU
A = cytnx.UniTensor.ones([2,2,2], device=cytnx.Device.cuda+0,\
                         name="A", labels=["alpha","beta","gamma"])
B = cytnx.UniTensor.ones([2,2], device=cytnx.Device.cuda+0,\
                         name="B", labels=["beta","delta"])
# Or, using .to
A = cytnx.UniTensor.ones([2,2,2]).to(cytnx.Device.cuda+0)\
                    .set_name("A").relabel(["alpha","beta","gamma"])
B = cytnx.UniTensor.ones([2,2]).to(cytnx.Device.cuda+0)\
                         .set_name("A").relabel(["beta","delta"])
# Contract the two UniTensors with the same API as for CPU
AB = cytnx.Contract(A, B)
\end{lstlisting}

Since tensor elements can be stored in the memory of a given CPU or GPU on multi-CPU/GPU systems, it is important that the user ensures tensors are on the same device when they are used together in contractions or linear algebra functions. In Cytnx, the object \lstinline{cytnx.Device} handles all device properties.
The \lstinline{Device} of a \lstinline{UniTensor} can be seen when printing the tensor by \lstinline{print_diagram} or \lstinline{print} (see \cref{sec:print}).

\subsection{Device status}
\paragraph{Number of threads.}\label{sec:threadnum}
To see how many threads can be used in the current program by Cytnx, one can check \lstinline{Device.Ncpus}.

\begin{lstlisting}[language=python] 
print(cytnx.Device.Ncpus)
\end{lstlisting}

\paragraph{Number of GPUs.}
The number of GPUs available to Cytnx in the current program can be checked with \lstinline{Device.Ngpus}.

\begin{lstlisting}[language=python] 
print(cytnx.Device.Ngpus)
\end{lstlisting}
If Cytnx is not compiled with CUDA, \lstinline{Device.Ngpus} returns \lstinline{0}.

\subsection{Initializing tensors on CPU and GPU}
For the GPU computations, Cytnx uses CUDA-based third-party libraries like cuBlas~\cite{lib_cuBlas}, cuSolver~\cite{lib_cuSolver}, cuTensor~\cite{lib_cuTensor} and cuQuantum~\cite{lib_cuQuantum}. The performance of tensor network algorithms implemented in Cytnx can be significantly accelerated when using GPUs, as shown in \cref{sec:benchmark}. Developers of tensor network algorithms can directly benefit from the GPU acceleration.

Users can easily transfer tensor network algorithms implemented in Cytnx from the CPU to GPU version with minimal code modifications. Cytnx provides two ways to create objects on GPU devices: either by initializing objects on a GPU directly or by using the \lstinline{to} method to convert objects stored in the CPU memory.

\paragraph{Initializing tensors on a GPU.}
\lstinline{UniTensor} objects with data stored in the GPU memory can be created with the argument \lstinline{device} in the initialization step. For example:
\begin{lstlisting}[language=python,label={code:GPU_UniTensor_declare}, caption={Create a \lstinline{UniTensor} on a CPU or a GPU.}]
# Create a UniTensor on CPU; initialized with zeros
uT = cytnx.UniTensor.zeros([3,4,5], device=cytnx.Device.cpu)\
                .set_name("UniTensor").relabel(["a","b","c"])
# Create a UniTensor on GPU; initialized with zeros
uTgpu = cytnx.UniTensor.zeros([3,4,5], device=cytnx.Device.cuda+0)\
                .set_name("UniTensor on GPU").relabel(["a","b","c"])
# Check device
uT.print_diagram()      # on CPU
uTgpu.print_diagram()   # on GPU
\end{lstlisting}
Other initialization methods like \lstinline{ones}, \lstinline{arange}, \lstinline{ones}, \lstinline{uniform}, \lstinline{normal}, or \lstinline{eye} can be used similarly.

\paragraph{Data Transfer between Devices.}
The library provides a convenient way to move the data of Cytnx objects between CPU and GPU with the \lstinline{to} or \lstinline{to_} methods. For example, we create a \lstinline{UniTensor} in the memory accessible by the CPU and transfer it to the GPU with gpu-id=0:
\begin{lstlisting}[language=python,
    caption={Transfer a \lstinline{UniTensor} between CPU and GPU.},
    label={code:UniTensorGPU}]
uT = cytnx.UniTensor.ones([2,2]).set_name("uT").relabel(["a","b"])
uTgpu = uT.to(cytnx.Device.cuda+0).set_name("uTgpu")
uT.print_diagram()    # on CPU
uTgpu.print_diagram() # on GPU
uT.to_(cytnx.Device.cuda)
uT.print_diagram()    # on GPU
\end{lstlisting}

If only specific parts of an algorithm are suitable for GPU simulation, it is recommended to use the \lstinline{to} method to transfer the objects between CPU and GPU.

\section{Benchmarks}
\label{sec:benchmark}

\begin{figure}
    \centering
    \includegraphics[scale=0.5]{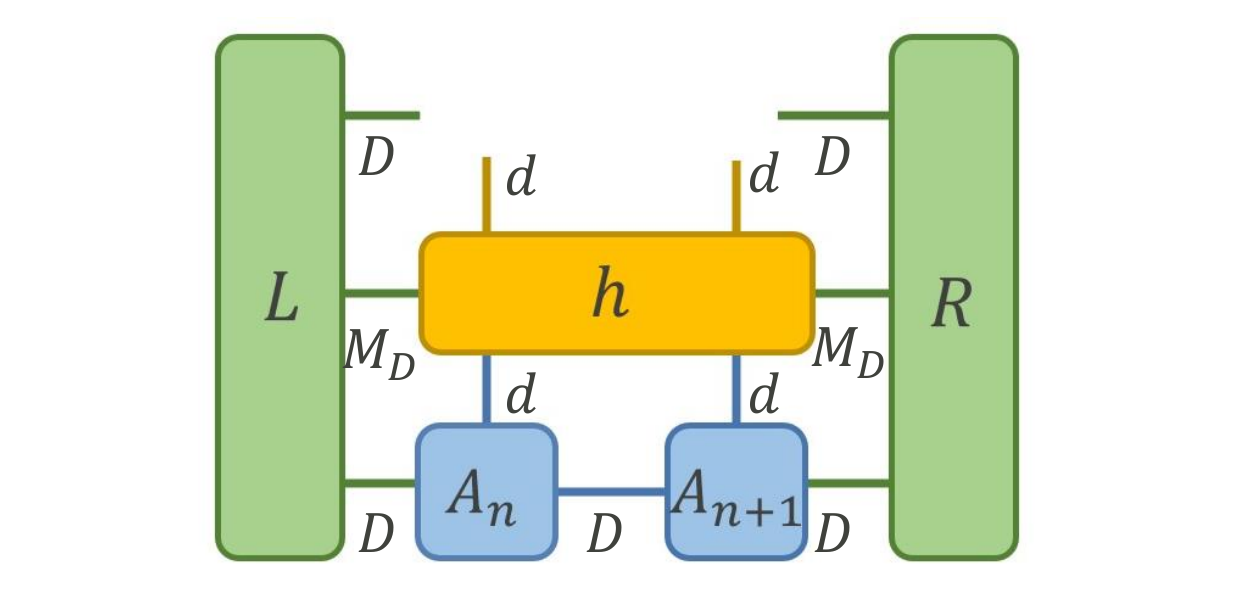}
    \caption{
    The tensor contraction which is used to update the $A_n$ and $A_{n+1}$ tensors during a DMRG sweep. 
    It involves the effective Hamiltonian composed of the left and right environment tensors ($L$ and $R$) and the two-site Hamiltonian tensor ($h$). The effective Hamiltonian is contracted with the effective wavefunction composed of the MPS tensors $A_n$ and $A_{n+1}$ at sites $n$ and $n+1$.
    The dimension of each index is indicated besides the bond, where $d$, $D$ and $M_D$ are the physical dimension, bond dimension and MPO bond dimension respectively. In our benchmark of the XX model: $d=2$, $M_D=4$ and $D$ is varied.
    }
    \label{fig:Hpsi_fig}
\end{figure}

We perform several benchmarks to test the performance of Cytnx.
As an example of a one-dimensional problem, we compare the execution times of a finite size density matrix renormalization group (DMRG) sweep and the main tensor contraction involved in it with Cytnx and ITensor\cite{itensor} on a CPU and on a GPU. Furthermore, as an example of a two-dimensional algorithm, we compare the execution times of the corner transfer matrix renormalization (CTMRG) algorithm in Cytnx and PEPS-Torch\cite{peps-torch} on a CPU.

\paragraph{Density matrix renormalization group.}
One of the main applications of tensor networks is to use them as a variational ansatz for a quantum many-body state. 
Here we benchmark the performance on a variational method, the 2-site DMRG algorithm~\cite{PhysRevLett.69.2863,SCHOLLWOCK201196}, which computes the ground state of a given Hamiltonian within a tensor network approximation. 
In particular, we consider the one-dimensional XX model with $N$ spins:
$$
\hat{H} = \sum^{N-1}_{j=1} ( \hat{S}^x_j \hat{S}^x_{j+1} + \hat{S}^y_{j} \hat{S}^y_{j+1} ).
$$
The Hamiltonian conserves the total spin $\sum_{j=1}^N S^z_j$ in the $z$ direction, which can be used as a quantum number. In a DMRG algorithm, one represents the ground state as a network of tensors $A_i$, where $i=1,\cdots N$. These tensors are updated sequentially from one boundary to the other in  a DMRG sweep. An essential step in the DMRG sweep is the update of two tensors $A_n$ and $A_{n+1}$, where a tensor contraction as sketched in \cref{fig:Hpsi_fig} is needed. In the figure, $d$, $D$, and $M_D$ correspond to the physical dimension, bond dimension and MPO bond dimension respectively. Since the bond dimension $D$ is related to the accuracy of the DMRG algorithm, we use $D$ as the control parameter in the benchmark. It is expected that the computational time scales as $D^3$ for large $D$. An implementation of DMRG using Cytnx can be found in the examples section in Cytnx's user guide~\cite{cytnxDoc}. We compare the executiont times with ITensor, which is implemented in both Julia and C++. In this benchmark, we use the C++ version of ITensor (v3.0.0).

\paragraph{CPU Benchmark.}

\begin{figure}
\centering
  \includegraphics[width=0.9\linewidth]{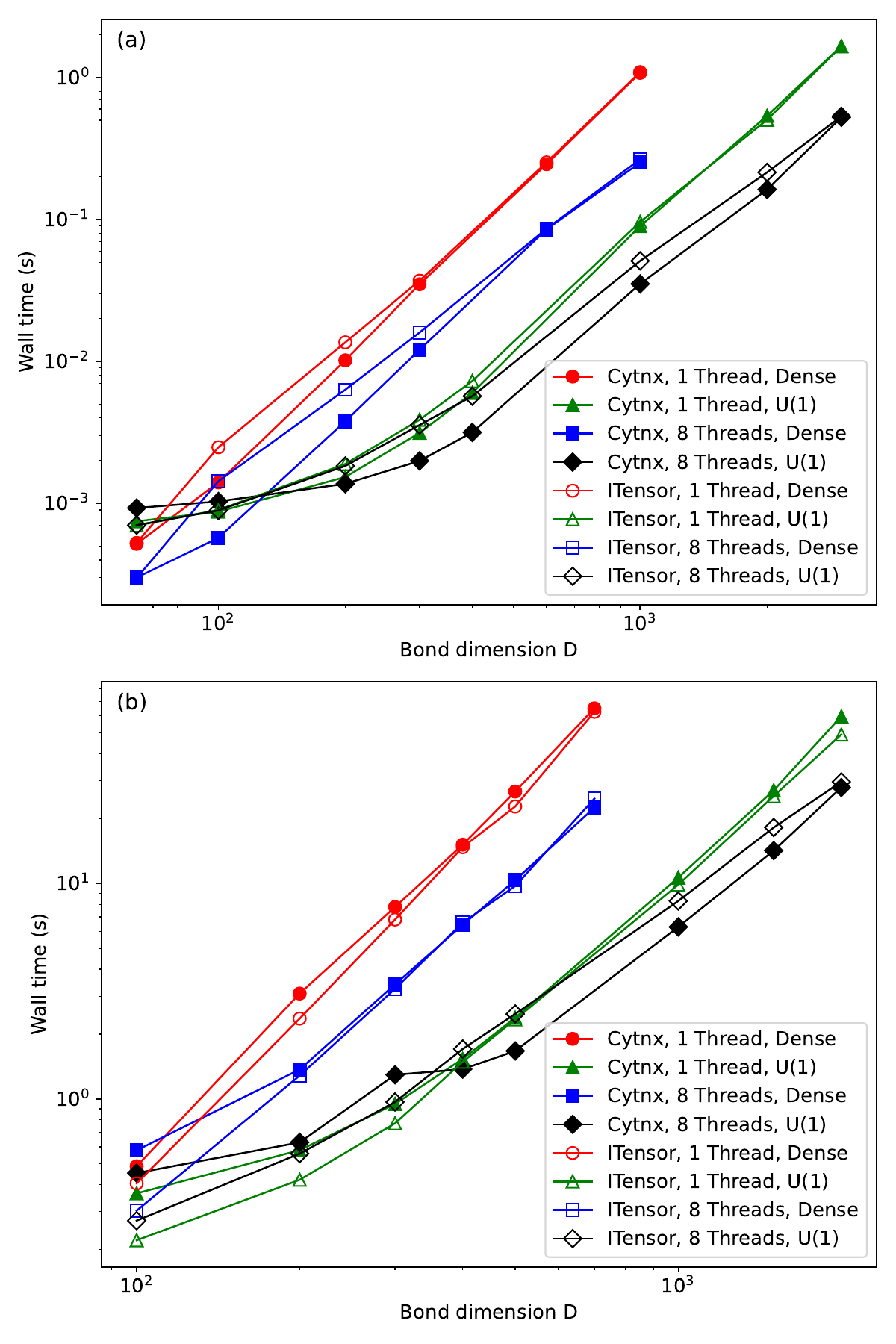}
  \caption{Computational time (wall time) for 
        (a) the tensor contraction sketched in \cref{fig:Hpsi_fig} and 
        (b) the DMRG sweep
        as a function of the bond dimensions $D$.
        Comparison between Cytnx and ITensor for dense and symmetric tensors, performed using one or eight threads on an Intel(R) Core i7-9700 CPU.} 
  \label{fig:benchmark_cpu}
\end{figure}

We first benchmark the performance on  an Intel(R) Core i7-9700 CPU. We perform the test with symmetric tensors which preserve the total spin, as well as unconstrained tensors without explicitly implemented symmetries. In \cref{fig:benchmark_cpu} (a), we show comparisons of wall times for the tensor contraction sketched in \cref{fig:Hpsi_fig}, as a function of $D$ on a log-log scale with Cytnx and ITensor. One can see that the performance of the two libraries is very similar. As expected, using symmetric tensors can reduce the computational costs except for very small bond dimensions. The parallelization with 8 threads reduces the execution time for the contraction compared to the single threaded run. \cref{fig:benchmark_cpu} (b) shows similar comparisons but for a whole DMRG sweep. Again, Cytnx and ITensor have a similar performance. We mention that the DMRG algorithm is implemented in slightly different ways in the two libraries, and additional observables are measured in ITensor. Therefore, it can only be concluded that the run-times of Cytnx are comparable to those of ITensor, even though one of the libraries might be slightly faster for certain tasks.

\begin{figure}
\centering
  \includegraphics[width=0.9\linewidth]{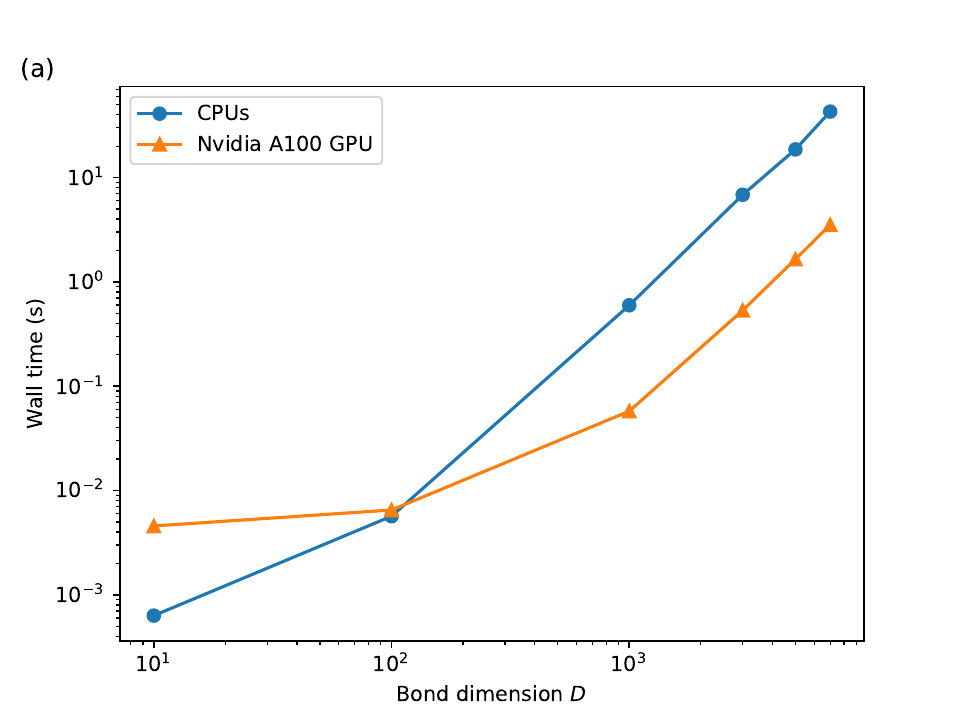}
  \includegraphics[width=0.9\linewidth]{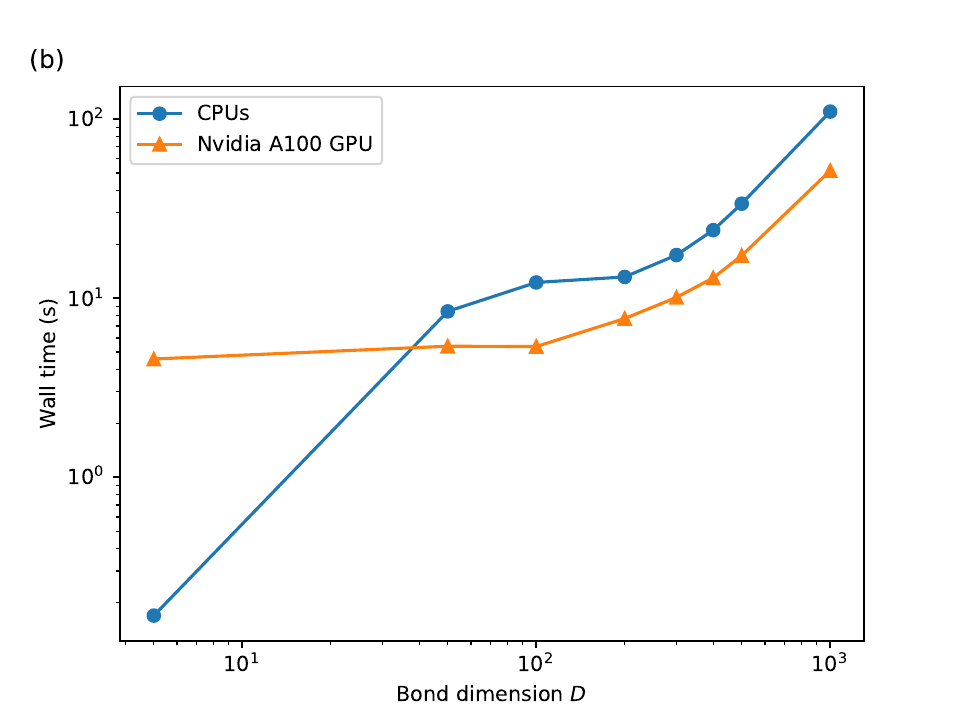}
  \caption{Computational time (wall time) for 
        (a) the tensor network contraction sketched in \cref{fig:Hpsi_fig} and 
        (b) a DMRG sweep with several bond dimensions of the matrix product state.
        Comparison between CPU and GPU calculations; execution times on an AMD EPYC 7742 CPU with 128 threads and a NVIDIA A100 (80GB).}
  \label{fig:benchmark_gpu}
\end{figure}

\paragraph{GPU Benchmark.}
We also benchmark the performance of Cyntx and ITensor on a GPU. The tests are executed on a machine with an AMD EPYC 7742 processor and one NVIDIA A100 GPU. For the CPU runs, the parallelization is realized through Intel MKL with 128 threads. 
Similar to the CPU benchmark, we test the performance of (a) the tensor contraction sketched in \cref{fig:Hpsi_fig} and (b) the DMRG sweep.
The benchmark results are shown in \cref{fig:benchmark_gpu}(a) and (b) respectively.
In these  benchmarks, one can observe a significant speedup for large bond dimensions when a GPU is used.
The execution times for the tensor contraction show that an 11$\times$ speedup can be reached for large bond dimensions. For a single iteration of a DMRG update, we gain a 2-3$\times$ speedup for large bond dimensions. We note that for $d=2, M_D=4, N=40$, one can run a finite-size DMRG on the NVIDIA A100 GPU (80G) up to $D\approx 6000$.

\paragraph{Corner transfer matrix.}

\begin{figure}
\centering
  \includegraphics[width=0.9\linewidth]{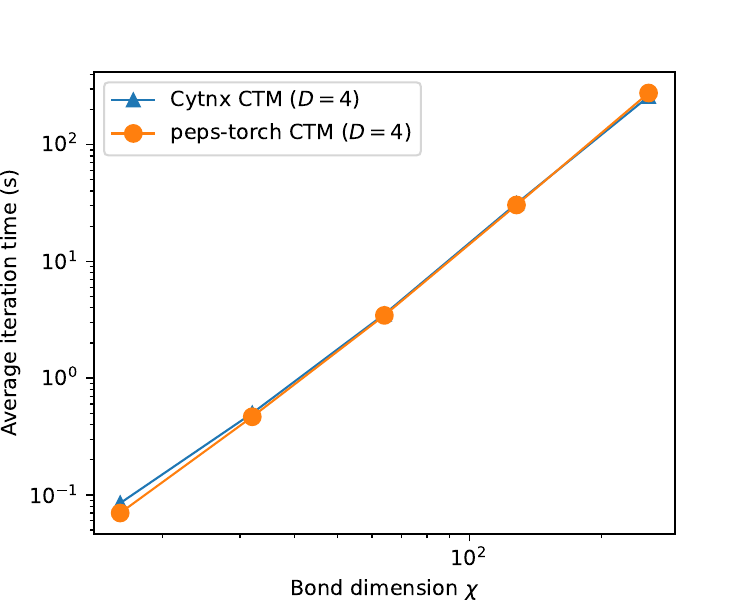}
  \caption{Computational time (wall time) for CTMRG to obtain converged corner tensor with bond dimension $\chi$ from an iPEPS of bond dimension $D$. Comparison between Cytnx and PEPS-torch; execution times on Intel Core i9-13900K CPU.}
  \label{fig:benchmark_ctm_cpu}
\end{figure}

Finally, we benchmark the corner transfer matrix renormalization group (CTMRG) algorithm~\cite{Nishino:1997kn}. In particular, we consider an iPEPS ansatz for a 2D quantum system and use CTMRG to compress the infinite tensor network representing the reduced density matrix of the ground state of the transverse field Ising model. The network is approximated by finite-dimensional corner tensors. The key ingredient is to iteratively enlarge the corner tensors by merging nearby tensors. Then, truncations are performed to obtain updated corner tenrosrs. This is repeated until convergence of the tensors. Two parameters define the systematic errors: the bond dimension $D$ of the iPEPS ansatz and the bond dimension $\chi$ of the corner tensor. Here, we compare the average execution time per iteration of the CTMRG algorithm to obtain the converged corner tensors by PEPS-Torch\cite{peps-torch} and the implementation in Cytnx on a CPU. As shown in \cref{fig:benchmark_ctm_cpu}, the average iteration times are very similar for the two implementations.

Summarizing, we conclude that Cytnx has a similar performance as other established tensor network libraries such as ITensor or PEPS-Torch. 
Cytnx allows users to run the most computationally expensive tasks on a GPU without much changes in the code by the user. This way, the computational time can be significantly reduced.

\section{Summary}\label{sec:summary}
We presented a library designed for high-dimensional tensor network algorithms. This library provides an easy-to-learn framework for new users to start developing their first tensor network code in a short period of time, yet comes with all the bells and whistles for advanced algorithms. This way, applications can quickly and conveniently be implemented and tested, but also be improved and optimized in a later step if desired. By design, the Cytnx Python and C++ interfaces follow the conventions in popular libraries for an intuitive usage. With almost identical APIs, users can easily switch between Python and C++ codes.
As a benchmark, we compare typical tensor network operations with ITensor and PEPS-Torch, and obtain similar performance. Following a similiar syntax as in PyTorch, users can benefit from GPU acceleration with only minimal changes of their code. In our benchmark, a significant boost of the performance can be achieved when tensor operations are executed on a GPU. 

There are several future directions for Cytnx. 
On the technical side, tensor network algorithms are often described as tensor diagrams, therefore a GUI interface that can automatically generate  code from a tensor diagram, similar to TensorTrace~\cite{evenbly2019tensortrace}, will be helpful. 
Automatic differentiation has revolutionized variationl optimization of complicated tensor networks such as PEPS ~\cite{autoDiff,autoDiff_CTM,autoDiff_iPEPS,autoDiff_CTMstable}, and provided simpler formulation to extract information about physical observables related to experiments, such as  spectral functions~\cite{10.21468/SciPostPhys.12.1.006}.
In the near future, Cytnx will support automatic differentiation through a PyTorch backend. 

Many physics problems involve fermions. In these cases, exchanging the order of two particles results in a sign flip because of the anti-commutation relation between fermionic operators. In tensor network algorithms, this can be implemented with symmetric tensors and specific tensors called swap-gates, which produce the correct sign changes~\cite{tensor_fermions,tensor_fermions_mera}. 
An alternative approach uses Grassman tensors~\cite{PhysRevB.88.115139,yosprakob2023grassmanntn}. In the future, an automated way to easily simulate fermionic systems will be implemented.

At the moment,  Cytnx supports only Abelian symmetries. With the current framework, it is easy to extend Cytnx to include symmetric {UniTensor}s with more complicated symmetries such as SU(2). This is planned for a future version. 

Regarding applications, several high-level tensor network algorithms such as (i)DMRG and iTEBD are implemented with Cytnx already~\cite{cytnxDoc}. We plan to extend this to provide a higher level abstraction in a spirit similar to the famous machine learning library Keras\cite{chollet2015keras}, so that physicists can focus on solving interesting physics problems. Through this, we hope to reduce the cognitive load for users and help develop more applications of tensor network algorithms in different fields of physics.
Going beyond physics, recently tensor network methods have found applications in fields such as machine learning, fluid dynamics, and quantum circuit simulation
\cite{NIPS2016_5314b967,PhysRevE.107.L012103,Chen_2021,Chen_2022,kiffner2023tensor,2023.Peddinti,liao2023simulation,2023fast,tindall2023efficient}. Due to the flexibility design and the versatility of Cytnx,
users can easily use Cytnx to develop applications in these directions and further enhance the synergy between tensor network and these fields. We provide an example of  quantum circuit simulation in App.~\ref{sec:qc} to demonstrate how this can be easily done using Cytnx.

\section{Acknowledgements}
We would like to thank the National Center for High-Performance Computing (NCHC) and NVIDIA for providing us access to the NVIDIA A100 GPU.
This work was supported by Taiwanese National Science and Technology Council (NSTC) Grant No. 113-2119-M-007-013.
C.-M.C. acknowledges the support by the NSTC under Grant No. 111-2112-M-110-006-MY3, and by the Yushan Young Scholar Program under the Ministry of Education (MOE) in Taiwan. 
We thank Ivana Gyro and Ian McCulloch for their significant contributions to the Cytnx library and their valuable suggestions.
We also thank Yu-Cheng Lin and Shih-Hao Hung for their help on the implementation of the GPU functionalities of Cytnx.


\appendix
\section{Tensor\label{sec:tensor}}

A tensor is typically represented in Cytnx by a \lstinline{UniTensor} object, which is explained in \cref{sec:uniten}. In addition to the data of the tensor elements, a  \lstinline{UniTensor} object also contains meta information for bonds, labels, name, and symmetries. Internally, the data of a \lstinline{UniTensor} is represented by one or more blocks. Each block contains a \lstinline{Tensor} object, which is is a multi-dimensional array containing the values of the tensor. All elements belong to the same datatype. The behavior and the API of \lstinline{Tensor} in Cytnx are very similar to the \lstinline{tensor} in PyTorch and the \lstinline{array} in NumPy. Typically, users do not have to deal with \lstinline{Tensor} objects and we recommend the users to use \lstinline{UniTensor} instead. However, we explain the underlying structure of a \lstinline{Tensor} for advanced users here. A more extensive description can be found in the user guide~\cite{cytnxDoc}.

\paragraph{Creating a Tensor.}\label{sec:tensor_create}

One can create a \lstinline{Tensor} through generators such as \lstinline{zero}, \lstinline{arange}, \lstinline{ones}:
\begin{lstlisting}[language=python, 
                   caption={Create a \lstinline{Tensor} by generators.}, 
                   label={code:TensorGenerators}]
# rank-3 Tensor of shape (3,4,5) with all elements set to zero.
A = cytnx.zeros([3,4,5])
# rank-1 Tensor containing the numbers [0,10) with step size 1
B = cytnx.arange(10)   
# rank-3 Tensor of shape (3,4,5) with all elements set to one.
C = cytnx.ones([3,4,5])
\end{lstlisting}
Moreover, one can create a randomly initialized \lstinline{Tensor}:
\begin{lstlisting}[language=python, caption={Create a randomly initialized \lstinline{Tensor}.}, label={code:RandomTensor}]
# Tensor of shape (3,4,5)
# All elements are distributed according to a normal distribution 
# around 0 with standard deviation 1
A = cytnx.random.normal([3,4,5], mean=0., std=1.)
# Tensor of shape (3,4,5)
# All elements are distributed uniformly between -1 and 1
B = cytnx.random.uniform([3,4,5], low=-1., high=1.)
\end{lstlisting}

\paragraph{Data types.}\label{sec:tensor_datatypes}
The default data type of the elements in a \lstinline{Tensor} is \lstinline{double} in C++ and \lstinline{float} in Python.
One can create a \lstinline{Tensor} with a different data type by providing an additional argument in the initialization function.
It is possible to convert a \lstinline{Tensor} of one data type to a different data type via \lstinline{Tensor.astype}.
The following example shows how to create a \lstinline{Tensor} with \lstinline{dtype=Type.Int64} and convert the data type to \lstinline{Type.Double}:

\begin{lstlisting}[language=python, caption={Data type of a \lstinline{Tensor} and data type conversion.}, label={code:TensorDataType}]
# Create a Tensor of 64bit integers
A = cytnx.ones([2,2], dtype=cytnx.Type.Int64)
# Change the dtype to Double
B = A.astype(cytnx.Type.Double)
# Print and check
print(A.dtype_str()) # Output: Int64
print(B.dtype_str()) # Output: Double (Float64)
\end{lstlisting}

\paragraph{Reshaping.}
The indices of a tensor can be combined or split, or brought in any other form of indices that match with the number of elements of the tensor. For this, the method \lstinline{reshape} can be used on a \lstinline{Tensor}. Suppose we want to create a rank-3 \lstinline{Tensor} with \lstinline{shape=(2,3,4)}. We can start with a rank-1 \lstinline{Tensor} with \lstinline{shape=(24)} initialized using \lstinline{arange}, and then reshape it to the desired shape. This can be done by using the \lstinline{Tensor.reshape} function. Note that calling \lstinline{Tensor.reshape} returns a new \lstinline{Tensor} object, and the shape of the original \lstinline{Tensor} object does not change. If one wants to change the shape of the original \lstinline{Tensor}, one should use \lstinline{Tensor.reshape_}.

\begin{lstlisting}[language=python, caption={Reshape a \lstinline{Tensor}.}, label={code:TensorReshape}]
# Create a Tensor with shape=(24), initialized with the numbers
# 0,1,2,...,23
A = cytnx.arange(24)
# Reshape the Tensor to shape=(2,3,4)
B = A.reshape(2,3,4)
# Print and check
print(A.shape()) # Output: [24]
print(B.shape()) # Output: [2, 3, 4]
\end{lstlisting}

\paragraph{Permutation. }
If the order of indices shall be changed, one can call \lstinline{permute}.
Consider a rank-3 \lstinline{Tensor} with \lstinline{shape=(2,3,4)} as an example. This time we want to permute the order of the \lstinline{Tensor} indices from \lstinline{(0,1,2)} to \lstinline{(1,2,0)}. This can be achieved with \lstinline{Tensor.permute}:

\begin{lstlisting}[language=python, caption={Permute the indices of a \lstinline{Tensor}.}, label={code:TensorPermute}]
# Create a Tensor with shape (2,3,4)
A = cytnx.arange(24).reshape(2,3,4)
# Permute the indices such that the new shape is (3,4,2)
B = A.permute(1,2,0)
# Print and Check
print(A.shape()) # Output: [2, 3, 4]
print(B.shape()) # Output: [3, 4, 2]
\end{lstlisting}

\paragraph{Converting to/from a NumPy array.}\label{sec:numpy}
A \lstinline{Tensor} can be  converted to an NumPy array, (see also \cref{sec:to_from_numpy}).
\begin{lstlisting}[language=python, caption={Convert a \lstinline{Tensor} to NumPy.}]
import numpy as np
# Create a Tensor
T = cytnx.ones([3,4])
# Convert to NumPy
nT = T.numpy()
print(type(nT)) # Output: <class 'numpy.ndarray'>
\end{lstlisting}
A conversion from NumPy to Cytnx \lstinline{Tensor} is possible similarly:
\begin{lstlisting}[language=python, caption={Convert a NumPy array to a \lstinline{Tensor}.}]
import numpy as np
# Create a NumPy array
nT = np.ones([3,4])
# Convert to Cytnx
T = cytnx.from_numpy(nT)
print(type(T))  # Output: <class 'cytnx.cytnx.Tensor'>
\end{lstlisting}

\paragraph{Linear Algebra.}
Cytnx provides several tensor operations and linear algebra functions like SVD, QR- and eigenvalue decomposition. These are explained in \cref{sec:decomp} for \lstinline{UniTensor} objects, but can be used in a similar way for a  \lstinline{Tensor}. However, since a \lstinline{Tensor} has no attribute \lstinline{rowrank}, one has to permute and reshape the tensor to bring it into the form of a matrix before calling linear algebra functions that expect matrices as input. See the user guide~\cite{cytnxDoc_LinearAlgebra} for a list of supported operations and more details.

\paragraph{Device.}
As explained in \cref{sec:gpu}, tensors can be stored either in the CPU or GPU memory. If tensors are located on the GPU, one can directly make use of the parallelization and speed up the code execution. The APIs of tensor operations and linear algebra functions work the same way independently of where the tensors are stored. The code examples in \cref{code:GPU_Tensor_declare,code:TensorGPU} demonstrate how to initialize a \lstinline{Tensor} on a specific device and how to move a \lstinline{Tensor} between devices.

\begin{lstlisting}[language=python,
                   caption={Create a \lstinline{Tensor} on a CPU or GPU.},
                   label={code:GPU_Tensor_declare}]
# Create a Tensor on CPU initialized with zeros
A = cytnx.zeros([3,4,5], device=cytnx.Device.cpu)
# Create a tensor on GPU  initialized with zeros
B = cytnx.zeros([3,4,5], device=cytnx.Device.cuda+0)
# Check device
print(A) # on CPU
print(B) # on GPU
\end{lstlisting}

\begin{lstlisting}[language=python,
    caption={Move a \lstinline{Tensor} between CPU and GPU.},
    label={code:TensorGPU}]
A = cytnx.ones([2,2])
B = A.to(cytnx.Device.cuda+0)
print(A) # on CPU
print(B) # on GPU
A.to_(cytnx.Device.cuda)
print(A) # on GPU
\end{lstlisting}

\section{Advanced usage of Cytnx }

\paragraph{Contiguous.}\label{sec:contiguous} 
Here, we explain a more subtle behavior in permuting index order. In Cytnx, the \lstinline{permute} operation does not move the elements in the memory immediately. Only the meta-data is changed. Permuting the index order means relocating the data in memory, which requires copying all the data in a \lstinline{Tensor} or \lstinline{UniTensor}. This can be expensive when the tensor dimensions are large. Instead, we follow an approach that is commonly used, for example in \lstinline{NumPy.array} and \lstinline{torch.tensor}:
When \lstinline{permute} is used, the order of the data is \emph{not} rearranged yet. The actual copying of data to the new layout in memory only happens when \lstinline{Tensor.contiguous} or \lstinline{UniTensor.contiguous} is called. With this design, one can permute the indices several times without significant costs in execution time, and call \lstinline{contiguous} only before an actual calculation is used which depends on the order of the elements -- for example, before a tensor contraction. In practice, users do not need to call the \lstinline{contiguous} command by themselves, as it is automatically done by Cytnx when needed.

We can use \lstinline{UniTensor.is_contiguous} or \lstinline{Tensor.is_contiguous} to check if a tensor is in contiguous form. One can force the contiguous status by calling \lstinline{Tensor.contiguous} or \lstinline{Tensor.contiguous_} (similarly for \lstinline{UniTensor}). However, the user does typically not need to do this explicitly as Cytnx automatically brings tensors to contiguous form when necessary.

\begin{lstlisting}[language=python, caption={Check contiguous form and make contigous.}]
# Create a Tensor with shape (2,3,4)
A = cytnx.arange(24).reshape(2,3,4)
# Check if it is contiguous
print(A.is_contiguous()) # Output: True
# Permute the tensor indices to a new shape (3,2,4)
A.permute_(1,0,2)
# Check if it is contiguous
print(A.is_contiguous()) # Output: False
# Make the tensor contiguous
A.contiguous_()
# Check if it is contiguous
print(A.is_contiguous()) # Output: True
\end{lstlisting}

\paragraph{Storage.}
\lstinline{Storage} is the low-level container that handles the memory allocation and the transfer between different devices. The tensor elements of a \lstinline{Tensor} object are stored inside a \lstinline{Storage} object. 
Typically, users do not directly need to access this object, but it might be useful in special cases. A \lstinline{Storage} object can be created in a similar way as a \lstinline{Tensor}. Note that \lstinline{Storage} does not have the concept of \lstinline{shape}, and is similar to the \lstinline{std::vector} in C++.
The code below in \cref{code:StorageCreate} shows how to create a \lstinline{Storage} with all elements set to zero, and how to change the data type.
One can call \lstinline{Storage.astype} to convert between different data types, similar to the conversion of a \lstinline{Tensor}. For details and available data types, see \cref{sec:tensor_datatypes}.

\begin{lstlisting}[language=Python, label={code:StorageCreate},
    caption={Create and print a \lstinline{Storage}.}]
# Create a Storage with 10 elements with data type Double
A = cytnx.Storage(10, dtype=cytnx.Type.Double)
# Set all the elements to zero
A.set_zeros()
# Change the data type to ComplexDouble
B = A.astype(cytnx.Type.ComplexDouble)
# Print the Storage
print(A)  # Part of output: dtype : Double (Float64)
print(B)  # Part of output: dtype : Complex Double (Complex Float64)
\end{lstlisting}


Internally, the data of a \lstinline{Tensor} is stored in a \lstinline{Storage} object. We can get the \lstinline{Storage} of a \lstinline{Tensor} using \lstinline{Tensor.storage}:

\begin{lstlisting}[language=Python,
    caption={Get \lstinline{Storage} from a \lstinline{Tensor}.}]
# Create a Tensor
A = cytnx.arange(10).reshape(2,5)
# Get its Storage
B = A.storage()
# Check objects
print(A)  # Part of output: dtype : Double (Float64)
print(B)  # Part of output: size  : 10
\end{lstlisting}

When a \lstinline{Tensor} is non-contiguous (see \cref{sec:contiguous}), its meta-data is detached from its memory. The latter is handled by a \lstinline{Storage}. Consequently, calling \lstinline{Tensor.storage} returns its current memory layout, not the permuted order of the \lstinline{Tensor}. The following code example demonstrates this behavior:

\begin{lstlisting}[language=Python,
    caption={Reorder elements of \lstinline{Storage} after \lstinline{permute}.}]
# Create a Tensor
A = cytnx.arange(8).reshape(2,2,2)
# Print its Storage
print(A.storage())        # Storage elements: [ 0 1 2 3 4 5 6 7 ]
# Make it non-contiguous
A.permute_(0,2,1)
print(A.is_contiguous())  # Output: False
# Note that the storage is not changed
print(A.storage())        # Storage elements: [ 0 1 2 3 4 5 6 7 ]
# Make it contiguous, thus, the elements are permuted in memory
A.contiguous_()
print(A.is_contiguous())  # Output: True
# Note that the storage now is changed
print(A.storage())        # Storage elements: [ 0 2 1 3 4 6 5 7 ]
\end{lstlisting}

\paragraph{OpenMP.}

If OpenMP is available, Cytnx uses it for the parallel execution of code on CPUs. If Cytnx is compiled without OpenMP, \lstinline{Device.Ncpus} (see \cref{sec:threadnum}) is always \lstinline{1}.
If OpenMP is enabled, one can change the environment variable \lstinline{OMP_NUM_THREADS} before executing the program to set a restriction on how many threads the program can use. For example, the following code example makes OpenMP and the internal functions of Cytnx use $16$ threads where possible:

\begin{lstlisting}[language=bash, caption={Set the number of threads used by OpenMP.}]
# In the command prompt, before executing a Cytnx program:
export OMP_NUM_THREADS=16
\end{lstlisting}

The speedup when using OpenMP and eight threads instead of only one thread can be seen in the benchmark in \cref{fig:benchmark_cpu}.

\section{Quantum Circuit Simulation\label{sec:qc}}

\begin{figure}[t]
\centering
  \includegraphics[width=0.7\linewidth]{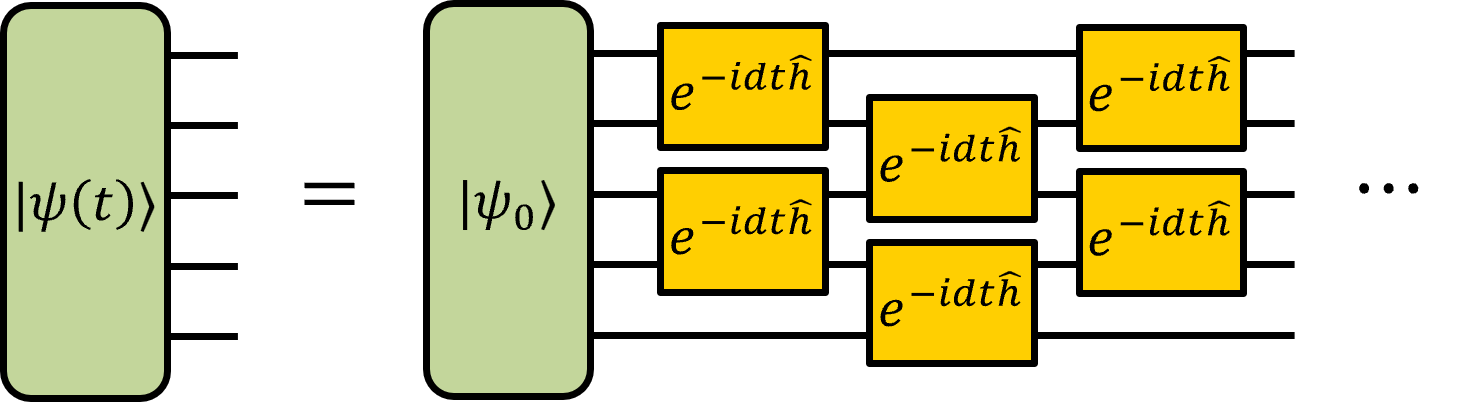}
  \caption{
        Construct the time-evolved state $|\psi(t)\rangle$ using a quantum circuit after applying the Trotter decomposition to the time-evolution operator, where $\hat{h}$ represents the local Hamiltonian of the nearest-neighbor Hamiltonian $\hat{H} = \sum_i{\hat{h}_i\hat{h}_{i+1}}$, and $|\psi_0\rangle$ represents the initial state.} 
  \label{fig:trotter_qc}
\end{figure}

One potential application of tensor networks is to simulate quantum circuits. 
Here, we present an example of using Cytnx to perform statevector simulations of quantum circuits. We consider the transverse-field Ising model (TFIM) with a longitudinal field,
\begin{equation}\label{eq:TFIM_with_longi}
    H=\sum_i J \sigma_i^z \sigma_{i+1}^z+h_x \sigma_i^x+h_z \sigma_i^z,
\end{equation}
where $h_x$ represents the transverse field strength and $h_z$ is the longitudinal field component. 
We want to find  the magnetization of the central site as a function of time. 
To construct the time-evolved state in the quantum circuit, we apply Trotter decomposition. This way, we obtain the real-time evolution unitary quantum gate, as illustrated in \cref{fig:trotter_qc}. After constructing the time-evolved sate $|\psi(t)\rangle$, we measure the magnetization along the $z$-axis of the central qubit, given by $\langle\sigma^z_{\text{central}}\rangle=
 \langle\psi(t)|\sigma^z_{N+1/2}|\psi(t)\rangle$, where $N$ is the number of sites. Note that in the Cyntx library, all quantum gates and states can be represented as \lstinline{UniTensor} objects.
 
 \begin{lstlisting}[language=python,label={code:qc_ex}, caption={An example of code used to create a real-time evolution unitary gate.}]
J, hx, hz = 1.0, 1.0, 3.0 
I = cytnx.eye(2)

# Create Pauli matrices as Tensor objects
sigma_z = cytnx.physics.pauli("z") # Pauli-Z matrix
sigma_x = cytnx.physics.pauli("x") # Pauli-X matrix
print(sigma_z)
print(sigma_x)

''' -------- Output ---------
Total elem: 4
type  : Complex Double (Complex Float64)
cytnx device: CPU
Shape : (2,2)
[[1.00000e+00+0.00000e+00j 0.00000e+00+0.00000e+00j ]
 [0.00000e+00+0.00000e+00j -1.00000e+00+0.00000e+00j ]]

Total elem: 4
type  : Complex Double (Complex Float64)
cytnx device: CPU
Shape : (2,2)
[[0.00000e+00+0.00000e+00j 1.00000e+00+0.00000e+00j ]
 [1.00000e+00+0.00000e+00j 0.00000e+00+0.00000e+00j ]]
'''

# Create a Tensor to represent the local Hamiltonian
H = J*(cytnx.linalg.Kron(sigma_z, sigma_z))
+ 0.5 * (hz*cytnx.linalg.Kron(sigma_z, I) + 
         hz*cytnx.linalg.Kron(I, sigma_z))
+ 0.5 * (hx*cytnx.linalg.Kron(sigma_x, I) + 
         hx*cytnx.linalg.Kron(I, sigma_x))
H.reshape_(2,2,2,2)
# Create a UniTensor to represent the local Hamiltonian
UT = cytnx.UniTensor(H, rowrank=2, labels=['in_up','in_bottom','out_up','out_bottom'])

# Construct the real-time evolution operator eH = exp(a * H + b)
dt = 0.1 
eH = cytnx.linalg.ExpM(UT, a = -dt*1.0j, b = 0)
eH.print_diagram()

''' -------- Output ---------
tensor Name : 
tensor Rank : 4
block_form  : False
is_diag     : False
on device   : cytnx device: CPU
                  ---------     
                 /         \    
   in_up     ____| 2     2 |____ out_up
                 |         |    
   in_bottom ____| 2     2 |____ out_bottom
                 \         /    
                  ---------     
'''

\end{lstlisting}
In Listing~\ref{code:qc_ex}, we demonstrate how to create the local Hamiltonian as shown in \cref{eq:TFIM_with_longi}, and then construct the corresponding real-time evolution gate. The Cytnx library offers a convenient way for users to create Pauli matrices and spin operators. For example, to generate the X-Pauli matrix, a user can simply call \lstinline{cytnx.physics.pauli("x")} which produces the X-Pauli matrix as a \lstinline{Tensor} object (see \cref{sec:tensor} for more details). If users want to create quantum gates directly as \lstinline{UniTensor} objects, they can  utilize the \lstinline{qgates} module, as seen in the following example.
\begin{lstlisting}[language=python,label={code:qc_gate_ex}, caption={}]
gate_z = cytnx.qgates.pauli_z()
gate_z.print_diagram()

''' -------- Output ---------
tensor Name : 
tensor Rank : 2
block_form  : False
is_diag     : False
on device   : cytnx device: CPU
          ---------     
         /         \    
   0 ____| 2     2 |____ 1
         \         /    
          ---------    
'''

\end{lstlisting}

\begin{figure}[t]
\centering
  \includegraphics[width=0.9\linewidth]{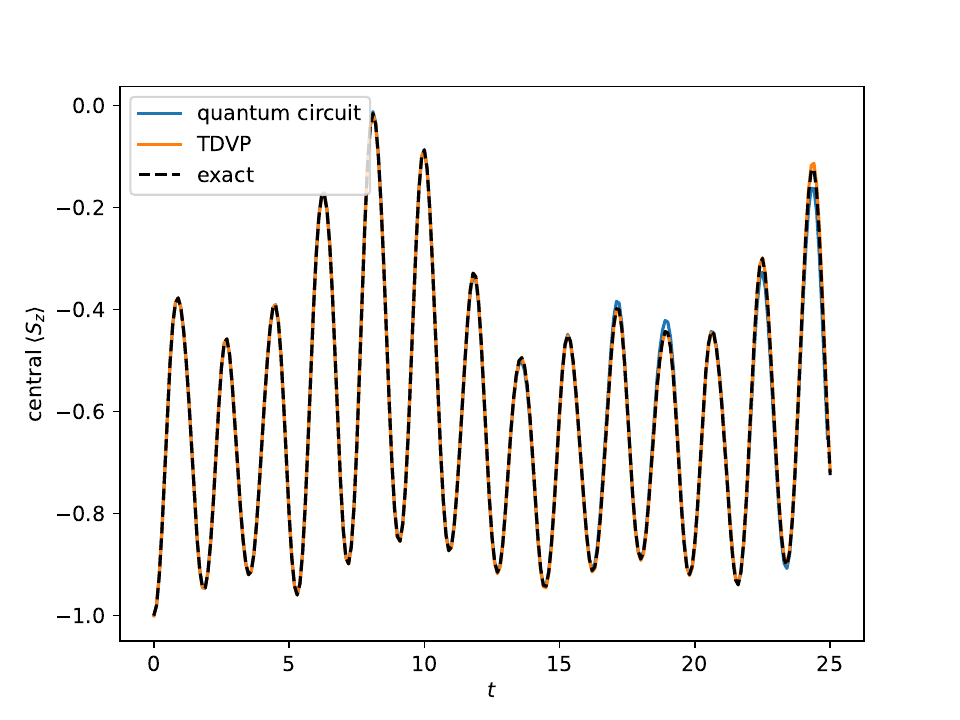}
  \caption{
         Simulation results for the magnetization along the z-axis at the central site, using quantum circuit simulation, exact diagonalization, and TDVP for a system with $N=11$ sites. The initial state $|\uparrow\uparrow
         \downarrow\downarrow\downarrow\downarrow\downarrow\downarrow\downarrow
         \uparrow\uparrow\rangle$ evolves under the Hamiltonian in Eq.\eqref{eq:TFIM_with_longi}, with parameters $J=h_x=1, h_z=3$, and the time step $dt=0.1$.  Simulations are performed using the Cytnx library.
         }
  \label{fig:E8_qc_t}
\end{figure}

The simulation results are shown in Fig.~\ref{fig:E8_qc_t}. We choose  a system with $N=11$ sites, and evolve under the Hamiltonian in Eq.\eqref{eq:TFIM_with_longi} with parameters $J=h_x=1$, $h_z = 3$, and the time step $dt=0.1$. Starting from the initial state $|\uparrow\uparrow
         \downarrow\downarrow\downarrow\downarrow\downarrow\downarrow\downarrow
         \uparrow\uparrow\rangle$, we perform the real-time evolution. Since $N=11$ so we measure the quantity $\langle\sigma^z_6\rangle$.
The blue line represents the statevector result from the quantum circuit simulation. For comparison, we also performed simulations using exact diagonalization (ED) and the time-dependent variational principle (TDVP) algorithm~\cite{finite_TDVP}, with results shown as the black-dashed line and orange line, respectively, in Fig.~\ref{fig:E8_qc_t}. All simulations, including those from the quantum circuit, exact diagonalization, and TDVP, are implemented using the Cytnx library.
 
\bibliography{cytnx.bib}

\nolinenumbers

\end{document}